\documentclass[twoside,11pt]{article}
\usepackage[preprint]{jmlr2e}
\usepackage[utf8]{inputenc}
\usepackage[T1]{fontenc}
\usepackage{amsmath,amssymb,amsfonts,booktabs,longtable,tabularx,array,multirow,enumitem,graphicx,xcolor,float,listings,pifont,adjustbox,textcomp}
\usepackage{cleveref}
\newcommand{\supplementlink}{\url{https://drive.google.com/drive/folders/1I_8AS1szTioU3lcNol2dRiHRXwXhIQaD?usp=sharing}}
\usepackage{newunicodechar}
\newunicodechar{→}{\ensuremath{\rightarrow}}

\hypersetup{hidelinks}
\graphicspath{{figures/}}
\usepackage{etoolbox}
\renewcommand{\topfraction}{0.90}

\renewcommand{\textfraction}{0.10}
\renewcommand{\floatpagefraction}{0.75}
\makeatletter
\let\SciSelectOriginalZTryFC\@ztryfc
\patchcmd{\@ztryfc}{\@testwrongwidth #1}{%
  \@testwrongwidth #1%
  \ifx\@flsucceed\@empty\else\@testtrue\fi
}{}{\PackageError{SciSelect layout}{Cannot apply one-float page rule}{Check the LaTeX kernel float macros.}}
\makeatother

\usepackage{afterpage}
\newcommand{\sciselect}{\textsc{SciSelect-Gym}}

\usepackage{lastpage}
\ShortHeadings{SciSelect-Gym}{Tang, Zhang, Fang, Ng and Chan}
\firstpageno{1}

\expandafter\long\expandafter\def\csname mainfloat@2\endcsname{%
\begin{figure}[t]
\centering
\includegraphics[width=\linewidth]{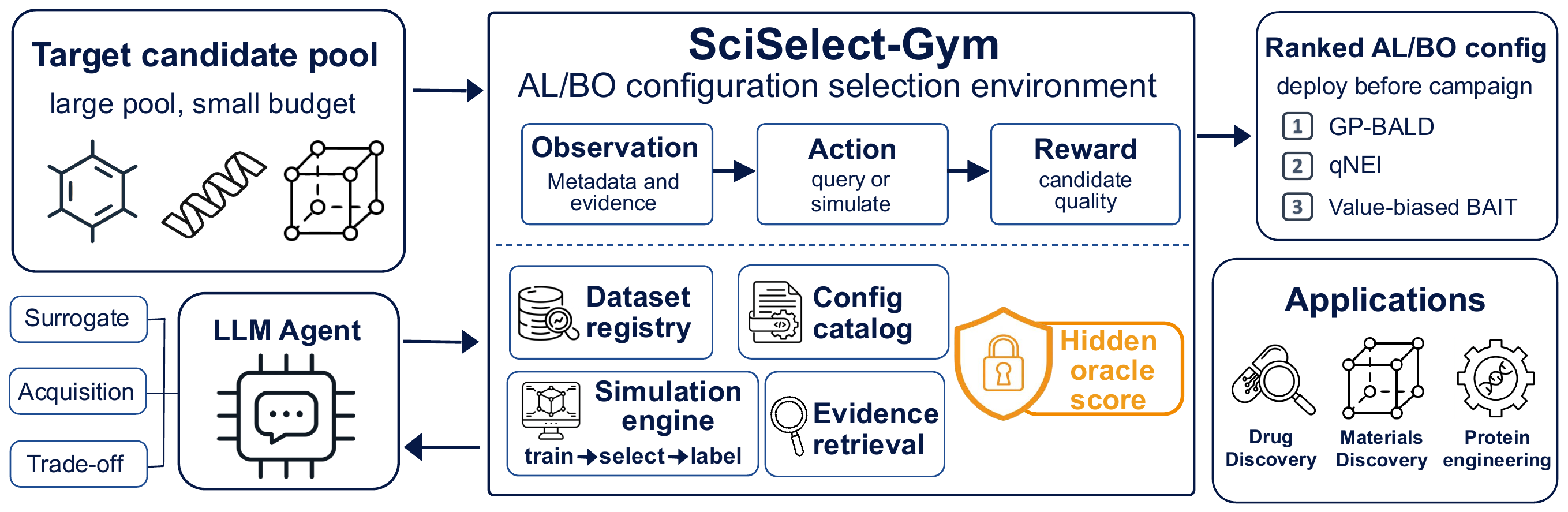}
\caption{The \sciselect{} environment for AL/BO configuration selection. An LLM agent proposes AL/BO configurations. Using dataset metadata and retrieved evidence, it runs each configuration as a simulated campaign (train → select → label) on the target candidate pool. Each run is rewarded by the quality of the candidates. The top-ranked one is deployed in the real experimental campaign.}
\label{fig:framework}
\end{figure}
}

\expandafter\long\expandafter\def\csname mainfloat@5\endcsname{%
\begin{table}[t]
\centering\footnotesize
\setlength{\tabcolsep}{4pt}\renewcommand{\arraystretch}{1.1}
\begin{tabular*}{\linewidth}{@{\extracolsep{\fill}}lrll@{}}
\toprule
Model & Count & Strategy family & Hyperparameter expansion \\
\midrule
XGB & 2 & Greedy mean; uniform random & None \\
XGB & 3 & Value-biased CoreSet & $t\in T$ \\
XGB & 3 & QBC (leaf-index) & $t\in T$ \\
XGB & 9 & QBC-diversity (leaf-index) & $(t,d)\in T^2$ \\\addlinespace[3pt]
NN & 17 & Greedy/random, CoreSet, QBC, QBC-diversity & XGB grid; MC-dropout QBC \\
NN & 3 & BALD & $t\in T$ \\
NN & 9 & Value-biased BAIT, BADGE, LCMD & $t\in T$ per method \\\addlinespace[3pt]
GP & 2 & Greedy mean; uniform random & None \\
GP & 6 & qEI, qNEI, qPI, GP-BALD, MES, GIBBON & None \\
GP & 3 & qUCB & $\beta\in\{0.5,2,8\}$ \\
GP & 3 & qEHVI, qNEHVI, qNParEGO (Pareto) & None \\\addlinespace[3pt]
VBLL & 17 & Greedy/random, CoreSet, QBC, QBC-diversity & XGB grid; predictive QBC \\
VBLL & 3 & BALD & $t\in T$ \\
\bottomrule
\end{tabular*}
\caption{All 80 supported configurations: 17 XGB, 29 NN, 14 GP and 20 VBLL; $T=\{0.1,1,10\}$. BindingDB uses 57 scalar configurations, joint/ADMET use 77, and three GP Pareto rules extend the predicted-XTT catalog to 80.}
\label{tab:configuration_summary}
\end{table}
}

\expandafter\long\expandafter\def\csname mainfloat@6\endcsname{%
\begin{table}[t]
\centering\footnotesize
\setlength{\tabcolsep}{4pt}\renewcommand{\arraystretch}{1.1}
\begin{tabular*}{\linewidth}{@{\extracolsep{\fill}}lrrrrr@{}}
\toprule
\multicolumn{6}{@{}l}{\textbf{(a) Dataset sizes: candidate records}} \\
\addlinespace[3pt]
Study & Datasets & Total & Median & IDR & 95\% range \\
\midrule
BindingDB & 100 & 331{,}187 & 2{,}602 & 1{,}140--6{,}106 & 913--8{,}634 \\
Joint binding--ADMET & 100 & 424{,}055 & 3{,}606 & 2{,}207--7{,}431 & 2{,}048--9{,}960 \\
Standalone ADMET & 34 & 144{,}903 & 4{,}652 & 422--7{,}207 & 366--7{,}932 \\
\bottomrule
\end{tabular*}
\par\vspace{7pt}
\begin{tabularx}{\linewidth}{@{}Xrlll@{}}
\toprule
\multicolumn{5}{@{}l}{\textbf{(b) Campaign settings}} \\
\addlinespace[3pt]
Study & Datasets & Configurations & Initialization & Batch \\
\midrule
BindingDB & 100 & 57 & 10\%, cap 200 & 100 \\
Joint, predicted XTT pools & 90 & 80 & 10\%, cap 200 & 100 \\
Joint, measured ADMET pools & 10 & 77 & 10\%, cap 200 & 100 \\
Objective preferences & 100 & 77 & 10\%, cap 200 & 100 \\
ADMET & 34 & 77 & 10\%, cap 200 & 20 \\
Initial-size sensitivity & 100 & 77 & 200, 400, 800 & 100 \\
Batch-size sensitivity & 100 & 77 & 200 & 20, 40, 80, 100, 160 \\
\bottomrule
\end{tabularx}
\caption{Dataset sizes and campaign settings. Size summaries use archived input records before featurization; totals count records across tasks, not unique molecules. IDR is the 10th--90th percentile interval; the 95\% range is the 2.5th--97.5th percentile interval across dataset sizes (rounded to whole records). Every setting uses five acquisition cycles and three random splits.}
\label{tab:extended_simulation_design_compact}
\end{table}
}

\expandafter\long\expandafter\def\csname mainfloat@7\endcsname{%
\begin{figure}[t]
\centering
\includegraphics[width=\linewidth]{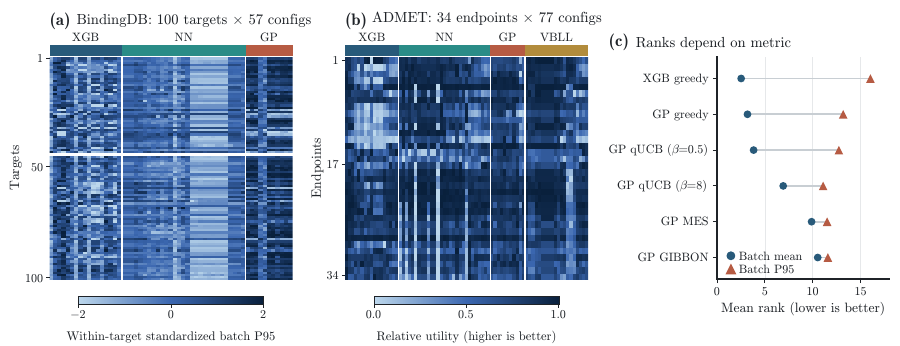}
\caption{BindingDB and ADMET configuration comparisons. (a,b) Within-dataset standardized batch P95 (BindingDB) and relative utility (ADMET) of every configuration. (c) BindingDB configuration ranks under batch mean and batch P95.}
\label{fig:configuration_heterogeneity}\label{fig:bindingdb_metric_choice}
\end{figure}
}

\expandafter\long\expandafter\def\csname mainfloat@8\endcsname{%
\begin{figure}[t]
\centering
\includegraphics[width=\linewidth]{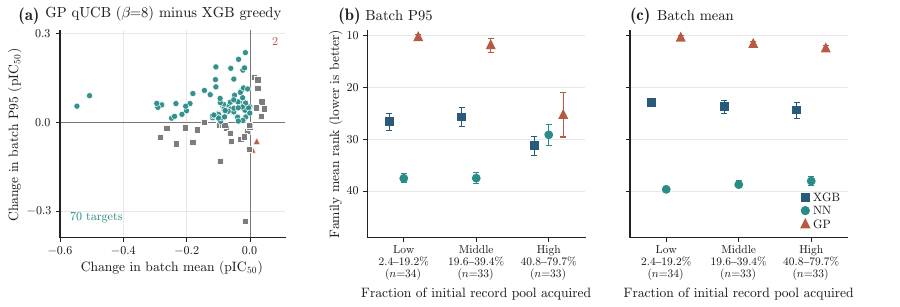}
\caption{BindingDB screening metrics and pool coverage. (a) Within-target changes in batch P95 and batch mean for GP qUCB ($\beta=8$) relative to XGB greedy exploitation. (b,c) Configuration ranks with the fraction of the candidate pool acquired, for batch P95 and batch mean.}
\label{fig:bindingdb_metric_targets}\label{fig:bindingdb_relative_budget}
\end{figure}
}

\expandafter\long\expandafter\def\csname mainfloat@9\endcsname{%
\begin{figure}[t]
\centering
\includegraphics[width=\linewidth]{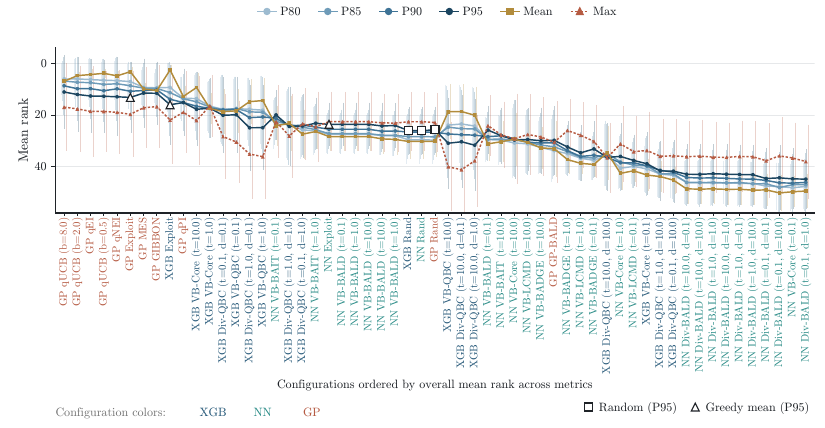}
\caption{BindingDB ranks under six batch statistics (lower is better), ordered by overall mean rank. Bars show between-target standard deviations over 100 datasets; metrics average five cycles and three seeds. Open squares/triangles on P95 mark random sampling/greedy exploitation. Label colors identify surrogate families.}
\label{fig:metric_rank_main}\label{fig:metric_rank}
\end{figure}
}

\expandafter\long\expandafter\def\csname mainfloat@10\endcsname{%
\begin{figure}[t]
\centering
\includegraphics[width=\linewidth]{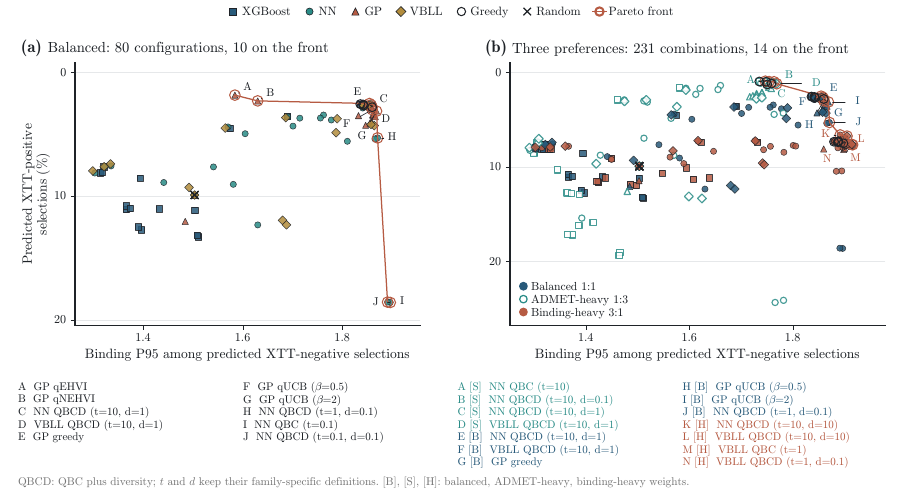}
\caption{Binding-toxicity outcomes on the 90 pools with predicted XTT labels. (a,b) Pareto fronts of dataset-averaged conditional binding P95 and percentage of selections predicted toxic for balanced acquisition (80 configurations) and for three objective preferences (231 combinations), with rings marking the 10- and 14-entry fronts.}
\label{fig:joint_pareto_details}\label{fig:joint_pareto}\label{fig:weight_pareto}\label{fig:weight_pareto_details}
\end{figure}
}

\expandafter\long\expandafter\def\csname mainfloat@11\endcsname{%
\begin{figure}[t]
\centering
\includegraphics[width=\linewidth]{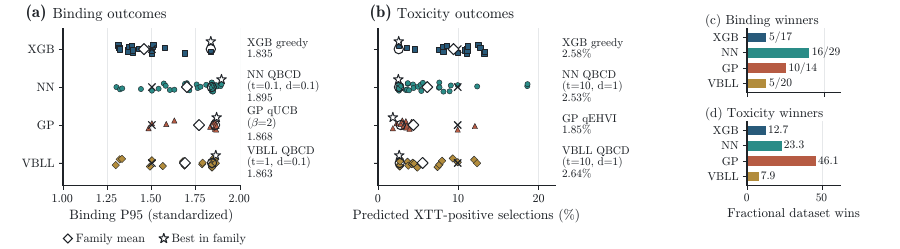}
\caption{Binding and toxicity outcomes by surrogate family on the 90 pools with predicted XTT labels. (a,b) Binding and toxicity outcomes. (c,d) Each family's dataset wins for each outcome, with credit shared among ties.}
\label{fig:multiobjective_landscape}
\end{figure}
}

\expandafter\long\expandafter\def\csname mainfloat@12\endcsname{%
\begin{figure}[t]
\centering
\includegraphics[width=\linewidth]{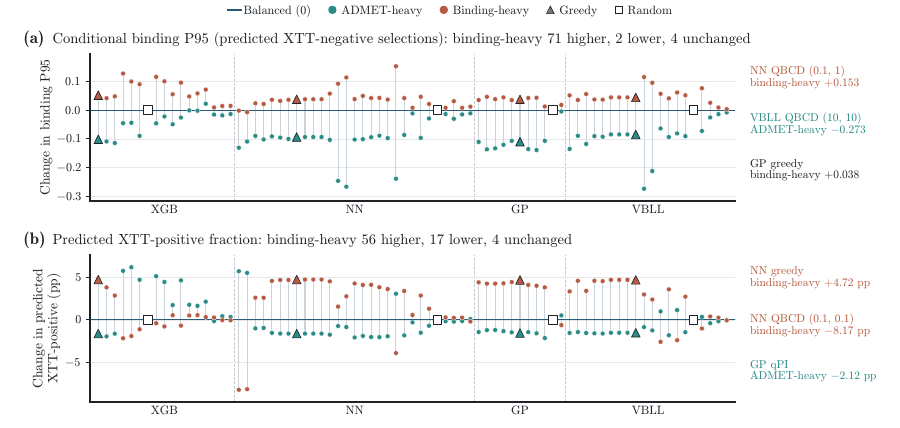}
\caption{Effects of the objective preference. (a,b) Changes from balanced acquisition on the 90 XTT pools in conditional binding P95 and the percentage of selections predicted toxic (in percentage points) for the 77 scalar configurations.}
\label{fig:weight_effects}\label{fig:weight_effects_toxicity}
\end{figure}
}

\expandafter\long\expandafter\def\csname mainfloat@13\endcsname{%
\begin{figure}[t]
\centering
\includegraphics[width=\linewidth]{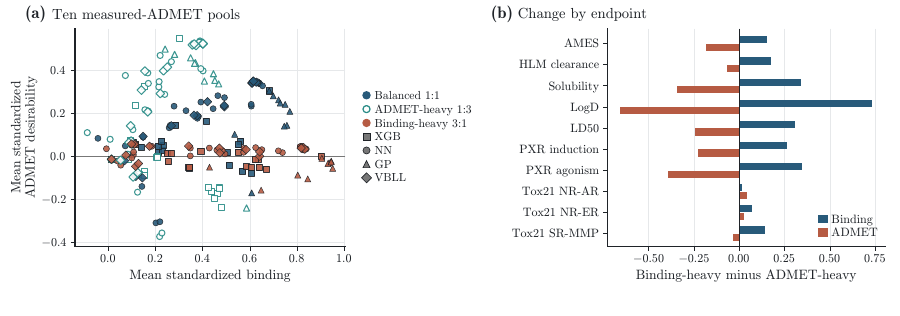}
\caption{Effects of the objective preference on the ten measured-ADMET pools. (a) Mean standardized binding and ADMET desirability for each configuration and preference. (b) Dataset-level change from ADMET-heavy to binding-heavy acquisition, averaged across configurations.}
\label{fig:measured_group_b_preferences}
\end{figure}
}

\expandafter\long\expandafter\def\csname mainfloat@14\endcsname{%
\begin{figure}[t]
\centering
\includegraphics[width=\linewidth]{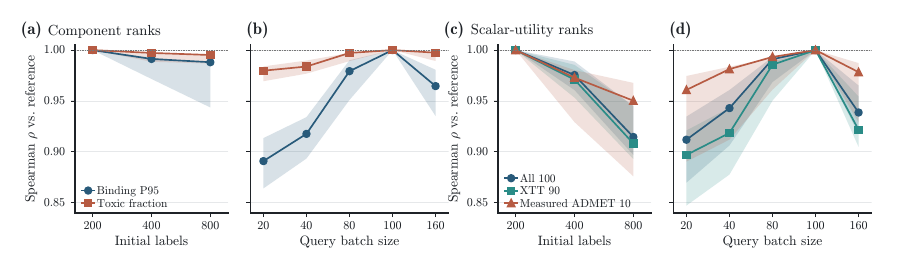}
\caption{Configuration ranks under different acquisition budgets. (a,b) Spearman rank correlations with the $(200,100)$ reference as the initial labeled-set size (a) and the batch size (b) vary, for conditional binding P95 and predicted-toxic fraction on the 90 XTT pools; (c,d) the same for batch P95 of the weighted binding-ADMET score on all 100 pools and each subset.}
\label{fig:al-knob-sensitivity}\label{fig:current_joint_scalar_budget}
\end{figure}
}

\expandafter\long\expandafter\def\csname mainfloat@15\endcsname{%
\begin{figure}[t]
\centering
\includegraphics[width=\linewidth]{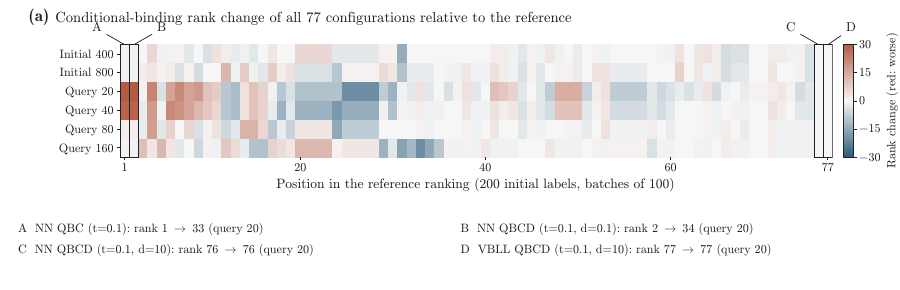}
\caption{Conditional-binding rank change of every configuration relative to the $(200,100)$ reference.}
\label{fig:budget_rank_changes}
\end{figure}
}

\expandafter\long\expandafter\def\csname mainfloat@16\endcsname{%
\begin{figure}[t]
\centering
\includegraphics[width=\linewidth]{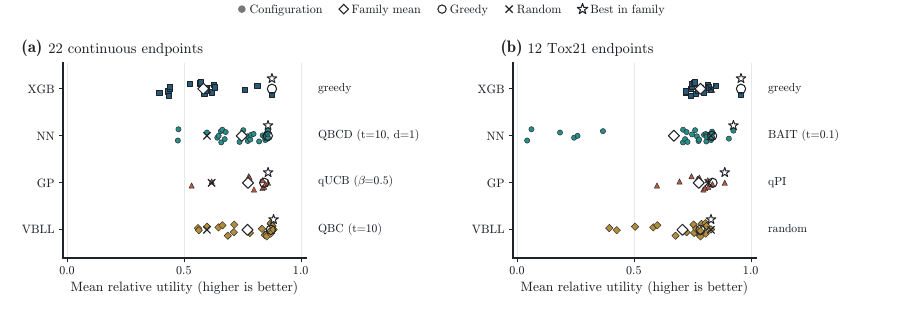}
\caption{ADMET configuration performance by endpoint type. (a,b) Mean relative utility of the 77 scalar configurations on the 22 continuous and 12 Tox21 endpoints, with each family's best rule starred.}
\label{fig:admet_endpoint_atlas}
\end{figure}
}

\expandafter\long\expandafter\def\csname mainfloat@17\endcsname{%
\begin{figure}[t]
\centering
\includegraphics[width=\linewidth]{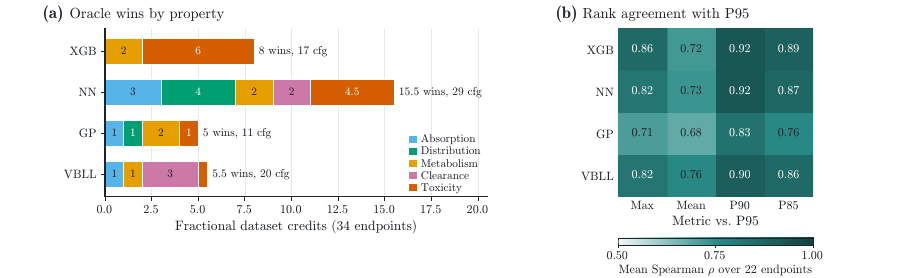}
\caption{ADMET property categories and screening metrics. (a) Property categories of the datasets on which each surrogate family achieves the highest utility. (b) Rank agreement of Max, Mean, P90 and P85 with P95.}
\label{fig:admet_property_metrics}
\end{figure}
}

\expandafter\long\expandafter\def\csname mainfloat@19\endcsname{%
\begin{figure}[t]
\centering
\includegraphics[width=\linewidth]{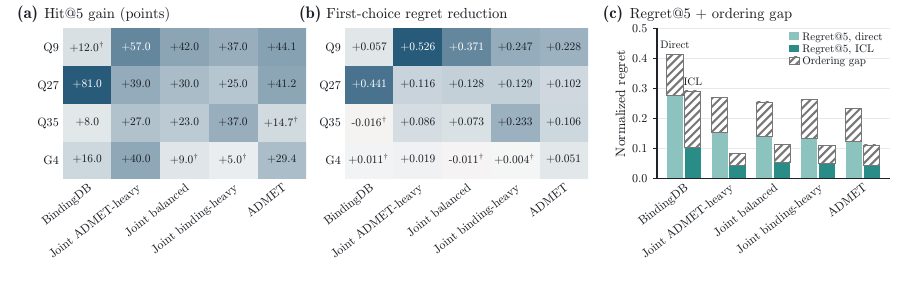}
\caption{Recommendation quality across the five settings. (a,b) Paired change from direct prediction to ICL with 20 sources in Hit@5 and first-choice regret, where positive values favor ICL and daggers mark 95\% intervals that include zero; (c) first-choice regret split into Regret@5 and the ordering gap.}
\label{fig:llm_usefulness}\label{fig:ordering_gap}
\end{figure}
}

\expandafter\long\expandafter\def\csname mainfloat@20\endcsname{%
\begin{table}[t]
\centering\footnotesize
\setlength{\tabcolsep}{1.7pt}
\begin{tabular}{@{}l@{\hspace{3pt}}rr@{\hspace{3pt}}rr@{\hspace{3pt}}rr@{\hspace{3pt}}rr@{\hspace{3pt}}rr@{}}
\toprule
& \multicolumn{2}{c}{BindingDB} & \multicolumn{2}{c}{ADMET} & \multicolumn{2}{c}{\shortstack{Joint\\balanced}} & \multicolumn{2}{c}{\shortstack{Joint\\binding-heavy}} & \multicolumn{2}{c}{\shortstack{Joint\\ADMET-heavy}} \\
\cmidrule(lr){2-3}\cmidrule(lr){4-5}\cmidrule(lr){6-7}\cmidrule(lr){8-9}\cmidrule(lr){10-11}
Method & Hit@5 & Regret & Hit@5 & Regret & Hit@5 & Regret & Hit@5 & Regret & Hit@5 & Regret \\
\midrule
\multicolumn{11}{l}{\textit{A. Reference methods}} \\
Random & 37.9 & 0.581 & 29.2 & 0.281 & 29.2 & 0.348 & 29.2 & 0.351 & 29.2 & 0.320 \\
Global Best & 74.0 & 0.291 & 41.2 & 0.097 & 53.0 & \textbf{0.082} & 62.0 & \textbf{0.092} & \textbf{82.0} & 0.055 \\
Frequency Voting & \textbf{85.0} & 0.377 & 61.8 & 0.097 & \textbf{63.0} & 0.138 & \textbf{72.0} & 0.126 & 76.0 & 0.055 \\
Most-Similar Source & 79.0 & 0.293 & 47.1 & 0.139 & 45.0 & 0.115 & 53.0 & 0.138 & 59.0 & 0.084 \\
K20 rank aggregation & 79.0 & \textbf{0.271} & \textbf{73.5} & \textbf{0.092} & 62.0 & 0.090 & \textbf{72.0} & 0.096 & 79.0 & \textbf{0.050} \\
\addlinespace[1.5pt]
\multicolumn{11}{l}{\textit{B. Direct prediction}} \\
Qwen3.5-9B & 58.0 & 0.396 & 17.6 & 0.361 & 8.0 & 0.519 & 16.0 & 0.377 & 2.0 & 0.681 \\
Qwen3.8-27B & 3.0 & 0.720 & 26.5 & 0.207 & 33.0 & 0.238 & 31.0 & 0.236 & 36.0 & 0.177 \\
Qwen3.6-35B-A3B & \textbf{75.0} & 0.276 & \textbf{47.1} & 0.200 & 36.0 & 0.174 & 27.0 & 0.339 & \textbf{46.0} & 0.147 \\
Gemma-4-26B-A4B-it & 69.0 & \textbf{0.270} & 26.5 & \textbf{0.170} & \textbf{56.0} & \textbf{0.093} & \textbf{57.0} & \textbf{0.104} & 34.0 & \textbf{0.084} \\
\addlinespace[1.5pt]
\multicolumn{11}{l}{\textit{C. In-context learning ($K=20$ similar sources)}} \\
Qwen3.5-9B & 70.0 & 0.340 & 61.8 & 0.132 & 50.0 & 0.148 & 53.0 & 0.130 & 59.0 & 0.155 \\
Qwen3.8-27B & 84.0 & 0.280 & \textbf{67.6} & 0.105 & 63.0 & 0.109 & 56.0 & 0.108 & \textbf{75.0} & \textbf{0.061} \\
Qwen3.6-35B-A3B & 83.0 & 0.291 & 61.8 & \textbf{0.094} & 59.0 & \textbf{0.101} & \textbf{64.0} & 0.105 & 73.0 & \textbf{0.061} \\
Gemma-4-26B-A4B-it & \textbf{85.0} & \textbf{0.260} & 55.9 & 0.119 & \textbf{65.0} & 0.104 & 62.0 & \textbf{0.099} & 74.0 & 0.064 \\
\bottomrule
\end{tabular}
\caption{Configuration recommendations across the five settings: Hit@5 (\%, higher is better) and first-choice normalized regret (lower is better). Panels compare reference methods, direct prediction, and ICL with 20 similarity-selected sources, and bold marks tied best values within each panel. Hit@1, Hit@3, mean rank and BestRk@5 are in \Cref{tab:llm_full_results,tab:llm_joint_results}.}
\label{tab:main_llm_summary}\label{tab:selection_summary}
\end{table}
}

\expandafter\long\expandafter\def\csname mainfloat@21\endcsname{%
\begin{figure}[t]
\centering
\includegraphics[width=\linewidth]{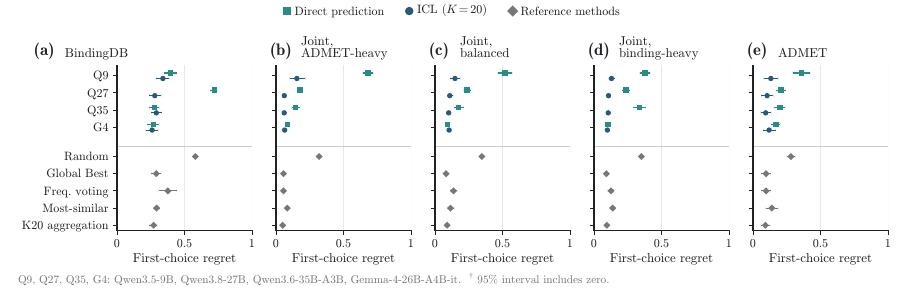}
\caption{First-choice regret of each language model with and without ICL and of the reference methods. Panels (a-e) show BindingDB, joint ADMET-heavy, joint balanced, joint binding-heavy and ADMET.}
\label{fig:llm_absolute}
\end{figure}
}

\expandafter\long\expandafter\def\csname mainfloat@22\endcsname{%
\begin{table}[t]
\centering\footnotesize\setlength{\tabcolsep}{2.2pt}
\begin{tabularx}{\linewidth}{@{}l*{5}{>{\centering\arraybackslash}X}@{}}
\toprule
Method & BindingDB & ADMET & \shortstack{Joint\\balanced} & \shortstack{Joint\\binding-heavy} & \shortstack{Joint\\ADMET-heavy} \\
\midrule
\addlinespace[2pt]
GB & 0.291 & 0.097 & 0.082 & 0.092 & 0.055 \\
FV & 0.377 & 0.097 & 0.138 & 0.126 & 0.055 \\
K20 & 0.271 & 0.092 & 0.090 & 0.096 & 0.050 \\
R-All & 0.250 & 0.093 & 0.089 & 0.092 & 0.056 \\
R-K20 & 0.237 & 0.084 & 0.103 & 0.089 & 0.053 \\
ICL-Q9 & 0.340 & 0.132 & 0.148 & 0.130 & 0.155 \\
ICL-Q27 & 0.280 & 0.105 & 0.109 & 0.108 & 0.061 \\
ICL-Q35 & 0.291 & 0.094 & 0.101 & 0.105 & 0.061 \\
ICL-G4 & 0.260 & 0.119 & 0.104 & 0.099 & 0.064 \\
\bottomrule
\end{tabularx}
\caption{Held-out configuration selection. Mean first-choice regret of reference methods, ridge selectors and ICL; lower is better.}
\label{tab:learned_comparison}
\end{table}
}

\expandafter\long\expandafter\def\csname mainfloat@23\endcsname{%
\begin{table}[t]
\centering\footnotesize\setlength{\tabcolsep}{2.2pt}
\begin{tabularx}{\linewidth}{@{}l*{5}{>{\centering\arraybackslash}X}@{}}
\toprule
Method & BindingDB & ADMET & \shortstack{Joint\\balanced} & \shortstack{Joint\\binding-heavy} & \shortstack{Joint\\ADMET-heavy} \\
\midrule
\addlinespace[2pt]
GB & 0.197 & 0.075 & 0.049 & 0.051 & 0.023 \\
FV & 0.097 & 0.051 & 0.036 & 0.035 & 0.022 \\
K20 & 0.154 & 0.034 & 0.043 & 0.032 & 0.024 \\
R-All & 0.119 & 0.050 & 0.045 & 0.045 & 0.023 \\
R-K20 & 0.102 & 0.038 & 0.041 & 0.040 & 0.022 \\
ICL-Q9 & 0.140 & 0.062 & 0.087 & 0.066 & 0.110 \\
ICL-Q27 & 0.097 & 0.035 & 0.042 & 0.048 & 0.026 \\
ICL-Q35 & 0.095 & 0.039 & 0.043 & 0.044 & 0.024 \\
ICL-G4 & 0.091 & 0.047 & 0.040 & 0.041 & 0.023 \\
\bottomrule
\end{tabularx}
\caption{Held-out configuration selection. Mean best-list regret (Regret@5) of reference methods, ridge selectors and ICL; lower is better.}
\label{tab:learned_best_list}
\end{table}
}

\expandafter\long\expandafter\def\csname mainfloat@24\endcsname{%
\begin{table}[t]
\centering\footnotesize
\setlength{\tabcolsep}{3pt}\renewcommand{\arraystretch}{1.15}
\begin{tabularx}{\linewidth}{@{}>{\raggedright\arraybackslash}p{0.18\linewidth}*{5}{>{\centering\arraybackslash}X}@{}}
\toprule
\multicolumn{6}{@{}l}{\textbf{(a) Recommendation ordering}} \\
\addlinespace[3pt]
Metric & BindingDB & ADMET & \shortstack{Joint\\balanced} & \shortstack{Joint\\binding-heavy} & \shortstack{Joint\\ADMET-heavy} \\
\midrule
\shortstack[l]{Source reranking\\$\Delta R_1$} & \shortstack{$-0.018$\\{\scriptsize$[-0.053, 0.015]$}} & \shortstack{$-0.014$\\{\scriptsize$[-0.036, 0.010]$}} & \shortstack{$-0.010$\\{\scriptsize$[-0.020, -0.001]$}} & \shortstack{$-0.009$\\{\scriptsize$[-0.017, -0.001]$}} & \shortstack{$-0.010$\\{\scriptsize$[-0.015, -0.005]$}} \\
\bottomrule
\end{tabularx}
\par\vspace{7pt}
\begin{tabularx}{\linewidth}{@{}>{\raggedright\arraybackslash}p{0.18\linewidth}*{5}{>{\centering\arraybackslash}X}@{}}
\toprule
\multicolumn{6}{@{}l}{\textbf{(b) Campaign repeatability: two-seed choice}} \\
\addlinespace[3pt]
Metric & BindingDB & ADMET & \shortstack{Joint\\balanced} & \shortstack{Joint\\binding-heavy} & \shortstack{Joint\\ADMET-heavy} \\
\midrule
Regret reduction & \shortstack{$0.070$\\{\scriptsize$[0.047, 0.093]$}} & \shortstack{$0.018$\\{\scriptsize$[-0.009, 0.046]$}} & \shortstack{$0.024$\\{\scriptsize$[0.015, 0.036]$}} & \shortstack{$0.021$\\{\scriptsize$[0.008, 0.035]$}} & \shortstack{$0.013$\\{\scriptsize$[0.007, 0.018]$}} \\
\addlinespace[4pt]
\shortstack[l]{Hindsight gap\\transferred (\%)} & \shortstack{$33.7$\\{\scriptsize$[24.4, 41.8]$}} & \shortstack{$23.7$\\{\scriptsize$[-14.4, 48.8]$}} & \shortstack{$30.2$\\{\scriptsize$[19.6, 40.7]$}} & \shortstack{$27.8$\\{\scriptsize$[11.6, 43.2]$}} & \shortstack{$24.4$\\{\scriptsize$[13.3, 35.8]$}} \\
\bottomrule
\end{tabularx}
\caption{Held-out configuration selection. Ordering and repeatability controls averaged over the four ICL models, with paired descriptive 95\% held-out-fold intervals in brackets.}
\label{tab:main_controls}
\end{table}
}

\expandafter\long\expandafter\def\csname mainfloat@25\endcsname{%
\begin{figure}[t]
\centering
\includegraphics[width=\linewidth]{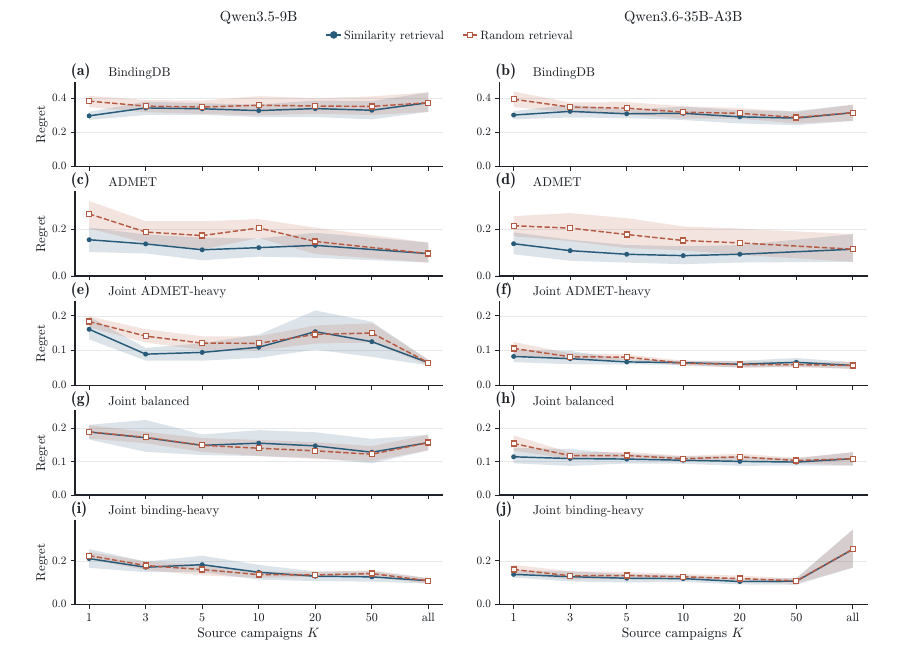}
\caption{Similarity versus random retrieval as the number of source campaigns $K$ grows, for Qwen3.5-9B (left) and Qwen3.6-35B-A3B (right). Rows show BindingDB, ADMET and the three joint preferences (a-j), and lower first-choice regret is better. Lines are means with descriptive 95\% held-out-fold bootstrap bands.}
\label{fig:llm_sources}\label{fig:llm_sources_joint}\label{fig:llm_sources_admet}\label{fig:llm_sources_joint_balanced}\label{fig:llm_sources_joint_binding}
\end{figure}
}

\expandafter\long\expandafter\def\csname mainfloat@26\endcsname{%
\begin{figure}[t]
\centering
\includegraphics[width=\linewidth]{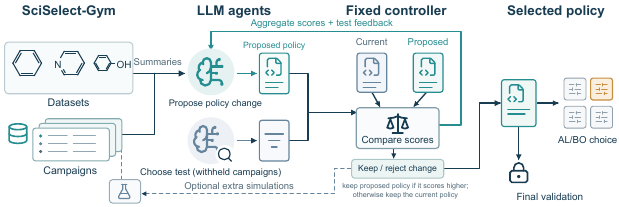}
\caption{Agent interaction with \sciselect{}. Two LLM roles propose a configuration-selection policy and choose a test based on withholding source campaigns; a fixed controller evaluates the proposal using campaign records or permitted simulations and returns aggregate feedback to the proposing LLM. Final validation is kept separate from policy development.}
\label{fig:agent_use_case}
\end{figure}
}

\newcommand{\PlaceMainPageFloat}{%
  \ifcsname mainfloat@\number\value{page}\endcsname
    \csname mainfloat@\number\value{page}\endcsname
  \fi
  \ifnum\value{page}<26\relax
    \afterpage{\PlaceMainPageFloat}%
  \fi}

\begin{document}
\title{SciSelect-Gym: An Environment and Benchmark for Active Learning and Bayesian Optimization Configuration Selection}
\author{\name Ying-Peng Tang \\
\name Tianle Zhang \\
\name Wanlong Fang \\
\name Ng Jing Xiang \\
\name Alvin Chan \email guoweialvin.chan@ntu.edu.sg \\
\addr 
Nanyang Technological University\\50 Nanyang Avenue, Singapore 639798\\Corresponding author: Alvin Chan}
\maketitle
\PlaceMainPageFloat
\begin{abstract}%
  In many scientific tasks, scientists repeatedly choose a small set of candidates to test, learn from the results, and use the observations to guide the next selection. Methods based on active learning~(AL) or Bayesian optimization~(BO) have been proposed for these iterative screening campaigns with a surrogate model. Yet, selecting the surrogate model and acquisition function for a new dataset remains challenging due to limited empirical evidence, while the effectiveness of large language models (LLMs) in making such decisions remains unclear. To address these gaps, we introduce \sciselect{}, an environment for simulating AL/BO data selection campaigns and evaluating configuration choices.
We conducted simulations of AL/BO data selection campaigns across several settings and more than 100 molecular datasets. 
Using completed simulation campaigns as context improves LLM recommendation shortlists on held-out datasets more consistently than it improves the quality of their first choice.
\sciselect{} provides a simulation platform and recommendation protocols for learning how to choose configurations before an acquisition campaign. An implemented agent use case connects these campaign records and simulation tools to LLM-generated configuration-selection policies. The code is provided as supplementary material.
\end{abstract}

\section{Introduction}
Many scientific discovery campaigns begin with the same practical dilemma: a team has a large pool of candidates, but only enough experimental or computational budget to test a small batch at a time. In drug discovery, for example, the candidates may be molecules for a target protein. After each batch is evaluated, the new measurements are used to select the next batch. Active learning (AL) and Bayesian optimization (BO) provide principled ways to run this loop \citep{settles2009active,shahriari2016taking,garnett2023bayesian}, using a surrogate model to guide the next selection. The acquisition rule determines how predictions and uncertainty are translated into a batch of candidates. The configuration is evaluated by the measured properties of the selected candidates, with surrogate prediction accuracy assessed separately. For screening, a useful comparison must also specify whether success means finding a few exceptional candidates, improving the average quality of a batch, or satisfying several property objectives.

The hard choice often comes before the loop starts. A scientist must decide which AL/BO configuration will govern the campaign, including the choice of surrogate model, acquisition rule, and the exploration-exploitation trade-off. This choice can determine whether early experimental rounds concentrate on useful regions of the search space or waste budget on uninformative candidates. However, this choice may be guided by convention, familiarity, or a small pilot study rather than a systematic comparison of configurations. A configuration that works well for one target may be less effective on another, and changing the acquisition budget or the relative importance of two properties can change the preferred choice. This makes configuration selection a learning problem in its own right: the input is a description of the new campaign, and the output is the configuration to use within that campaign. Comparing configurations on completed datasets provides the performance records needed to study this choice.

Large language models (LLMs) can use dataset descriptions to recommend AL/BO configurations. Evaluating these recommendations requires campaign records that show how each configuration performs. Completed campaigns provide examples of which configurations worked for different targets, molecular libraries, and objectives. A standardized environment is needed to evaluate whether LLMs can use these examples and simulation feedback to make better configuration choices before an expensive discovery campaign begins. The relevant evidence includes both the properties of a source dataset and the outcomes obtained by the configurations evaluated on that dataset. A model must decide which past campaigns are informative for the new task and how to use their results. Recommendation quality also has two practical meanings. A shortlist can help a researcher identify alternatives for a pilot study, whereas a campaign that immediately adopts the first recommendation depends on the quality of that single choice. Evaluating both makes it possible to determine how historical evidence changes these decisions.

To this end, we introduce \sciselect{}, an interactive environment for LLM agents to select and evaluate AL/BO configurations. Offline simulations record the property scores of candidates selected by each configuration. Agents can inspect target dataset descriptions, retrieve completed source campaigns, request additional source simulations, and rank configurations for a new target. Completed source campaigns provide evidence for selecting a configuration for the target campaign.

To provide completed campaigns for the environment, we conduct four studies using drug discovery as an application domain. In drug discovery, identifying strong target binders is a first step, motivating single-objective comparisons on 100 BindingDB-derived datasets \citep{gilson2016bindingdb}. Drug discovery also requires evaluating absorption, distribution, metabolism, excretion, and toxicity (ADMET). Our cross-property study covers 34 ADMET endpoints to examine how configuration performance varies across molecular properties beyond binding. These properties must also be considered together: a strong binder may be toxic, creating competing objectives for candidate selection. We therefore evaluate 100 joint binding-ADMET datasets with varying objective weights to examine how the preferred configuration changes with these priorities. Finally, projects differ in their available measurements and experimental budgets. We vary the initial labeled-set and query batch sizes to test whether configuration rankings remain stable under different query settings. Together, these studies examine how configuration choice depends on the target, property objectives, and budget. Although evaluated here in drug discovery, \sciselect{} could extend to other scientific discovery domains with suitable candidate datasets and objectives. The resulting campaign records retain the candidates selected at each cycle and their screening scores, enabling comparisons against the same completed experiments.
We then use completed BindingDB, joint binding-ADMET, and ADMET campaigns to compare direct prediction and in-context learning (ICL) for four LLMs. These comparisons evaluate how campaign records help LLMs identify useful configurations and place the strongest recommendation first. Across the four models and five recommendation settings, adding source campaigns improves mean shortlist quality in all 20 comparisons and reduces mean first-choice regret in 18. Source-based aggregation and ridge regression provide competitive reference methods. The ICL study and reference methods compare different ways of using the same campaign evidence. Controls that reorder the same five recommendations isolate the effect of list ordering; evaluation on a held-out campaign seed tests whether selecting the best listed configuration on previous campaign runs improves performance on a new run. These results make the resource useful for studying both the transfer of campaign evidence and the decision to deploy one configuration.

\section{Related Work}
\label{sec:related}

\paragraph{Scientific datasets and acquisition benchmarks.}
Scientific machine-learning benchmarks define datasets, objectives, metrics, and reference methods so that approaches can be compared on the same tasks. MoleculeNet and the Therapeutics Data Commons organize molecular property prediction and drug-discovery tasks, including ADMET and target interactions~\citep{wu2018moleculenet,huang2021tdc}. Related resources examine transfer, distribution shift, and activity cliffs~\citep{stanley2021fsmol,ji2023drugood,vantilborg2022activity,DBLP:conf/nips/0001F0C22}, while GuacaMol and Molecular Sets (MOSES) evaluate molecular generation and goal-directed design~\citep{brown2019guacamol,polykovskiy2020moses}. For sequential acquisition, AL and BO choose candidates using a model updated from previous measurements~\citep{settles2009active,warmuth2003active,shahriari2016taking,garnett2023bayesian}. Benchmarks compare query strategies and learning algorithms across datasets and budgets, including batch-mode active learning and ALPBench~\citep{holzmuller2023bmdal,margraf2024alpbench}. HPOBench standardizes hyperparameter-optimization evaluations, and Design-Bench studies offline model-based optimization~\citep{eggensesperger2021hpobench,trabucco2022designbench}. Studies in materials and molecular discovery further compare acquisition policies and surrogate models on scientific data~\citep{liang2021benchmarking,wang2022benchmarkingal,gorantla2024benchmarking,yin2023evaluating}. These studies establish how acquisition procedures perform. \sciselect{} uses such campaign outcomes to study the preceding decision: choosing a complete AL/BO configuration for a new dataset and screening objective.

\paragraph{Algorithm selection and learning from previous evaluations.}
Algorithm selection maps information about a problem instance to an algorithm or configuration~\citep{rice1976algorithm,hutter2019automated}. ASlib supports this task with instance features and algorithm-performance records~\citep{bischl2016aslib}; HPO-B makes historical evaluations and fixed transfer protocols available for hyperparameter optimization~\citep{arango2021hpob}. Optimization packages such as SMAC3 and BoFire provide configurable procedures that can be deployed after this choice~\citep{lindauer2022smac3,duerholt2025bofire}. Language models can also supply representations for selection: AS-LLM uses language-model features to learn algorithm representations for supervised algorithm selection~\citep{wu2024asllm}. \sciselect{} applies algorithm selection to scientific acquisition campaigns with fixed budgets. Each source dataset supplies its description and the outcomes of the tested configurations, and target datasets are held out when fitting or prompting a selector. This supports comparisons among methods that aggregate source configuration ranks, supervised metadata-based selectors, and prompted LLMs under the same information restrictions. Evaluating both the ordered shortlist and its first entry then distinguishes finding useful alternatives from choosing the configuration to deploy.

\paragraph{LLM-assisted optimization and interactive environments.}
LLMs and adaptive selection methods can use observations from an ongoing campaign to guide optimization. LLAMBO uses language models for warm starting, surrogate modeling, and candidate proposal~\citep{liu2024llambo}. TAILOR switches among active-learning strategies, BOOST evaluates kernel-acquisition pairs using observed target points, and LMABO selects acquisition functions from the current optimization state~\citep{zhang2023tailor,park2026boost,ngo2026lmabo}. These methods adapt decisions as target observations become available. The recommendation task in \sciselect{} instead uses completed source campaigns to choose a configuration before observing the target campaign's outcomes. Interactive benchmarks provide another relevant comparison: Memory Gym evaluates agent memory, while Science-Gym studies experiment selection and equation discovery~\citep{pleines2025memorygym,cerrato2026sciencegym}. The \sciselect{} interface makes scientific configuration-selection tasks available through dataset inspection, source retrieval, optional source simulations, and recommendation scoring. Its empirical studies compare direct prediction with source-informed recommendations, and the controls for recommendation ordering and campaign repeatability examine the quality of the first configuration selected from each shortlist.

\section{Simulation Environment and Campaign Results}
\label{sec:simulation_results}

We first examine how the screening objective changes the preferred configuration, then compare binding-ADMET trade-offs, acquisition budgets and transfer across ADMET endpoints.

\subsection{Common Experimental Design}
\label{sec:benchmark_overview}

Let $D=\{(x_i,y_i)\}_{i=1}^{n}$ denote a target scientific dataset, where $x_i$ describes a candidate and its optional context, and $y_i$ is its measured or predicted property score. In drug discovery, for example, $x_i$ may represent a ligand together with a fixed protein target and $y_i$ may be a transformed binding score; in materials or catalyst discovery, $x_i$ may encode a candidate composition and $y_i$ an experimentally measured property.

At the start of an acquisition campaign, the environment splits the data into an initially labeled set $L_0$, an unlabeled pool $U_0$, and a held-out diagnostic set $H$. At cycle $t$, a surrogate model is trained on $L_t$, an acquisition rule selects a batch $B_t \subset U_t$, and the simulation reveals the recorded measured or predicted labels for that batch. The labeled and unlabeled sets are then updated before the next cycle.
The labels can be continuous, binary, or multivariate, covering regression, classification, and multi-objective selection.

The benchmark covers three drug-discovery settings: binding scores on 100 BindingDB datasets, joint binding and ADMET objectives on 100 candidate pools, and 34 individual ADMET endpoints. BindingDB configurations combine XGBoost (XGB)~\citep{chen2016xgboost}, neural networks (NN), or Gaussian processes (GP)~\citep{rasmussen2006gaussian} with acquisition rules. The joint and ADMET studies also include variational Bayesian last-layer models (VBLL)~\citep{harrison2024vbll}; their inputs use molecular fingerprints without protein features. BindingDB uses 57 configurations: 17 XGB, 29 NN and 11 GP. The extended catalog for scalar objectives adds 20 VBLL configurations, giving 77. Each scalar configuration fits a single property score or a weighted sum of property scores. We compare complete configurations, including their representations and fitting procedures. Appendix~\ref{app:configs} defines the acquisition rules and parameter grids.

\Cref{tab:configuration_summary} summarizes all supported acquisition configurations.
The catalog includes random sampling, greedy exploitation, committee-based and diversity-based rules, and GP acquisition rules based on confidence bounds, improvement or information gain. For example, expected improvement (EI), noisy EI (NEI) and probability of improvement (PI) favor candidates likely to improve on previous outcomes, while batch active learning via information matrices (BAIT) seeks informative batches~\citep{jones1998efficient,ash2021bait}. CoreSet and the other batch-mode active-learning rules add coverage and diversity to the comparison.

Each campaign runs for five acquisition cycles with three random data splits. Before the initialization cap is applied, 10\% of records initialize the model, 70\% form the acquisition pool, and 20\% are reserved for surrogate diagnostics. We average repetitions within a dataset before comparing datasets. Batch mean is the mean property score of the selected candidates, and the 95th percentile (P95) of their scores emphasizes candidates with high scores. These statistics are computed within each batch and then averaged over the stated cycles. The first study shows why this choice of metric changes the preferred configuration; subsequent studies examine competing property objectives, acquisition budgets and transfer across ADMET endpoints.

\Cref{tab:extended_simulation_design_compact}(a) summarizes dataset sizes across the three study collections.
The following comparisons introduce the datasets, objectives and budgets needed to interpret each finding (\Cref{tab:extended_simulation_design_compact}). Appendix~\ref{sec:methods} gives the complete simulation protocols and the metrics used in each study. Model implementations, execution options and runnable examples accompany the paper in Supplementary Software. The five supplementary packages are available for review at \supplementlink{}; Appendix~\ref{app:availability} describes their contents and use.

\subsection{Single-Objective Selection: BindingDB}

\subsubsection{Configuration Rankings across Targets}
\label{sec:results}

The single-objective study evaluates 57 configurations on 100 target-specific BindingDB datasets of half-maximal inhibitory concentration (IC$_{50}$) \citep{gilson2016bindingdb}. The benchmark uses pIC$_{50}$ as its binding score: the negative base-10 logarithm of IC$_{50}$ in molar units, with larger values indicating greater inhibitory potency. Each campaign initializes with 10\% of the featurized records, capped at 200, and queries 100 retained assay records per cycle; repeated observations of a structure are kept. Greedy exploitation selects candidates with the highest predicted means; figures abbreviate this rule as greedy. The configurations use Morgan molecular fingerprints and ESM-2 protein representations; neural networks start from a shared KIBA-pretrained model. Appendix~\ref{app:bindingdb_protocol} defines the fixed target cohort and record-level sampling.

\label{sec:rq1}

We report the relative performance of all configurations across datasets in \Cref{fig:configuration_heterogeneity}(a,b). Each entry averages the dataset metric over five acquisition cycles and three split seeds. In panel (a), each target's P95 is standardized across configurations and the colors saturate at $\pm2$; in panel (b), relative utility is scaled within each endpoint from zero for the worst tested configuration to one for the best. For continuous endpoints, utility is batch P95 after transforming the property so that higher values are preferred; for binary endpoints, utility is the fraction of selected molecules labeled inactive. Surrogate families group the columns. Within the tested BindingDB catalog, XGB and GP configurations generally achieve higher batch P95 than NN configurations, although their relative performance varies across targets. Differences within each surrogate family further show that acquisition choice matters. ADMET exhibits more varied patterns across endpoints, with strong configurations appearing in all four surrogate families. Together, these results support selecting AL/BO configurations according to the target property and budget rather than applying one configuration across datasets. \Cref{fig:bindingdb_relative_budget}(b,c) groups results by the fraction of the candidate pool acquired. The comparisons use each surrogate with its tested representation, initialization, hyperparameters, and acquisition rules.

Acquisition labels name the rule and its coefficients. In the comparisons below, GP upper confidence bound (UCB) selection balances predicted mean and uncertainty through $\beta$; larger $\beta$ gives uncertainty more weight. Query by committee (QBC) uses disagreement among model predictions to guide selection. The NN and VBLL implementations use dropout predictions, adding their min-max-normalized variance to $t$ times their normalized mean. The XGB configurations stored under the QBC label instead use the mean and dispersion of terminal-leaf identifiers across trees. The QBC-diversity variant adds $d$ times the normalized distance to the labeled and already selected candidates. A $q$ prefix (qUCB, qNEI, qEHVI) denotes a GP acquisition function defined for $q$ candidates. Here, these functions score each candidate individually ($q=1$), and the highest-scoring candidates form the batch. Appendix~\ref{app:configs} gives the full score for each surrogate family.

\subsubsection{Screening Metrics and Pool Coverage}

The evaluation metric defines what a successful screening campaign should achieve. Batch mean rewards high average quality across the selected records; an upper-tail percentile emphasizes the stronger candidates, while the maximum measures the single best selection. These goals can favor different configurations (\Cref{fig:bindingdb_metric_choice}(c) and \Cref{fig:bindingdb_metric_targets}(a)). Upper-tail percentiles give similar broad orderings, but batch mean and maximum can change which acquisition rules lead. The metric should therefore match the intended screening goal when comparing configurations or evaluating an agent's recommendations; \Cref{fig:metric_rank_main} gives the full rankings.

\begin{samepage}
\Cref{fig:metric_rank_main} shows closely aligned P80--P95 rankings across all 57 configurations (Spearman $\rho=0.942$--$0.997$), compared with $\rho=0.678$ for mean versus maximum. GP qUCB $(\beta=8)$ leads under P85, P90 and P95; value-biased XGB CoreSet $(t=1)$ leads under maximum score (average rank 16.46). Selecting strong batches and finding one extreme candidate can therefore favor different acquisition rules.\par
\end{samepage}

The batch-mean comparisons here average quality over all five acquisition cycles. The recommendation study in Section~\ref{sec:llm_study} instead uses the final batch's mean to assess quality at the end of the campaign. Appendix~\ref{app:stability} compares these evaluation horizons directly.

The value of configuration selection also depends on the evaluation horizon. In a test of repeatability across campaign splits, the configuration with the highest average utility on two recorded splits of each target is selected and evaluated on the third. Relative to XGB greedy exploitation, this procedure reduces mean normalized regret by 0.096 for final-batch mean and 0.261 for final-batch P95. Here regret is the loss relative to the best tested configuration divided by the held-out split's best-to-worst utility range. For mean quality averaged over all five cycles, XGB greedy exploitation already has regret 0.020; the paired regret difference (selected configuration minus XGB greedy exploitation) is 0.004 (descriptive 95\% interval $[-0.0002,0.008]$). Thus, the configuration decision depends on whether a campaign prioritizes mean selected score across acquisition cycles or the quality of its final batch (Appendix~\ref{app:stability}).

\Cref{fig:bindingdb_metric_choice}(c) ranks all 57 configurations within each target, assigning tied configurations their best shared rank, then averages ranks over 100 targets. Target metrics average five cycles and three seeds. Its six displayed configurations are the union of the top three under batch mean and P95. XGB greedy exploitation has average rank 2.52 under batch mean but 16.01 under P95. GP qUCB $(\beta=8)$ has higher P95 but lower mean than XGB greedy exploitation on 70 targets (teal in \Cref{fig:bindingdb_metric_targets}(a)), the reverse on two (vermilion), and other outcomes on the remaining targets (gray). Thus, the change in preferred configuration reflects differences within targets as well as across-target rankings.

We group the 100 BindingDB datasets into thirds by acquired-pool fraction, $500/[N-\lfloor0.2N\rfloor-\min(\lfloor0.1N\rfloor,200)]$, where $N$ is the number of featurized records (\Cref{fig:bindingdb_relative_budget}(b,c)). The mean P95 rank across GP configurations changes from 10.18 in the lowest-coverage group to 25.22 in the highest, while the mean across NN configurations changes from 37.53 to 29.13. The family ordering by batch mean remains GP, XGB and NN across all three groups. The differences between configuration ranks therefore vary with pool coverage across these datasets. Because the coverage groups contain different targets, this comparison describes an association between coverage and rank. Family mean ranks use average ranks for ties, and the error bars are descriptive 95\% intervals from 10,000 within-group dataset-bootstrap resamples. Acquired records may share molecular structures.

We compare batch P95 across acquisition cycles in \Cref{tab:all57-p95-cycle,tab:all57-p95-cycle-cont} (Appendix~\ref{app:single_results}). GP qUCB $(\beta=8)$ decreases from 9.176 to 8.486, whereas NN QBC-diversity $(t=0.1,d=1)$ increases from 8.033 to 8.417; random sampling stays near 8.57. Depletion of candidates with high property scores before the fifth cycle is one possible explanation for this pattern.

\subsection{Joint Selection: Binding and ADMET}

\subsubsection{Binding-Toxicity Trade-offs}
\label{sec:joint_binding_safety}

The joint study compares 100 candidate pools with binding and ADMET objectives. Because public datasets often lack both measurements for the same molecules, we supplement available measurements with pretrained property predictions. Of the 100 pools, 89 pair measured binding with XTT cytotoxicity predictions from TxGemma-27B-Predict~\citep{wang2025txgemma}, ten pair measured ADMET endpoints with Boltz-2 binding predictions~\citep{passaro2025boltz2}, and one ZINC subset of 9{,}111 candidates has both objectives predicted~\citep{irwin2012zinc}. XTT measures cell viability, so the 90 pools with XTT predictions support a comparison of predicted cytotoxicity under the same assay definition. The ten measured ADMET pools are assessed using their endpoint-specific directions. Each is paired with a preselected protein outside the BindingDB-derived joint cohort; the ZINC pool uses WWP2-N. Predicted binding uses the negative Boltz-2 affinity score.

For the 77 scalar configurations, we standardize binding and ADMET desirability as $z_b$ and $z_s$ and fit a surrogate to their weighted sum. Balanced acquisition uses coefficients $(1,1)$. Three additional GP configurations use expected hypervolume improvement (qEHVI), its noisy variant (qNEHVI), or noisy expected improvement with random scalarization (qNParEGO)~\citep{daulton2020qehvi,daulton2021qnehvi,knowles2006parego}.

These configurations fit a separate GP to each objective and use hypervolume improvement or random scalarizations to seek trade-offs, whereas scalar configurations fit one model to a fixed weighted objective. The implemented GP rules score candidates individually and select the highest-scoring batch. These three configurations are compared on the 90 pools with predicted XTT labels. Each campaign starts with 10\% of the dataset, capped at 200 labels, and acquires five batches of 100. All joint comparisons use full-pool property means and standard deviations to set a common scale within each dataset. We call the preference that places greater weight on ADMET desirability ADMET-heavy.

For the 90 datasets with predicted XTT labels, we report two outcomes: the fraction of selected molecules with predicted XTT toxicity scores of at least 0.5, and conditional binding P95, defined as binding P95 among selected molecules below that toxicity threshold. Binding is standardized within each candidate pool, so its scale expresses relative binding quality. We average the batch summaries over five cycles and three random splits, then weight datasets equally. The ten measured ADMET pools use their own property objectives, with endpoint-specific evaluation criteria.

Configurations with similar binding quality can select very different proportions of predicted toxic molecules (\Cref{fig:multiobjective_landscape}(a,b)). NN QBC-diversity $(t=0.1,d=0.1)$ achieves conditional binding P95 of 1.895 with 18.59\% of selections predicted toxic, whereas GP qUCB $(\beta=2)$ achieves 1.868 with 3.14\%. Binding P95 is computed among the predicted non-toxic molecules selected by each configuration.
The ten-entry Pareto front of dataset-averaged outcomes shows the available trade-offs (\Cref{fig:joint_pareto}(a)). GP qEHVI has the lowest predicted-toxic fraction, 1.85\%, at conditional binding P95 of 1.583; NN QBC-diversity $(t=10,d=1)$ raises P95 to 1.854 at 2.53\% toxicity. Color and shape identify surrogate families. Letters identify configurations on the Pareto front, with their names given beneath each panel. Appendix~\ref{app:configs} defines the coefficients $t$ and $d$; XGB QBC combines means and dispersions of terminal-leaf identifiers.

The binding statistic also changes which configurations lie on the Pareto front. Under balanced acquisition, GP qEHVI remains the Pareto-front configuration with the lowest predicted-toxic fraction for the 90th percentile (P90), 85th percentile (P85), maximum and mean of conditional binding scores. The high-binding end shifts from NN QBC-diversity for P90/P85 to VBLL QBC-diversity for the maximum and VBLL greedy exploitation for the mean (Appendix~\ref{app:component_metrics}). Choosing a trade-off therefore requires specifying which part of the binding distribution matters.

The number of datasets on which each surrogate family ranks first differs between outcomes (\Cref{fig:multiobjective_landscape}(c,d)). GP achieves the lowest predicted-toxic fraction most often, with 46.1 dataset wins after dividing credit equally among tied winners. Across these 90 datasets, 36 configurations achieve the highest conditional binding P95 on at least one dataset. The preferred AL/BO configuration thus depends on both the dataset and the desired outcome.

\subsubsection{Objective Preference Weights}

Binding-heavy settings usually select molecules with stronger binding, while ADMET-heavy settings usually reduce predicted toxicity. The coefficient pairs give the binding weight first and the ADMET weight second. Balanced acquisition uses $(1,1)$, ADMET-heavy acquisition uses $(0.25,0.75)$, and binding-heavy acquisition uses $(0.75,0.25)$. Comparisons with the balanced setting therefore vary both the relative weights and the scale of the weighted score used to train the surrogate. Across all 231 configuration-weight combinations, binding-heavy settings account for most of the highest-binding entries, while the 24 lowest-toxicity entries are ADMET-heavy. The 14-entry Pareto front spans conditional binding P95 of 1.746 at 0.95\% toxicity to 1.915 at 7.65\% (\Cref{fig:weight_pareto}(b)). Color and marker fill indicate the preference, shape indicates the surrogate family, and B, S and H in the key denote balanced, ADMET-heavy and binding-heavy weights. The preference weights shift the selected outcomes, but the size of this shift depends on the acquisition rule. For the full 100-dataset scalar comparison, each selected molecule receives a weighted sum of standardized binding and ADMET desirability. We take P95 within each batch and average over cycles and random splits. The leading configuration changes with preference: GP qUCB leads under balanced and binding-heavy weights, and VBLL QBC-diversity under ADMET-heavy weights. On the ten measured ADMET datasets, binding-heavy weights also increase mean binding in all ten pools, while reducing mean ADMET desirability in eight, relative to ADMET-heavy weights and after averaging the 77 scalar configurations (\Cref{fig:measured_group_b_preferences}(a,b)).

Relative to balanced weights, binding-heavy weights usually raise both binding quality and predicted toxicity (\Cref{fig:weight_effects}(a,b)). Binding-heavy weights improve conditional binding P95 for 71 of 77 configurations and increase predicted toxicity for 56. Responses remain rule-dependent: toxicity rises by 4.71 percentage points for XGB greedy exploitation but falls by 8.17 points for NN QBC-diversity $(t=0.1,d=0.1)$. Both panels share the same configuration order within each surrogate family.

\subsubsection{Measured ADMET Objectives}
\label{sec:measured_admet_main}

The ten pools with measured ADMET labels pair different biological endpoints with predicted binding: kinetic solubility, human microsomal clearance, the logarithm of the distribution coefficient (LogD), pregnane X receptor (PXR) potency and single-concentration fold change, Ames mutagenicity, median lethal dose (LD50), and three Toxicology in the 21st Century (Tox21) activity labels. We transform each ADMET outcome so that higher values represent greater desirability under the specified screening objective, then standardize it. These joint tasks maximize kinetic solubility and minimize the other recorded endpoint scores; standalone LogD instead rewards proximity to 2. Appendix~\ref{app:measured_group_b} gives the endpoint definitions, including the encoded LD50 scale.

For these datasets we report the mean standardized binding scores and ADMET desirability in each selected batch, then average cycles, seeds, and datasets. The preference weights trade binding against the measured ADMET outcomes (\Cref{fig:measured_group_b_preferences}(a,b)). Each point in panel (a) is one scalar configuration under one preference, and panel (b) shows the dataset-level change from ADMET-heavy to binding-heavy acquisition, averaged across configurations. Averaged across the 77 scalar configurations and ten datasets, moving from ADMET-heavy to binding-heavy weights increases mean binding from 0.246 to 0.499 and decreases mean ADMET desirability from 0.216 to 0.006. Binding increases in all ten datasets; ADMET desirability decreases in eight, with the Tox21 androgen-receptor (NR-AR) and estrogen-receptor (NR-ER) assays as exceptions. The standardized values express each outcome relative to its candidate pool's mean, in units of that pool's standard deviation.

\subsubsection{Acquisition Budget Sensitivity}

The seven-setting sweep varies initial labeled-set size over $\{200,400,800\}$ and batch size over $\{20,40,80,100,160\}$, using its own fixed-initialization $(200,100)$ reference. Fixed-size initialization reshuffles the labeled and acquisition sets while retaining the diagnostic set, so this reference differs from the main capped-initialization split. The leading configurations can change even when overall rankings remain similar (\Cref{fig:al-knob-sensitivity}(a,b)). On the 90 XTT datasets, conditional binding ranks at batch size 20 correlate strongly with the reference (Spearman $\rho=0.891$), yet retain only two of its top ten configurations when cutoff ties are included. Toxicity ranks are more stable ($\rho=0.980$).

In contrast, increasing initialization to 400 or 800 retains all ten configurations with the highest conditional binding P95, with rank correlations of 0.991 and 0.988. This stability concerns conditional binding on the 90 XTT pools; the scalar-utility comparison below also evaluates both the binding and ADMET outcomes of every selected molecule. With five cycles fixed, changing batch size also changes total acquisition budget. \Cref{fig:current_joint_scalar_budget}(c,d) reports the scalar-utility comparison for all 100 datasets. Bands in \Cref{fig:al-knob-sensitivity} are descriptive 95\% paired-dataset bootstrap intervals, and the batch statistics average five cycles and three seeds.

High overall rank correlation can hide large changes among the leading configurations (\Cref{fig:budget_rank_changes}). At batch size 20, NN QBC-diversity $(t=0.1,d=0.1)$ drops from rank 2 to 34 and NN QBC $(t=0.1)$ from 1 to 33, while two configurations with large diversity weights remain at ranks 76 and 77 (\Cref{fig:budget_rank_changes}). In that panel, red marks worse and blue better ranks, and A-D label the four displayed configurations.

For the seven-setting budget study we fix evaluation weights to $(0.5,0.5)$ and compare the 77 scalar configurations with the sweep's $(200,100)$ reference setting (\Cref{fig:current_joint_scalar_budget}(c,d)). Using catalog order to resolve top-ten cutoff ties, batch size 20 yields rank correlation 0.912 across all 100 datasets but retains only one of the reference setting's top-ten configurations; initialization 800 yields correlation 0.915 and retains four. For the ten measured endpoints alone, batch size 20 yields correlation 0.961 and retains nine. Unlike the conditional-binding comparison, this analysis scores both objectives for every selected molecule.

Batch and accumulated-set quality respond differently to budget changes. For XGB greedy exploitation, increasing batch size from 20 to 160 raises the predicted-toxic fraction in new batches from 1.33\% to 3.09\%, while the predicted-toxic fraction in the accumulated labeled set falls from 8.18\% to 5.10\%. Larger acquisitions reduce the proportion of initially sampled candidates in the accumulated labeled set. The accumulated set can therefore have a lower predicted-toxic fraction even as that fraction rises in newly selected batches.

\subsection{Cross-Property Selection: ADMET}

\subsubsection{Configuration Performance across ADMET Endpoints}
\label{sec:admet_endpoints}

The 34-endpoint ADMET study spans absorption, distribution, metabolism, clearance, and toxicity. It evaluates 77 configurations on 22 continuous endpoints from OpenADMET~\citep{openadmet2026asap,openadmet2025expansionrx,openadmet2026octant,openadmet2026pxr} and 12 binary Tox21 endpoints~\citep{huang2016tox21,wu2018moleculenet}. Runs initialize with 10\% of the dataset, capped at 200 labels, and acquire five batches of 20. Continuous properties are transformed so that higher values match the intended objective, and scored by batch P95. For example, plasma protein-binding utility rewards percent unbound, and LogD utility rewards proximity to 2. Binary endpoints use the fraction of selected molecules labeled inactive. Utilities average five cycles and three random splits. Relative utility scales these values within each endpoint so that zero is the worst tested configuration and one is the best.

Configuration performance varies by endpoint type (\Cref{fig:admet_endpoint_atlas}(a,b)). Stars mark each surrogate family's best rule and diamonds its family mean. On continuous endpoints, each family's best configuration has similar mean relative utility, 0.858-0.881. On Tox21, XGB greedy exploitation leads at 0.956, while NN spans 0.040-0.923 across acquisition rules. Both the molecular property type and the acquisition rule affect which configuration performs best.
XGB greedy exploitation improves the inactive fraction over random sampling on most Tox21 endpoints, selecting 96.56\% inactive molecules versus 92.78\% for random sampling. Its inactive fraction is higher on ten endpoints, equal on one, and lower by 0.33 percentage points on NR-AR. The 3.78-percentage-point gain occurs against a high inactive fraction under random sampling.

Dataset wins grouped by property category show property-specific strengths (\Cref{fig:admet_property_metrics}(a)). NN receives 15.5 of 34 dataset wins after sharing credit among ties and wins in all five categories. VBLL receives three of five clearance dataset wins; XGB receives six of twelve toxicity dataset wins. The leading surrogate family varies across properties. Plot labels report dataset wins with credit shared among ties.

Upper-tail metrics agree more closely with P95 than batch mean does (\Cref{fig:admet_property_metrics}(b)). Panel (b) shows within-family Spearman correlations of Max, Mean, P90 and P85 ranks with P95, averaged over the 22 continuous endpoints using average ranks for ties. Across families, P90 has the highest mean correlation with P95 ($\rho=0.826$-$0.921$), and batch mean the lowest ($0.676$-$0.761$). Selection for the mean score and selection for upper-tail scores can therefore favor different configurations.

\subsubsection{Transfer between ADMET Endpoints}

To assess transfer between properties, we choose the configuration with the highest average relative utility on the source endpoints and evaluate it on the held-out endpoint. Normalized regret measures the loss relative to the best tested configuration, divided by that endpoint's best-to-worst utility range; lower regret is better. Within-family comparisons select among configurations from one surrogate family and compute regret relative to the full catalog.

The average performance of a default configuration can differ substantially from its performance on individual properties. Across all 34 ADMET endpoints, leave-one-endpoint-out selection always chooses XGB greedy exploitation, with mean regret 0.097 but regret 0.469 on cytochrome P450 2J2 (CYP2J2) substrate remaining. Average performance should be considered alongside endpoint-specific losses.

To test configuration transfer within each surrogate family, we select a configuration using all but one endpoint of the same type and score it on the held-out endpoint. Continuous-property mean regret is 0.119 for VBLL, 0.126 for XGB, 0.142 for GP, and 0.197 for NN. For Tox21, XGB has the lowest point estimate at 0.044, followed by NN at 0.077, GP at 0.114, and VBLL at 0.207. The relative transfer performance of the surrogate families therefore depends on the endpoint type.

Configuration choices based on upper-tail scores are particularly property-dependent. Across the 22 continuous endpoints, 64 of 231 endpoint pairs have disjoint P95 top-ten lists, versus ten under batch mean.

Configuration rankings agree more across properties under batch mean than under P95. Among the 22 continuous properties, mean pairwise Spearman correlation of the 77 configuration ranks across the 231 endpoint pairs is 0.582 for batch mean and 0.336 for P95. These are descriptive comparisons within the supplied collection, with each endpoint contributing to several pairs.

With batches of 20, linearly interpolated P95 depends on the two largest selected outcomes. Appendix~\ref{app:admet_endpoint_transfer} reports how this upper-tail emphasis affects agreement between repeated campaigns.

Across repeated campaigns on each of the 22 continuous endpoints, the mean rank correlation between split seeds is 0.854 for batch mean and 0.654 for P95. The paired P95-minus-mean difference is $-0.200$ (descriptive 95\% interval $[-0.248,-0.155]$). Upper-tail selection therefore calls for checking campaign repeatability alongside transfer between endpoints.

\section{Using Completed Campaigns to Recommend Configurations}
\label{sec:llm_study}

The acquisition studies show that the target property, metric and budget affect configuration choice. We next test whether outcomes from completed campaigns help recommend configurations for a new dataset. Direct prediction gives an LLM the target dataset description, configuration catalog and screening objective. In-context learning (ICL) adds descriptions and the top 15 configurations from each of 20 related source datasets. Each method returns an ordered list of five configurations.

\subsection{Evaluation Protocol}

Dataset descriptions summarize molecular properties, candidate-pool size and available protein context. Retrieval compares molecular fingerprints, protein sequences and dataset sizes to identify related source campaigns. The full descriptor list and similarity formula are in Appendix~\ref{app:similarity}.

We use an LLM-generated prompt specific to this task and a similarity rule for source retrieval, both fixed before recommendation evaluation. Direct prediction and ICL use the same target description, catalog, evaluation objective and generation settings, so the comparison tests the contribution of source-campaign evidence.

We evaluate Qwen3.5-9B~\citep{qwen2026qwen35}, Qwen3.8-27B~\citep{qwen2026qwen38}, Qwen3.6-35B-A3B~\citep{qwen2026qwen36}, and Gemma-4-26B-A4B-it~\citep{google2026gemma4} on BindingDB, standalone ADMET and three joint preferences: ADMET-heavy, balanced and binding-heavy. Each collection is divided into ten folds. One fold supplies target datasets for evaluation, and the other nine supply source campaigns; all preferences for the same joint pool remain in the same fold. Target campaign outcomes are excluded from prompts, retrieval and selector fitting. The model identities are reported with the results; the scientific evaluation protocol is in Appendix~\ref{app:llm_protocol}, and the prompts and inference settings accompany Supplementary Software.

Recommendation scores match the screening objective. BindingDB uses the mean binding score of the final acquisition batch. Joint selection uses P95 of each molecule's weighted standardized binding and ADMET score, averaged over five cycles. For standalone ADMET, continuous endpoints use P95 after transforming the property so that higher values are preferred, and binary endpoints use the inactive fraction. All three utilities average the three campaign splits; Appendix~\ref{sec:utility_regret} gives their full definitions.

For every recommendation setting, normalized regret is the selected configuration's utility loss relative to the best tested configuration, divided by the catalog's best-to-worst utility range; regret is zero when all configurations have equal utility. First-choice regret measures this loss for the first recommendation. Regret@5 measures the loss of the best configuration in the five-entry list, evaluated using the completed target campaigns. Their difference, $R_1-R_5$, is the ordering gap. Hit@5 is one if the list contains any of the target's five highest-scoring configurations and zero otherwise; catalog order resolves ties in that reference ranking. BestRk@5 gives the best rank in the list, assigning tied configurations their best shared rank. Invalid responses receive regret one, Hit@5 zero, and rank equal to the catalog size. The main comparison includes 12 invalid responses among 3,472 predictions. Reported 95\% intervals resample held-out folds while keeping comparisons paired.

The full BindingDB comparison also reports Hit@1 and Hit@3: Hit@$k$ checks whether the first $k$ recommendations overlap the target's $k$ highest-scoring configurations. Mean rank averages the first recommendation's rank across targets, using the same best shared rank for ties as BestRk@5.

The reference methods use the same source-target separation. Global Best transfers configurations with the best average source ranks; K20 rank aggregation averages the same 20 source lists supplied to ICL, assigning rank 16 to configurations absent from a source's top 15 and using catalog order for ties. Frequency Voting favors configurations that most often rank first on source datasets; Most-Similar Source transfers the nearest source's ranking, and Random samples configurations uniformly. Ridge regression learns to predict configuration ranks from dataset descriptors, using either those 20 sources or all eligible sources. This comparison tests how effectively each method uses previous campaign evidence.

\subsection{Direct Prediction and In-Context Learning}
\label{sec:rq2}

Completed campaigns help LLMs identify promising configurations for a new dataset. Hit@5 records whether the five recommendations include at least one of the five highest-scoring configurations on the target dataset; BestRk@5 is the best rank among those recommendations. Shortlist metrics assess whether the list contains a highly ranked configuration and how well its best entry performs (per-dataset values are in Supplementary Data 7). Mean Hit@5 and BestRk@5 improve for all four models across the five settings (\Cref{tab:main_llm_summary,tab:llm_full_results,tab:llm_joint_results,tab:llm_joint_safety}, \Cref{fig:llm_usefulness}(a,b)). In \Cref{tab:main_llm_summary}, each entry is a descriptive mean over held-out datasets (100 BindingDB targets, 34 ADMET endpoints and 100 joint pools per preference), including penalties for invalid lists; bold marks tied best values at the displayed precision within each panel. In the tables and figures, Q9, Q27, Q35 and G4 denote Qwen3.5-9B, Qwen3.8-27B, Qwen3.6-35B-A3B and Gemma-4-26B-A4B-it. When configurations with identical recorded utilities, such as NN greedy exploitation and its mean-only variants, are grouped as ties, ICL still has lower BestRk@5 than direct prediction in all 20 comparisons (Appendix~\ref{app:jmlr_diagnostics}). On ADMET, for example, Qwen3.5-9B's Hit@5 rises from 17.6\% to 61.8\%, while its first-choice regret falls from 0.361 to 0.132. Direct prediction and ICL receive the same target description and objective, so this comparison shows the value of adding empirical evidence about how configurations performed on related tasks.

ICL also reduces mean first-choice regret in 18 of the 20 model-setting comparisons. Qwen3.6-35B-A3B on BindingDB and Gemma-4-26B-A4B-it on balanced joint selection improve their shortlists without reducing first-choice regret. Campaign evidence helps the models find useful configurations more consistently than it helps them put the strongest choice first. The counts summarize mean differences; paired intervals report their uncertainty and include zero in some comparisons (Appendix~\ref{app:llm_results}).

Simple source-based baselines also use campaign evidence effectively. On ADMET, K20 rank aggregation achieves mean first-choice regret 0.092 from the same source rankings supplied to ICL, compared with 0.094-0.132 for the four LLMs. Global Best also has lower mean first-choice regret than every ICL model on balanced joint selection. Together with the gains over direct prediction, these comparisons show the value of completed campaigns and establish practical reference points for improving how LLMs use this evidence. \Cref{fig:llm_absolute}(a-e) shows first-choice regret for each language model with and without ICL and for the reference methods; error bars are descriptive 95\% intervals from 10,000 held-out-fold block bootstrap resamples, and the joint panels cover the ADMET-heavy, balanced and binding-heavy preferences.

The LLM experiments establish reference performance for an LLM-generated prompt specific to this task and a fixed retrieval rule. Aggregating the same source rankings matches or exceeds several ICL conditions, providing a concrete target for improving the use of campaign evidence. This motivates developing prompts, retrieval strategies and configuration-selection policies for the task; \Cref{sec:agent_use_case} describes an agent system for developing such policies.

\subsection{Learned Selectors}
\label{sec:learned_results}

We compare ridge regression selectors based on dataset metadata with source-rank aggregation methods and the same four ICL models (\Cref{tab:learned_comparison}). Each method returns five configurations. K20 ridge regression and K20 rank aggregation receive the same source descriptions and top-15 lists as ICL; all-source ridge regression and Global Best use every eligible source dataset's complete rankings. Ridge regression uses numerical dataset metadata, whereas ICL also interprets the descriptions and objective prose. Entries in \Cref{tab:learned_comparison,tab:learned_best_list} are means across held-out datasets, and Regret@5 in \Cref{tab:learned_best_list} uses target outcomes to identify the best entry for evaluation.

In \Cref{tab:learned_comparison,tab:learned_best_list}, GB denotes Global Best, FV Frequency Voting, K20 K20 rank aggregation, R-All all-source ridge, and R-K20 K20 ridge; ICL prefixes the four model abbreviations defined above.

K20 ridge regression has mean first-choice regret 0.237 on BindingDB and 0.084 on ADMET, compared with 0.271 and 0.092 for K20 rank aggregation. Paired differences (ridge minus aggregation) are $-0.033$ [$-0.063,-0.003$] and $-0.008$ [$-0.028,0.008$], respectively, using descriptive 95\% held-out-fold intervals. Its BindingDB mean is also below the four ICL means, 0.260-0.340. The ordering differs across objectives: Global Best has the lowest mean first-choice regret in the balanced joint comparison, 0.082, whereas K20 rank aggregation leads on ADMET-heavy joint selection, 0.050. A learned mapping from metadata to configuration ranks is therefore a useful reference, while aggregation of source configuration ranks remains competitive. Paired differences and their descriptive intervals are reported in \Cref{tab:learned_paired} (Appendix~\ref{app:jmlr_diagnostics}).

The utility of a shortlist and its ordering reveal different aspects of selection quality (\Cref{fig:ordering_gap}(c)). Averaging the four models within each BindingDB target, ICL reduces first-choice regret from 0.416 to 0.293 and Regret@5 from 0.279 to 0.106. The ordering gap consequently increases from 0.137 to 0.187. In the other four settings, mean ICL ordering gaps are 0.040-0.067. The improvement in the best listed configuration can therefore exceed the improvement in the first recommendation.

\subsection{Recommendation Ordering with Fixed List Membership}
\label{sec:source_rerank}

To isolate ordering, we compare each original list with a uniform shuffle and a reordering using K20 rank aggregation on the same 20 source rankings. List membership is unchanged, and neither control uses target outcomes.

Holding the five recommendations fixed tests whether their ordering uses the available evidence effectively (\Cref{tab:main_controls}). On BindingDB, model-averaged ICL first-choice regret is 0.293, compared with 0.296 for uniformly shuffling the same lists and 0.275 for source-based reranking. Source reranking lowers mean first-choice regret in 19 of 20 model-setting comparisons; Gemma-4-26B-A4B-it on BindingDB has a small mean increase. On ADMET, all four original orderings have lower regret than uniform ordering, with advantages of 0.039-0.055. Five of the 20 per-model paired intervals lie wholly below zero (\Cref{tab:same_list_rerank} in Appendix~\ref{app:ordering_controls}); the pooled intervals in \Cref{tab:main_controls} quantify uncertainty in the mean differences. Negative $\Delta R_1$ in \Cref{tab:main_controls} favors source-based reranking, and the comparison uses utility averaged over the three campaign splits. Empirical source ranks can therefore help order alternatives even after the language model has constructed the shortlist.

\subsection{Repeatability across Campaign Splits}

We select a configuration using two recorded campaign splits and evaluate it on the third, rotating the held-out split. Using outcomes from two splits of the target dataset tests the repeatability of configuration selection across splits. Source-based selectors retain their original recommendations. The regret scale is computed separately for the held-out split.

Repeated campaigns test whether the best configuration on previous runs also performs well on a new run (\Cref{tab:seed_oracle_stability} in Appendix~\ref{app:seed_stability}). For BindingDB, averaging models within target, the original ICL first recommendation has regret 0.334 under held-out-seed scoring. Selecting a list entry on the other two seeds reduces this to 0.265; selecting the best list entry using the held-out split itself gives regret 0.127. The two-seed selection therefore achieves 33.7\% of the mean regret reduction available from choosing the best entry on the held-out split. The corresponding fractions are 23.7-30.2\% for the other settings, with the ADMET gain interval including zero. K20 ridge regression retains lower BindingDB regret than K20 rank aggregation under this scoring, 0.287 versus 0.313. The controls quantify how much of the potential regret reduction persists across campaign splits. In \Cref{tab:main_controls}, a positive reduction favors the two-seed choice over the original ICL first entry using each held-out split's utility range for normalization, and the last column divides that mean reduction by the mean regret difference between the original first entry and the best list entry on the held-out split under the same scoring.

The configurations available for recommendation define the selection task. On 44 of the 100 BindingDB targets, the best recorded entry is one of twelve XGB configurations with scores based on terminal-leaf identifiers, and the other four settings have at most six such targets. Removing these twelve from the original recommendation lists, without refilling them, leaves ICL with lower regret than direct prediction for the best remaining list entry in 18 of the 20 model-setting comparisons; Gemma-4-26B-A4B-it and Qwen3.6-35B-A3B on BindingDB are the two exceptions (Appendix~\ref{app:catalog_sensitivity}). Together, the controls let researchers distinguish source evidence, list ordering and campaign repeatability when comparing recommendation methods.

\subsection{Source Retrieval and Context Size}
\label{sec:rq3}

We compare similarity-based and random retrieval for Qwen3.5-9B and Qwen3.6-35B-A3B, using $K\in\{1,3,5,10,20,50,\mathrm{all}\}$ source campaigns for BindingDB and joint selection, and omitting $K=50$ for ADMET. Using all eligible sources gives 90 BindingDB datasets or joint pools, or 30-31 ADMET endpoints. The all-source condition uses the same ordering for both retrieval methods; Appendix~\ref{app:similarity} defines the similarity measure and sampling protocol. Points in \Cref{fig:llm_sources} average repetitions within each target and then average across targets. Random retrieval uses three source-selection seeds. The all-source condition reuses one prediction for both retrieval methods. Bands are descriptive 95\% intervals from 10,000 held-out-fold bootstrap resamples; numerical results are in \Cref{tab:icl_ablation,tab:icl_ablation_admet,tab:icl_ablation_joint_balanced,tab:icl_ablation_joint_binding,tab:icl_ablation_joint_safety} (Appendix~\ref{app:llm_results}).

Choosing a related campaign is useful when the prompt contains few examples. With one ADMET source, Qwen3.5-9B has first-choice regret 0.157 under similarity retrieval, compared with 0.268 under random retrieval (\Cref{fig:llm_sources}(c,d), \Cref{tab:icl_ablation_admet}). Similarity retrieval also gives lower first-choice regret with one source for both models on BindingDB and across the joint objectives, although the difference for Qwen3.5-9B under balanced joint selection is less than 0.001 (\Cref{fig:llm_sources_joint}(e-j)). A prompt containing one relevant source campaign can therefore improve configuration recommendations.

Adding more campaigns has a different effect across scientific objectives. Under similarity retrieval, Qwen3.5-9B attains its lowest observed BindingDB first-choice regret with the most similar source, whereas its lowest ADMET first-choice regret uses all sources. The all-source condition presents examples in a fixed order rather than similarity order, so it changes both the number and order of examples. The pattern also depends on the model: Qwen3.6-35B-A3B reaches its lowest observed ADMET first-choice regret with ten sources, and first-choice regret rises when all sources are included. These comparisons show why context size and source selection need to be considered together. How many campaigns to include and how to combine their evidence remain important choices for the recommendation procedure.

\subsection{Agent Use Case: Developing a Configuration-Selection Policy}
\label{sec:agent_use_case}

Our ICL experiments use an LLM-generated prompt specific to this task with similarity-based source retrieval. The baseline and context-size comparisons motivate optimizing which source campaigns to retrieve and how to use their results. A reusable configuration-selection policy specifies how to turn this campaign evidence into a recommendation.

The AL/BO Skill Evolver automates the development of such a configuration-selection policy through repeated policy proposals and tests using \sciselect{}'s BindingDB campaign records (\Cref{fig:agent_use_case}). Inspired by the proposal-and-evaluation loop of AIDE$^2$~\citep{srikanth2026recursive}, one LLM proposes a policy change and another chooses a test that withholds selected source campaigns. A policy change can alter which completed campaigns the policy uses, how much weight each receives, or how their configuration scores are combined. Evaluation feedback guides the next proposal, allowing the system to revise the policy automatically.

A fixed controller runs the current and proposed policies, compares their scores, and keeps or rejects the change. Policies receive the evaluation dataset's description and source-campaign outcomes; outcomes from the evaluation dataset are reserved for scoring. The controller returns aggregate scores and test feedback and can request additional simulations when the protocol permits them. For BindingDB, the agent targets upper-tail quality and scores the selected configuration by P95 of the final acquisition batch; the direct-prediction and ICL study evaluates average final-batch quality.

The implementation connects to \sciselect{}'s simulation toolkit to load the dataset registry, configuration catalog, completed outcomes, and molecular and protein similarities. For each evaluation dataset, the controller supplies its description and source-campaign outcomes to the policy, which selects a configuration without receiving that dataset's outcomes. The controller then scores the selected configuration using its recorded outcomes on the evaluation dataset. For example, the starting policy selects similar source datasets, weights their configuration ranks by similarity, and chooses the configuration with the lowest weighted mean rank. Each proposed policy is stored as an executable specification with matching human-readable instructions, so the same rule can be tested across datasets.
Policy development revises the configuration-selection policy while keeping the controller and scoring rules fixed. During development, the controller compares the current and proposed policies on tests using source campaigns and on the datasets used for policy development. It returns aggregate scores and test feedback to the proposing LLM. The test-selecting LLM chooses its test before these scores are revealed. Final validation is conducted separately from the policy-development feedback loop. The supplementary source code includes a runnable example of retrieval, simulation requests, and recommendation scoring through the \sciselect{} interface.

\section{Discussion and Conclusion}
\label{sec:discussion}

\sciselect{} provides an interactive environment and benchmark for comparing AL/BO configurations and evaluating how LLMs use completed campaigns to recommend configurations for new datasets. Our simulations cover 100 BindingDB datasets, 100 joint binding-ADMET datasets, and 34 ADMET endpoints.

The results show that effective configuration choices depend on both the dataset and the screening objective. A configuration that selects batches with high average quality may differ from one that identifies the strongest candidates. Similar binding quality can also accompany substantially different fractions of predicted toxic molecules. Objective weights, acquisition budgets, and endpoint definitions further affect the preferred AL/BO configuration. These findings support matching configuration choice to the intended metric, objective preference, and budget on the evaluated tasks.

Completed campaigns improve LLM recommendation shortlists more consistently than their first choices. Related examples help when the prompt contains few source campaigns, and the effect of adding more sources depends on the model and objective. The shared comparison with source-based baselines separates the value of campaign evidence from the ability of a language model to use it. The competitive performance of source-based baselines motivates prompts, retrieval strategies and agent workflows designed for AL/BO configuration selection. Shortlist quality and first-choice regret therefore provide complementary criteria for selecting a configuration to deploy. The learned-selector comparison makes this distinction explicit: metadata can improve first-choice selection in some settings, while a strong five-entry list can retain a substantial ordering gap. Reranking fixed lists using source campaigns and evaluating on held-out campaign splits explain this distinction: source evidence can improve ordering, and part of the regret reduction available from choosing the best list entry persists across repeated campaigns. The catalog sensitivity analysis identifies which conclusions depend on the available AL/BO configurations.

\clearpage
\appendix

\makeatletter
\let\@ztryfc\SciSelectOriginalZTryFC
\makeatother
\setcounter{topnumber}{5}
\setcounter{totalnumber}{5}
\renewcommand{\topfraction}{0.95}
\renewcommand{\textfraction}{0.05}
\renewcommand{\floatpagefraction}{0.70}
\setlength{\floatsep}{8pt plus 2pt minus 2pt}
\setlength{\textfloatsep}{8pt plus 2pt minus 2pt}

\setcounter{figure}{0}
\setcounter{table}{0}
\renewcommand{\thefigure}{S\arabic{figure}}
\renewcommand{\thetable}{S\arabic{table}}
\renewcommand{\theHfigure}{S\arabic{figure}}
\renewcommand{\theHtable}{S\arabic{table}}
\section{Campaign Design and Simulation Protocols}
\label{sec:methods}
\label{sec:bench}
\label{sec:bench-bdb}
\label{app:benchmark_spec}

\subsection{BindingDB Campaigns}
\label{app:bindingdb_protocol}

The BindingDB study uses 100 target-specific IC$_{50}$ datasets from public interaction records~\citep{gilson2016bindingdb}, listed in \Cref{tab:dataset_list}. Each candidate is an assay record; repeated measurements of a molecule remain separate. Positive numeric concentrations are converted to pIC$_{50}$ as $9-\log_{10}(\mathrm{IC}_{50}[\mathrm{nM}])$. After feature extraction, random splits allocate 10\% of records to initialization, 70\% to acquisition and 20\% to surrogate diagnostics. Initialization is capped at 200 records, with excess records returned to the acquisition pool. Each of 57 configurations acquires five batches of 100 records, using three split seeds.

BindingDB configurations combine Morgan molecular fingerprints with ESM-2 protein-language-model representations \citep{lin2023evolutionary}. Neural-network configurations initialize from a shared model pretrained on KIBA kinase-binding data \citep{tang2014making} before target-specific fitting.

\subsection{Joint Binding-ADMET Campaigns}

The 100 joint pools combine measured and predicted properties: 89 pair measured BindingDB scores with TxGemma-27B-Predict XTT cytotoxicity predictions~\citep{wang2025txgemma}; ten pair measured ADMET endpoints with Boltz-2 binding predictions~\citep{passaro2025boltz2}; and one ZINC subset of 9{,}111 candidates has both objectives predicted~\citep{irwin2012zinc}. XTT is a cell-viability assay. Predicted binding uses the negative Boltz-2 affinity score.

The 89 BindingDB-derived pools occupy equally spaced positions in the ligand-count ordering of targets with 2{,}000--20{,}000 unique ligands. Repeated measurements of each canonical molecular structure are aggregated by their median. The ten measured-ADMET pools use fixed endpoint-protein pairings outside this cohort: proteins have at least 100 distinct molecular entries and occupy ten equally spaced sequence-length ranks between 250 and 450 residues after duplicate UniProt accessions are removed. Joint pools contain 2{,}005--11{,}117 rows. Supplementary Inputs supplies the candidate tables; \Cref{tab:extended_simulation_design_compact} gives the budgets and configuration sets.

\subsection{Standalone ADMET Campaigns}

The ADMET study uses 34 endpoints, with 351--10{,}870 input rows, and 77 configurations. Records sharing a canonical simplified molecular-input line-entry system (SMILES) string are combined by median label. Non-finite labels are discarded; binary endpoints retain only binary values after aggregation. Higher utility is preferred: LogD (logarithm of the distribution coefficient) uses $-|y-2|$, minimized endpoints are negated, and Tox21 uses $1-y$ for inactive/active labels $y=0/1$. The random 10/70/20 initialization/acquisition/diagnostic split caps initialization at 200 and returns excess records to acquisition. Each campaign acquires five batches of 20, using three split seeds.

\subsection{Evaluation Utilities and Horizons}

\Cref{tab:evaluation_map} specifies the property statistic, acquisition cycles and catalog used in each comparison. P95 denotes the 95th percentile within a selected batch; P$x$ denotes its $x$th percentile. Utilities use measured or predicted labels, oriented so that larger values are preferred.

\begin{table}[t]
\centering\footnotesize
\setlength{\tabcolsep}{3pt}

\begin{tabularx}{\linewidth}{@{}p{.28\linewidth}Xp{.10\linewidth}r@{}}
\toprule
Experiment & Dataset-level utility & Cycles & Catalog \\
\midrule
BindingDB comparison & Batch P95; other statistics analyzed separately & C1-5 & 57 \\
Joint XTT component comparison (90 datasets) & Conditional binding P95 and predicted-toxic fraction, separately & C1-5 & 80 \\
XTT preferences / budgets (90 datasets) & Same component summaries & C1-5 & 77 \\
Measured joint components (10 datasets) & Mean standardized binding and ADMET desirability & C1-5 & 77 \\
Joint scalar comparison / budgets (100 datasets) & Batch P95 of weighted standardized binding and ADMET scores & C1-5 & 77 \\
ADMET comparison / recommendation & Batch P95 of transformed property scores (continuous); inactive fraction (binary) & C1-5 & 77 \\
BindingDB recommendation & Batch mean & C5 & 57 \\
Joint recommendation (three preferences) & Batch P95 of weighted standardized binding and ADMET scores & C1-5 & 77 \\
\bottomrule
\end{tabularx}
\caption{Metrics and acquisition cycles used in each study. Each utility is averaged over three split seeds. C5 denotes the final cycle; C1-5 denotes the mean of the statistics calculated separately for each of the five acquisition batches.}
\label{tab:evaluation_map}
\end{table}

\label{sec:utility_regret}

For selected batch $B$,
\begin{equation}
    \mathrm{P95}(B)=\mathrm{Quantile}_{0.95}\bigl(\{y_i:i\in B\}\bigr).
\end{equation}
We use linear quantile interpolation. Utility $M_{d,c}$ averages the study's batch statistic over the specified cycles and three split seeds. The empirical oracle is the best tested configuration under this utility, $c_d^\star=\arg\max_{c\in\mathcal C}M_{d,c}$, and normalized regret is
\begin{equation}
 \mathrm{Regret}(d,\hat c)=
 \frac{M_{d,c_d^\star}-M_{d,\hat c}}
 {M_{d,c_d^\star}-\min_{c\in\mathcal C}M_{d,c}},
\end{equation}
with zero regret when all configurations have equal utility. Comparisons weight datasets equally.

\section{Benchmark Datasets}
\label{app:datasets}
\label{app:dataset_registry}

Counts refer to dataset instances or endpoint tasks. The BindingDB and BindingDB-derived joint collections share 33 UniProt accessions, and all three joint preferences use the same pools. Supplementary Source Data records this overlap.

\paragraph{BindingDB targets.}
\Cref{tab:dataset_list} lists the 100 targets; their candidate tables are in Supplementary Inputs.

\begin{table}[t]
\centering\scriptsize
\setlength{\tabcolsep}{2pt}\renewcommand{\arraystretch}{1.0}
\begin{tabular}{@{}rl@{\hspace{8pt}}rl@{\hspace{8pt}}rl@{\hspace{8pt}}rl@{\hspace{8pt}}rl@{\hspace{8pt}}rl@{\hspace{8pt}}rl@{\hspace{8pt}}rl@{}}
\toprule
{\color{gray}1} & \texttt{C7C422} & {\color{gray}14} & \texttt{O15530} & {\color{gray}27} & \texttt{O75460} & {\color{gray}40} & \texttt{P00746} & {\color{gray}53} & \texttt{P04626} & {\color{gray}66} & \texttt{P07333} & {\color{gray}79} & \texttt{P08514} & {\color{gray}92} & \texttt{P10635} \\
{\color{gray}2} & \texttt{O00329} & {\color{gray}15} & \texttt{O35433} & {\color{gray}28} & \texttt{O75762} & {\color{gray}41} & \texttt{P00748} & {\color{gray}54} & \texttt{P04629} & {\color{gray}67} & \texttt{P07339} & {\color{gray}80} & \texttt{P08581} & {\color{gray}93} & \texttt{P10721} \\
{\color{gray}3} & \texttt{O00408} & {\color{gray}16} & \texttt{O42275} & {\color{gray}29} & \texttt{O75874} & {\color{gray}42} & \texttt{P00797} & {\color{gray}55} & \texttt{P05067} & {\color{gray}68} & \texttt{P07711} & {\color{gray}81} & \texttt{P08684} & {\color{gray}94} & \texttt{P11309} \\
{\color{gray}4} & \texttt{O00459} & {\color{gray}17} & \texttt{O42713} & {\color{gray}30} & \texttt{O75907} & {\color{gray}43} & \texttt{P00918} & {\color{gray}56} & \texttt{P05106} & {\color{gray}69} & \texttt{P07900} & {\color{gray}82} & \texttt{P09382} & {\color{gray}95} & \texttt{P11362} \\
{\color{gray}5} & \texttt{O00519} & {\color{gray}18} & \texttt{O43613} & {\color{gray}31} & \texttt{O76074} & {\color{gray}44} & \texttt{P01111} & {\color{gray}57} & \texttt{P05177} & {\color{gray}70} & \texttt{P07943} & {\color{gray}83} & \texttt{P09619} & {\color{gray}96} & \texttt{P11511} \\
{\color{gray}6} & \texttt{O00763} & {\color{gray}19} & \texttt{O43614} & {\color{gray}32} & \texttt{O94782} & {\color{gray}45} & \texttt{P01375} & {\color{gray}58} & \texttt{P05979} & {\color{gray}71} & \texttt{P07949} & {\color{gray}84} & \texttt{P09874} & {\color{gray}97} & \texttt{P11597} \\
{\color{gray}7} & \texttt{O14649} & {\color{gray}20} & \texttt{O43683} & {\color{gray}33} & \texttt{O94925} & {\color{gray}46} & \texttt{P03372} & {\color{gray}59} & \texttt{P06213} & {\color{gray}72} & \texttt{P08069} & {\color{gray}85} & \texttt{P09917} & {\color{gray}98} & \texttt{P11712} \\
{\color{gray}8} & \texttt{O14684} & {\color{gray}21} & \texttt{O60341} & {\color{gray}34} & \texttt{O95180} & {\color{gray}47} & \texttt{P03951} & {\color{gray}60} & \texttt{P06239} & {\color{gray}73} & \texttt{P08183} & {\color{gray}86} & \texttt{P09960} & {\color{gray}99} & \texttt{P11802} \\
{\color{gray}9} & \texttt{O14744} & {\color{gray}22} & \texttt{O60563} & {\color{gray}35} & \texttt{O95271} & {\color{gray}48} & \texttt{P03952} & {\color{gray}61} & \texttt{P06241} & {\color{gray}74} & \texttt{P08235} & {\color{gray}87} & \texttt{P0C6X7} & {\color{gray}100} & \texttt{P12931} \\
{\color{gray}10} & \texttt{O14757} & {\color{gray}23} & \texttt{O60885} & {\color{gray}36} & \texttt{P00374} & {\color{gray}49} & \texttt{P03956} & {\color{gray}62} & \texttt{P06276} & {\color{gray}75} & \texttt{P08238} & {\color{gray}88} & \texttt{P0DTC1} &  &  \\
{\color{gray}11} & \texttt{O14920} & {\color{gray}24} & \texttt{O70536} & {\color{gray}37} & \texttt{P00519} & {\color{gray}50} & \texttt{P04049} & {\color{gray}63} & \texttt{P06401} & {\color{gray}76} & \texttt{P08246} & {\color{gray}89} & \texttt{P0DTD1} &  &  \\
{\color{gray}12} & \texttt{O14965} & {\color{gray}25} & \texttt{O75116} & {\color{gray}38} & \texttt{P00734} & {\color{gray}51} & \texttt{P04150} & {\color{gray}64} & \texttt{P06493} & {\color{gray}77} & \texttt{P08253} & {\color{gray}90} & \texttt{P10275} &  &  \\
{\color{gray}13} & \texttt{O15379} & {\color{gray}26} & \texttt{O75417} & {\color{gray}39} & \texttt{P00742} & {\color{gray}52} & \texttt{P04585} & {\color{gray}65} & \texttt{P06870} & {\color{gray}78} & \texttt{P08254} & {\color{gray}91} & \texttt{P10415} &  &  \\
\bottomrule
\end{tabular}
\caption{Protein accessions for the 100 BindingDB datasets.}
\label{tab:dataset_list}
\end{table}

\subsection{Configuration-Selection Task}

For target dataset $d$, the configuration recommender receives dataset descriptions, the catalog $\mathcal C$, and optional source-campaign outcomes, then returns an ordered list of $k$ configurations:
\begin{equation}
    \hat C_d=(\hat c_{d,1},\ldots,\hat c_{d,k}),\qquad \hat c_{d,j}\in\mathcal C.
\end{equation}
Source campaigns come from the training folds; outcomes from the held-out target score the final recommendation. The reported direct-prediction and ICL comparisons use completed source campaigns as fixed context.

\section{Compared Acquisition Rules}
\label{app:configs}
\label{app:top_configs}

Acquisition abbreviations are defined as follows: Bayesian active learning by disagreement (BALD), batch-mode deep active learning (BMDAL), batch active learning via information matrices (BAIT), batch active learning by diverse gradient embeddings (BADGE), largest cluster maximum distance (LCMD), probability of improvement (PI), max-value entropy search (MES), and general-purpose information-based Bayesian optimization (GIBBON). MC denotes Monte Carlo. \Cref{tab:config_space} lists the 57 BindingDB configurations under the labels C01-C57 used in the figures; within each row, labels follow the listed rule or coefficient order.

\begin{table}[t]
\centering\footnotesize
\setlength{\tabcolsep}{4pt}\renewcommand{\arraystretch}{1.05}
\begin{tabular}{@{}llrll@{}}
\toprule
Model & Labels & $n$ & Acquisition rule & Parameters \\
\midrule
XGB & C01, C02 & 2 & Greedy mean; uniform random & --- \\
 & C03, C05, C07 & 3 & Value-biased CoreSet & $t=0.1,1,10$ \\
 & C04, C06, C08 & 3 & QBC (leaf-index mean and dispersion) & $t=0.1,1,10$ \\
 & C09--C17 & 9 & QBC-diversity (leaf-index) & $(t,d)\in T^2$ \\
\addlinespace[2pt]
NN & C18, C19 & 2 & Greedy mean; uniform random & --- \\
 & C20, C22, C24 & 3 & Value-biased CoreSet & $t=0.1,1,10$ \\
 & C21, C23, C25 & 3 & QBC (MC dropout) & $t=0.1,1,10$ \\
 & C26--C34 & 9 & QBC-diversity (MC dropout) & $(t,d)\in T^2$ \\
 & C35, C36, C37 & 3 & BALD & $t=0.1,1,10$ \\
 & C38, C39, C40 & 3 & Value-biased BAIT & $t=0.1,1,10$ \\
 & C41, C42, C43 & 3 & Value-biased BADGE & $t=0.1,1,10$ \\
 & C44, C45, C46 & 3 & Value-biased LCMD & $t=0.1,1,10$ \\
\addlinespace[2pt]
GP & C47, C48 & 2 & Greedy mean; uniform random & --- \\
 & C49, C50, C51 & 3 & qEI; qNEI; qPI & --- \\
 & C52, C53, C54 & 3 & GP-BALD; MES; GIBBON & --- \\
 & C55, C56, C57 & 3 & qUCB & $\beta=0.5,2,8$ \\
\bottomrule
\end{tabular}
\caption{The 57 BindingDB configurations, labelled C01--C57 in the figures, with their acquisition rules and parameter grids. The set $T=\{0.1,1,10\}$ is crossed for QBC-diversity, with $d$ running through $T$ before $t$ advances. The coefficient $t$ weights the mean or exploitation term, $d$ weights diversity and $\beta$ weights uncertainty in UCB.}
\label{tab:main_configuration_labels}\label{tab:config_space}
\end{table}

BindingDB has 17 XGBoost (XGB)~\citep{chen2016xgboost}, 29 neural-network (NN), and 11 Gaussian-process (GP)~\citep{rasmussen2006gaussian} configurations. The scalar ADMET and joint catalogs add 20 variational Bayesian last-layer (VBLL) configurations~\citep{harrison2024vbll}. \Cref{tab:config_space} gives the BindingDB rules and coefficient grids.

The catalog includes CoreSet~\citep{sener2018active}, query by committee (QBC)~\citep{seung1992query}, and BMDAL~\citep{holzmuller2023bmdal} rules based on information matrices (BAIT), gradient diversity (BADGE), and cluster distances (LCMD)~\citep{ash2020badge,ash2021bait}. GP rules include probability and expected improvement~\citep{kushner1964new,jones1998efficient}, upper confidence bounds~\citep{srinivas2010gaussian}, BALD~\citep{houlsby2011bayesian}, MES~\citep{wang2017max}, and GIBBON~\citep{moss2021gibbon}.

XGB configurations labeled QBC use terminal-leaf identifiers across trees. For candidate $x$, let $\ell_r(x)$ be its identifier in tree $r$, $v(x)=\mathrm{Var}_r[\ell_r(x)]$ and $m(x)=\mathrm{Mean}_r[\ell_r(x)]$. Their score is
\begin{equation}
 a(x)=\mathcal N(v)(x)+t\mathcal N(m)(x),
\end{equation}
where $\mathcal N$ min-max scales current candidate values to $[0,1]$, returning one for a constant vector. QBC-diversity (QBCD) adds $d\mathcal N(\delta)(x)$, where $\delta$ is the minimum distance to labeled and already selected batch members, normalized over remaining candidates. Thus $t$ weights leaf identifiers, not property predictions; XGB greedy exploitation uses predicted property means. Appendix~\ref{app:catalog_sensitivity} tests removal of these twelve leaf-statistic configurations.

NN/VBLL QBC uses the same score with property predictions from MC dropout~\citep{gal2016dropout}; mean-only variants rank by raw predictive mean. Within a preselected shortlist, CoreSet and BMDAL hybrids minimize the selection-rule rank plus $t$ times the greedy-exploitation rank. Coefficients therefore have family-specific meanings. VBLL adds greedy exploitation, random sampling, three CoreSet, three QBC, nine QBC-diversity and three mean-only entries.

GP acquisitions score candidates individually ($q=1$) and take the highest scores to form a batch. Balanced joint selection adds qEHVI, qNEHVI and qNParEGO~\citep{daulton2020qehvi,daulton2021qnehvi,knowles2006parego}; preference and recommendation comparisons use the 77 scalar configurations.

\section{Single-Objective Data Selection: Additional Results}

\begin{figure}[t]
\centering
\includegraphics[width=\linewidth]{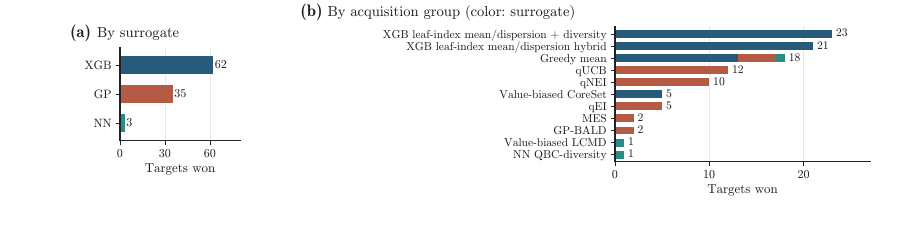}
\caption{Number of BindingDB targets on which each (a) surrogate family and (b) acquisition-rule group ranks first.}
\label{fig:bindingdb_winners}
\end{figure}

\Cref{fig:bindingdb_winners}(a,b) counts the targets on which each surrogate family and acquisition-rule group ranks first using final-cycle batch mean after seed averaging, with ties resolved in catalog order. Numerical cycle-wise P95 values are in \Cref{tab:all57-p95-cycle,tab:all57-p95-cycle-cont}.

\label{app:single_results}
\subsection{Split Variability and Tied Outcomes}\label{app:stability}

We select the configuration with highest average utility on two campaign splits and evaluate it on the third, rotating the held-out split and resolving ties by catalog order. Regret uses the held-out split's utility range. Rotations are averaged within target, then across 100 targets. Five-cycle mean averages all 500 acquired observations; five-cycle P95 averages the five batch percentiles.

Two-split selection improves final-batch quality more than five-cycle mean quality. Paired regret differences from XGB greedy exploitation (selected minus XGB) are $-0.096$ [$-0.144,-0.051$], $-0.261$ [$-0.321,-0.201$], $0.004$ [$-0.0002,0.008$] for five-cycle mean, and $-0.139$ [$-0.183,-0.098$] for five-cycle P95. Brackets give descriptive 95\% intervals from 10,000 target-bootstrap resamples. XGB's five-cycle mean regret is already 0.020.

The 57 configurations yield 53 distinct sequences of batch mean/P95 across 1,500 target-seed-cycle combinations. NN greedy exploitation and its three mean-only variants coincide, as do XGB and NN random sampling. Grouping these sequences leaves 34 groups that attain the highest final-batch P95 on at least one target and reduces the mean number within 5\% normalized regret from 3.84 to 3.68. Target P01111 has equal P95 for every configuration and regret zero. Tie-aware Hit@1 is unchanged for all 12 main methods with saved lists; Random's analytic Hit@1 changes from 1.754\% to 1.807\%. Appendix~\ref{app:llm_sensitivity} extends the analysis to aggregate recommendation utilities across all five settings.

\subsection{Complete Cycle Results}

GP configurations occupy nine of the top ten positions by five-cycle mean P95 (\Cref{tab:all57-p95-cycle,tab:all57-p95-cycle-cont}). GP qUCB ($\beta=8$) leads at 8.849; GP MES, GIBBON, qUCB ($\beta=2$), qNEI and qEI are within roughly 0.011 pIC$_{50}$. The top 20 comprises nine GP, nine XGB and two NN configurations.

Leading GP and XGB configurations tend to acquire high-potency compounds early, with batch P95 declining over later cycles. Several diversity-based XGB and NN configurations improve or remain steadier but have lower five-cycle averages. The best five-cycle mean P95 exceeds the worst by about 0.57 pIC$_{50}$ and XGB Random by about 0.28.

\begin{table}[t]
\centering\scriptsize\setlength{\tabcolsep}{2.6pt}\renewcommand{\arraystretch}{0.96}
\begin{tabular}{@{}rllp{2.08in}ccccc@{}}
\toprule
Rank & Code & Pos. & Configuration & Cycle 1 & Cycle 2 & Cycle 3 & Cycle 4 & Cycle 5 \\
\midrule
1 & C57 & 1 & GP qUCB ($\beta{=}8.0$) & 9.18$\pm$0.94 & 9.10$\pm$0.86 & 8.85$\pm$0.91 & 8.63$\pm$0.92 & 8.49$\pm$0.97 \\
2 & C53 & 7 & GP MES & 9.16$\pm$0.97 & 9.08$\pm$0.91 & 8.86$\pm$0.90 & 8.66$\pm$0.92 & 8.47$\pm$0.93 \\
3 & C54 & 8 & GP GIBBON & 9.16$\pm$0.94 & 9.07$\pm$0.89 & 8.85$\pm$0.92 & 8.64$\pm$0.94 & 8.50$\pm$0.96 \\
4 & C56 & 2 & GP qUCB ($\beta{=}2.0$) & 9.20$\pm$0.93 & 9.09$\pm$0.89 & 8.86$\pm$0.90 & 8.62$\pm$0.89 & 8.44$\pm$0.97 \\
5 & C50 & 5 & GP qNEI & 9.16$\pm$0.94 & 9.11$\pm$0.87 & 8.85$\pm$0.93 & 8.62$\pm$0.95 & 8.46$\pm$0.95 \\
6 & C49 & 3 & GP qEI & 9.19$\pm$0.94 & 9.12$\pm$0.88 & 8.85$\pm$0.94 & 8.60$\pm$0.92 & 8.44$\pm$0.98 \\
7 & C55 & 4 & GP qUCB ($\beta{=}0.5$) & 9.22$\pm$0.93 & 9.10$\pm$0.89 & 8.83$\pm$0.92 & 8.60$\pm$0.93 & 8.42$\pm$0.97 \\
8 & C47 & 6 & GP Greedy mean & 9.24$\pm$0.91 & 9.08$\pm$0.89 & 8.82$\pm$0.92 & 8.60$\pm$0.91 & 8.42$\pm$0.98 \\
9 & C01 & 9 & XGB Greedy mean & 9.30$\pm$0.85 & 9.00$\pm$0.90 & 8.77$\pm$0.95 & 8.55$\pm$0.97 & 8.42$\pm$0.97 \\
10 & C51 & 10 & GP qPI & 9.17$\pm$0.95 & 9.04$\pm$0.93 & 8.74$\pm$0.96 & 8.58$\pm$0.94 & 8.46$\pm$0.91 \\
11 & C07 & 11 & XGB val.-biased CoreSet (t=10.0) & 9.09$\pm$0.83 & 8.85$\pm$0.86 & 8.68$\pm$0.86 & 8.52$\pm$0.89 & 8.39$\pm$0.98 \\
12 & C09 & 13 & XGB leaf-index + div. (t=0.1, d=0.1) & 8.61$\pm$0.86 & 8.71$\pm$0.87 & 8.76$\pm$0.95 & 8.69$\pm$0.97 & 8.58$\pm$1.04 \\
13 & C04 & 14 & XGB leaf-index QBC (t=0.1) & 8.62$\pm$0.86 & 8.70$\pm$0.90 & 8.75$\pm$0.94 & 8.69$\pm$0.98 & 8.57$\pm$1.03 \\
14 & C05 & 12 & XGB val.-biased CoreSet (t=1.0) & 8.79$\pm$0.83 & 8.71$\pm$0.83 & 8.66$\pm$0.81 & 8.58$\pm$0.85 & 8.53$\pm$0.84 \\
15 & C38 & 17 & NN val.-biased BAIT (t=0.1) & 8.65$\pm$0.78 & 8.63$\pm$0.79 & 8.63$\pm$0.80 & 8.60$\pm$0.82 & 8.60$\pm$0.84 \\
16 & C10 & 19 & XGB leaf-index + div. (t=0.1, d=1.0) & 8.45$\pm$0.79 & 8.56$\pm$0.90 & 8.65$\pm$0.94 & 8.69$\pm$1.00 & 8.68$\pm$1.00 \\
17 & C39 & 20 & NN val.-biased BAIT (t=1.0) & 8.61$\pm$0.82 & 8.61$\pm$0.81 & 8.58$\pm$0.80 & 8.58$\pm$0.80 & 8.58$\pm$0.82 \\
18 & C06 & 16 & XGB leaf-index QBC (t=1.0) & 8.58$\pm$0.99 & 8.62$\pm$1.04 & 8.63$\pm$1.00 & 8.61$\pm$1.02 & 8.51$\pm$1.06 \\
19 & C12 & 15 & XGB leaf-index + div. (t=1.0, d=0.1) & 8.58$\pm$1.00 & 8.64$\pm$1.03 & 8.66$\pm$1.04 & 8.58$\pm$1.03 & 8.48$\pm$1.00 \\
20 & C13 & 18 & XGB leaf-index + div. (t=1.0, d=1.0) & 8.42$\pm$0.93 & 8.56$\pm$0.98 & 8.63$\pm$1.01 & 8.64$\pm$1.04 & 8.64$\pm$1.04 \\
21 & C18 & 21 & NN Greedy mean & 8.59$\pm$0.82 & 8.59$\pm$0.80 & 8.57$\pm$0.82 & 8.55$\pm$0.82 & 8.55$\pm$0.81 \\
22 & C35 & 22 & NN mean-only (t=0.1) & 8.59$\pm$0.82 & 8.59$\pm$0.80 & 8.57$\pm$0.82 & 8.55$\pm$0.82 & 8.55$\pm$0.81 \\
23 & C36 & 23 & NN mean-only (t=1.0) & 8.59$\pm$0.82 & 8.59$\pm$0.80 & 8.57$\pm$0.82 & 8.55$\pm$0.82 & 8.55$\pm$0.81 \\
24 & C37 & 24 & NN mean-only (t=10.0) & 8.59$\pm$0.82 & 8.59$\pm$0.80 & 8.57$\pm$0.82 & 8.55$\pm$0.82 & 8.55$\pm$0.81 \\
25 & C23 & 26 & NN val.-biased QBC (t=1.0) & 8.59$\pm$0.82 & 8.58$\pm$0.81 & 8.57$\pm$0.82 & 8.56$\pm$0.82 & 8.55$\pm$0.81 \\
26 & C25 & 25 & NN val.-biased QBC (t=10.0) & 8.59$\pm$0.82 & 8.59$\pm$0.80 & 8.57$\pm$0.82 & 8.56$\pm$0.82 & 8.55$\pm$0.81 \\
27 & C21 & 33 & NN val.-biased QBC (t=0.1) & 8.59$\pm$0.82 & 8.58$\pm$0.81 & 8.57$\pm$0.83 & 8.56$\pm$0.82 & 8.54$\pm$0.81 \\
28 & C48 & 29 & GP Uniform random & 8.57$\pm$0.82 & 8.58$\pm$0.83 & 8.56$\pm$0.81 & 8.57$\pm$0.82 & 8.56$\pm$0.83 \\
29 & C02 & 27 & XGB Uniform random & 8.57$\pm$0.82 & 8.58$\pm$0.83 & 8.57$\pm$0.81 & 8.57$\pm$0.81 & 8.55$\pm$0.82 \\
\bottomrule
\end{tabular}
\caption{Batch P95 on BindingDB in each acquisition cycle: mean $\pm$ between-dataset standard deviation after seed averaging, for all 57 configurations. Rank orders the five-cycle mean, and Pos.\ is the position of the configuration in \Cref{fig:metric_rank}. Codes follow \Cref{tab:main_configuration_labels}. Ranks 1--29.}
\label{tab:all57-p95-cycle}\label{tab:bindingdb_plot_lookup}
\end{table}

\begin{table}[t]
\centering\scriptsize\setlength{\tabcolsep}{2.6pt}\renewcommand{\arraystretch}{0.96}
\begin{tabular}{@{}rllp{2.08in}ccccc@{}}
\toprule
Rank & Code & Pos. & Configuration & Cycle 1 & Cycle 2 & Cycle 3 & Cycle 4 & Cycle 5 \\
\midrule
30 & C19 & 28 & NN Uniform random & 8.57$\pm$0.82 & 8.58$\pm$0.83 & 8.57$\pm$0.81 & 8.57$\pm$0.81 & 8.55$\pm$0.82 \\
31 & C40 & 34 & NN val.-biased BAIT (t=10.0) & 8.55$\pm$0.81 & 8.55$\pm$0.84 & 8.56$\pm$0.81 & 8.58$\pm$0.83 & 8.56$\pm$0.83 \\
32 & C52 & 38 & GP GP-BALD & 8.57$\pm$0.83 & 8.55$\pm$0.82 & 8.51$\pm$0.78 & 8.50$\pm$0.80 & 8.51$\pm$0.80 \\
33 & C43 & 37 & NN val.-biased BADGE (t=10.0) & 8.54$\pm$0.84 & 8.51$\pm$0.82 & 8.51$\pm$0.80 & 8.53$\pm$0.80 & 8.55$\pm$0.81 \\
34 & C46 & 36 & NN val.-biased LCMD (t=10.0) & 8.50$\pm$0.80 & 8.53$\pm$0.80 & 8.55$\pm$0.83 & 8.51$\pm$0.80 & 8.54$\pm$0.82 \\
35 & C24 & 35 & NN val.-biased CoreSet (t=10.0) & 8.50$\pm$0.80 & 8.51$\pm$0.82 & 8.50$\pm$0.85 & 8.53$\pm$0.85 & 8.53$\pm$0.84 \\
36 & C42 & 39 & NN val.-biased BADGE (t=1.0) & 8.52$\pm$0.83 & 8.52$\pm$0.83 & 8.49$\pm$0.79 & 8.50$\pm$0.84 & 8.52$\pm$0.81 \\
37 & C41 & 41 & NN val.-biased BADGE (t=0.1) & 8.50$\pm$0.83 & 8.51$\pm$0.80 & 8.48$\pm$0.82 & 8.49$\pm$0.81 & 8.50$\pm$0.83 \\
38 & C45 & 40 & NN val.-biased LCMD (t=1.0) & 8.46$\pm$0.81 & 8.48$\pm$0.80 & 8.49$\pm$0.81 & 8.50$\pm$0.80 & 8.52$\pm$0.81 \\
39 & C15 & 31 & XGB leaf-index + div. (t=10.0, d=0.1) & 8.44$\pm$0.98 & 8.50$\pm$1.01 & 8.51$\pm$1.08 & 8.48$\pm$1.05 & 8.50$\pm$1.00 \\
40 & C08 & 30 & XGB leaf-index QBC (t=10.0) & 8.44$\pm$0.99 & 8.48$\pm$1.01 & 8.48$\pm$1.05 & 8.50$\pm$1.03 & 8.49$\pm$1.01 \\
41 & C16 & 32 & XGB leaf-index + div. (t=10.0, d=1.0) & 8.44$\pm$0.98 & 8.46$\pm$1.04 & 8.52$\pm$1.07 & 8.50$\pm$1.05 & 8.47$\pm$1.03 \\
42 & C44 & 44 & NN val.-biased LCMD (t=0.1) & 8.41$\pm$0.80 & 8.46$\pm$0.83 & 8.47$\pm$0.81 & 8.48$\pm$0.81 & 8.50$\pm$0.83 \\
43 & C22 & 43 & NN val.-biased CoreSet (t=1.0) & 8.36$\pm$0.79 & 8.39$\pm$0.75 & 8.42$\pm$0.79 & 8.45$\pm$0.78 & 8.46$\pm$0.82 \\
44 & C17 & 42 & XGB leaf-index + div. (t=10.0, d=10.0) & 8.20$\pm$0.83 & 8.36$\pm$0.88 & 8.45$\pm$0.90 & 8.50$\pm$0.93 & 8.56$\pm$0.92 \\
45 & C03 & 45 & XGB val.-biased CoreSet (t=0.1) & 8.19$\pm$0.79 & 8.29$\pm$0.79 & 8.36$\pm$0.82 & 8.40$\pm$0.81 & 8.43$\pm$0.81 \\
46 & C14 & 46 & XGB leaf-index + div. (t=1.0, d=10.0) & 8.07$\pm$0.76 & 8.26$\pm$0.83 & 8.36$\pm$0.83 & 8.41$\pm$0.86 & 8.49$\pm$0.83 \\
47 & C11 & 47 & XGB leaf-index + div. (t=0.1, d=10.0) & 8.05$\pm$0.75 & 8.27$\pm$0.81 & 8.36$\pm$0.82 & 8.41$\pm$0.86 & 8.47$\pm$0.82 \\
48 & C30 & 50 & NN QBC-diversity (t=1.0, d=1.0) & 8.03$\pm$0.75 & 8.24$\pm$0.80 & 8.32$\pm$0.79 & 8.39$\pm$0.83 & 8.43$\pm$0.79 \\
49 & C31 & 53 & NN QBC-diversity (t=1.0, d=10.0) & 8.03$\pm$0.74 & 8.23$\pm$0.81 & 8.32$\pm$0.80 & 8.39$\pm$0.83 & 8.43$\pm$0.79 \\
50 & C29 & 52 & NN QBC-diversity (t=1.0, d=0.1) & 8.03$\pm$0.75 & 8.24$\pm$0.80 & 8.32$\pm$0.79 & 8.38$\pm$0.84 & 8.43$\pm$0.79 \\
51 & C32 & 48 & NN QBC-diversity (t=10.0, d=0.1) & 8.03$\pm$0.75 & 8.24$\pm$0.80 & 8.32$\pm$0.80 & 8.38$\pm$0.84 & 8.42$\pm$0.80 \\
52 & C28 & 55 & NN QBC-diversity (t=0.1, d=10.0) & 8.03$\pm$0.75 & 8.23$\pm$0.80 & 8.31$\pm$0.81 & 8.40$\pm$0.82 & 8.42$\pm$0.80 \\
53 & C33 & 51 & NN QBC-diversity (t=10.0, d=1.0) & 8.03$\pm$0.74 & 8.24$\pm$0.80 & 8.32$\pm$0.79 & 8.38$\pm$0.84 & 8.42$\pm$0.80 \\
54 & C26 & 54 & NN QBC-diversity (t=0.1, d=0.1) & 8.04$\pm$0.75 & 8.24$\pm$0.81 & 8.32$\pm$0.79 & 8.38$\pm$0.83 & 8.42$\pm$0.79 \\
55 & C34 & 49 & NN QBC-diversity (t=10.0, d=10.0) & 8.03$\pm$0.74 & 8.24$\pm$0.80 & 8.31$\pm$0.80 & 8.39$\pm$0.84 & 8.42$\pm$0.80 \\
56 & C20 & 56 & NN val.-biased CoreSet (t=0.1) & 8.03$\pm$0.75 & 8.24$\pm$0.81 & 8.32$\pm$0.81 & 8.39$\pm$0.83 & 8.43$\pm$0.80 \\
57 & C27 & 57 & NN QBC-diversity (t=0.1, d=1.0) & 8.03$\pm$0.75 & 8.24$\pm$0.81 & 8.32$\pm$0.80 & 8.39$\pm$0.82 & 8.42$\pm$0.79 \\
\bottomrule
\end{tabular}
\caption{Batch P95 on BindingDB in each acquisition cycle: mean $\pm$ between-dataset standard deviation after seed averaging, for all 57 configurations. Rank orders the five-cycle mean, and Pos.\ is the position of the configuration in \Cref{fig:metric_rank}. Codes follow \Cref{tab:main_configuration_labels}. Ranks 30--57.}
\label{tab:all57-p95-cycle-cont}
\end{table}

\section{Multi-Objective Data Selection: Additional Results}

\label{app:extended_simulations}

\subsection{Joint Objectives and Component Metrics}

Let $b_i$ and $s_i$ denote binding and ADMET desirability, oriented so that higher values are preferred. Predicted XTT toxicity $p_i$ gives $s_i=1-p_i$; measured endpoints follow \Cref{app:measured_group_b}. Scalar campaigns standardize both objectives using full-pool means and population standard deviations before splitting. Weights are ordered binding/ADMET: balanced acquisition uses $(1,1)$ and evaluation $(0.5,0.5)$; ADMET-heavy and binding-heavy campaigns use $(0.25,0.75)$ and $(0.75,0.25)$ for both. Each of the 77 scalar configurations runs separately under each preference.

The three multi-objective GP configurations fit one GP per objective, standardized on current labeled data. Their reference point is 0.1 below the smallest standardized labeled value in each objective. The qEHVI and qNEHVI entries use log expected and log noisy expected hypervolume improvement. qNParEGO takes each candidate's largest log noisy expected improvement over eight random augmented-Chebyshev scalarizations. All three score candidates individually ($q=1$) and select the highest-scoring batch.

For each batch $B_t$ on the 90 pools with XTT predictions,
\begin{align}
 B_t^{\mathrm{NT}}&=\{i\in B_t:p_i<0.5\},\\
 T(B_t)&=\frac{100}{|B_t|}\sum_{i\in B_t}\mathbb{1}[p_i\geq0.5],\\
 Q(B_t)&=\mathrm{Quantile}_{0.95}\{z_b(i):i\in B_t^{\mathrm{NT}}\}.
\end{align}
Here $T$ is the predicted-toxic percentage and $Q$ is conditional binding P95, using full-pool standardized binding $z_b$. We average batch statistics over five cycles, then three split seeds, then equally across datasets. Scalar P95 instead evaluates every selected molecule's weighted score $w_bz_b+w_sz_s$ across all 100 joint pools.

Every main and preference batch contains a predicted non-toxic molecule. In the budget sweep, 33 of 727{,}650 batches contain none: 23, five, two, two and one at batch sizes 20, 40, 80, 100 and 160, respectively, all with initialization size 200. These batches are omitted from within-seed conditional-binding averages; all batches contribute to predicted-toxic fractions.

\subsection{Measured ADMET Endpoints}
\label{app:measured_group_b}

The ten measured-ADMET pools pair measured endpoints with predicted Boltz-2 binding. Their objectives minimize PXR pEC$_{50}$, PXR single-concentration log$_2$ fold change, Ames mutagenicity, three Tox21 activity labels, LD50$_{\mathrm{Zhu}}$, LogD and human microsomal clearance, and maximize kinetic solubility. LD50$_{\mathrm{Zhu}}$ is a negative logarithm of molar dose, so lower encoded values indicate higher lethal doses. Joint LogD minimization differs from the standalone objective $-|y-2|$.

ExpansionRx human microsomal clearance is recorded in mL/min/kg~\citep{openadmet2025expansionrx}. Historical joint prompts mislabeled it as $\mu$L/min/mg.

Across all 100 pools, the highest mean scalar P95 is attained by GP qUCB $(\beta=2)$ under balanced weights, GP qUCB $(\beta=8)$ under binding-heavy weights, and VBLL QBC-diversity $(t=10,d=1)$ under ADMET-heavy weights. These are within-preference comparisons because the utility changes with the weights.

\subsection{Dataset-Level Winners and Binding Sources}
\label{app:joint_winners}

\begin{figure}[t]
\centering
\includegraphics[width=\linewidth]{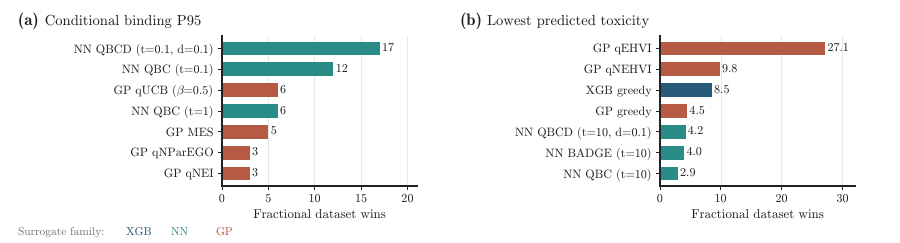}
\caption{The seven most frequent winners on 90 XTT datasets for (a) conditional binding P95 and (b) lowest predicted-toxic fraction. Ties share one dataset credit.}
\label{fig:joint_top_winners}
\end{figure}

The seven configurations that rank first most often differ between conditional binding P95 and predicted-toxic fraction (\Cref{fig:joint_top_winners}). NN configurations receive 48 of 90 dataset wins for conditional binding P95; GP qEHVI receives 27.1 wins for the lowest predicted-toxic fraction. Each dataset contributes one win, divided equally among tied configurations.

\label{app:weight_changes}

\subsection{Acquisition Budget Sensitivity}
\label{app:al_knob_sensitivity}

Increasing the initial labeled-set size to 400 or 800 retains all ten leading configurations under conditional binding P95 on the 90 XTT datasets (\Cref{tab:budget_all}).

The seven budget settings use 90 XTT pools and 77 scalar configurations. Fixed-size initialization reshuffles the combined initial and acquisition pools while retaining the diagnostic set. Comparisons therefore use the sweep's own $(200,100)$ reference rather than the main study's capped-initialization split.

\begin{table}[t]
\centering\footnotesize

\begin{tabular}{@{}rrrrr@{}}
\toprule
Initial size & Batch & Binding rank $\rho$ & Toxicity rank $\rho$ & Top-ten retained\\
\midrule
200 & 100 & 1.000 & 1.000 & 10\\
400 & 100 & 0.991 & 0.997 & 10\\
800 & 100 & 0.988 & 0.995 & 10\\
200 & 20 & 0.891 & 0.980 & 2\\
200 & 40 & 0.918 & 0.984 & 2\\
200 & 80 & 0.979 & 0.997 & 6\\
200 & 160 & 0.965 & 0.997 & 7\\
\bottomrule
\end{tabular}
\caption{Seven budget settings on the 90 XTT pools relative to the $(200,100)$ reference, with initial labeled-set size first and batch size second. Correlations use all 77 configurations; top-ten sets include all configurations tied at the cutoff.}
\label{tab:budget_all}
\end{table}

\subsection{Binding Metrics and the Binding-Toxicity Pareto Front}

\label{app:component_metrics}

\begin{figure}[t]
\centering
\includegraphics[width=\linewidth]{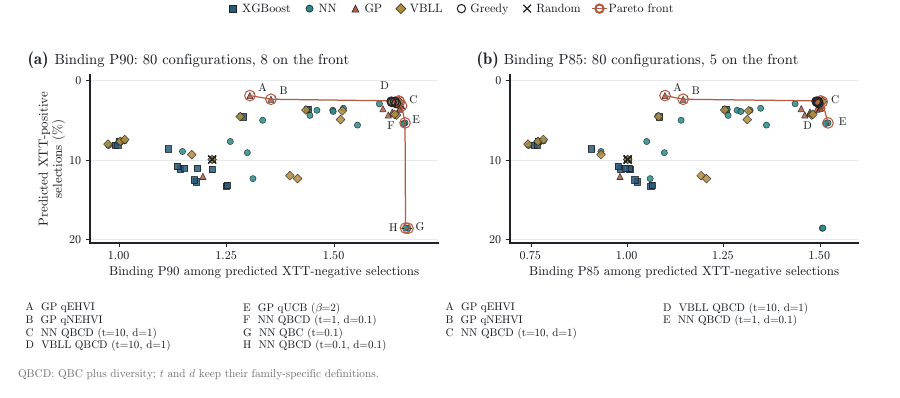}
\caption{Balanced binding-toxicity comparison on the 90 XTT datasets. Panels (a,b) use P90 and P85 binding among predicted non-toxic selections and retain all 80 configurations. Letters identify configurations on the Pareto front, with their names given beneath each panel.}
\label{fig:appendix_main_safe_p90}\label{fig:appendix_main_safe_p85}
\end{figure}

\begin{figure}[t]
\centering
\includegraphics[width=\linewidth]{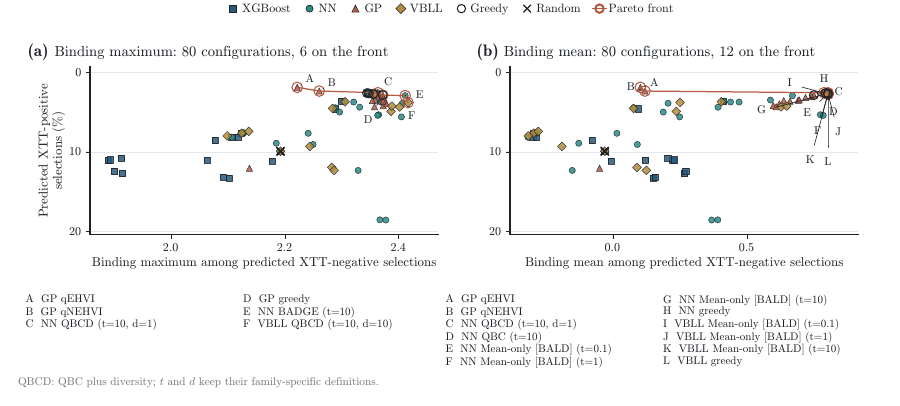}
\caption{Balanced binding-toxicity comparison on the 90 XTT datasets. Panels (a,b) use maximum and mean binding among predicted non-toxic selections and retain all 80 configurations. Letters identify configurations on the Pareto front, with their names given beneath each panel.}
\label{fig:appendix_main_safe_max}\label{fig:appendix_main_safe_mean}
\end{figure}

Changing the binding statistic changes the binding-toxicity Pareto front (\Cref{fig:appendix_main_safe_p90,fig:appendix_main_safe_max}). With balanced acquisition, P90, P85, maximum and mean binding among predicted non-toxic selections yield eight, five, six and twelve Pareto-front entries, respectively; the twelve mean-binding entries occupy six distinct coordinate pairs. GP qEHVI has the lowest predicted-toxic fraction on every front. The highest-binding configuration shifts from NN QBC-diversity for P90/P85 to VBLL QBC-diversity for the maximum and VBLL greedy exploitation for the mean. All panels use the same five-cycle, three-seed, 90-dataset aggregation.

\begin{figure}[t]
\centering
\includegraphics[width=\linewidth]{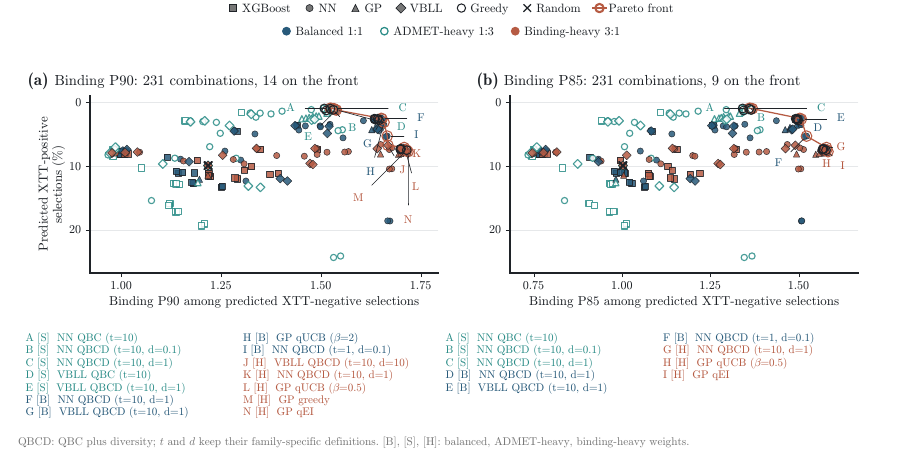}
\caption{Three-preference binding-toxicity comparison on the 90 XTT datasets. Panels (a,b) use P90 and P85 binding among predicted non-toxic selections and retain all 231 configuration-weight combinations. Letters identify configurations on the Pareto front, with their names given beneath each panel.}
\label{fig:appendix_weights_safe_p90}\label{fig:appendix_weights_safe_p85}
\end{figure}

\begin{figure}[t]
\centering
\includegraphics[width=\linewidth]{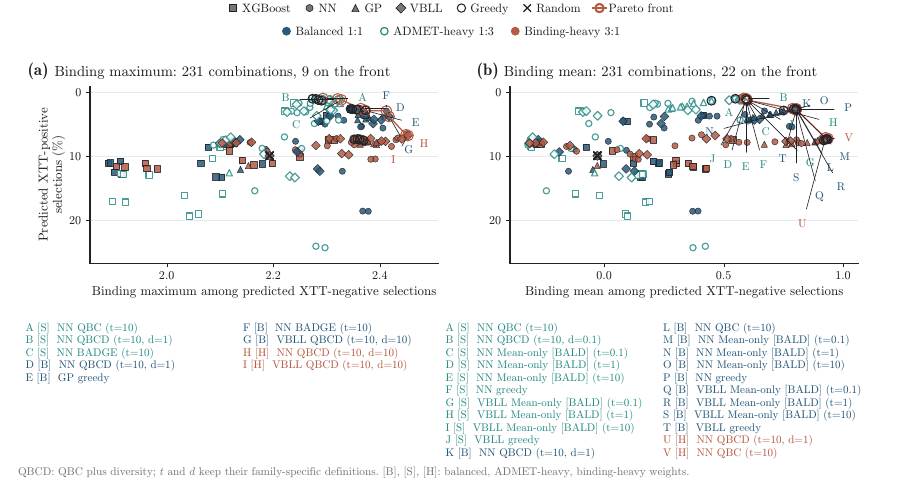}
\caption{Three-preference binding-toxicity comparison on the 90 XTT datasets. Panels (a,b) use maximum and mean binding among predicted non-toxic selections and retain all 231 configuration-weight combinations. Letters identify configurations on the Pareto front, with their names given beneath each panel.}
\label{fig:appendix_weights_safe_max}\label{fig:appendix_weights_safe_mean}
\end{figure}

Across the three preferences and 77 scalar configurations, P90, P85, maximum and mean binding yield 14, nine, nine and 22 Pareto-front entries (\Cref{fig:appendix_weights_safe_p90,fig:appendix_weights_safe_max}); the 22 mean-binding entries occupy ten distinct coordinate pairs. ADMET-heavy NN QBC $(t=10)$ has the lowest predicted-toxic fraction on every front. The highest-binding entries are GP qEI for P90/P85, VBLL QBC-diversity for the maximum and NN QBC $(t=10)$ for the mean. Both acquisition preference and evaluation statistic therefore affect the trade-off.

\section{Cross-Property ADMET Study: Additional Results}
\label{app:admet_results}

Assay labels in \Cref{tab:admet_endpoint_lookup} use MDR1 for multidrug resistance protein 1, MDCKII for Madin-Darby canine kidney II cells, Papp for apparent permeability, and A/B for apical/basolateral transport. CYP3A4 denotes cytochrome P450 3A4; EC$_{50}$ is half-maximal effective concentration, Emax maximal effect, and CL$_{\mathrm{int}}$ intrinsic clearance. Prefixes NR and SR denote nuclear-receptor and stress-response assays; AR is androgen receptor, LBD ligand-binding domain, AhR aryl hydrocarbon receptor, ER estrogen receptor, PPAR-$\gamma$ peroxisome proliferator-activated receptor gamma, ARE antioxidant response element, ATAD5 ATPase family AAA domain-containing protein 5, HSE heat-shock element, and MMP mitochondrial membrane potential. Potencies pIC$_{50}$ and pEC$_{50}$ are negative base-10 logarithms of molar concentrations. Units for E02, E18 and E19 follow the ASAP source assay definitions~\citep{openadmet2026asap}.

\begin{table}[t]
\centering\scriptsize
\setlength{\tabcolsep}{2.5pt}\renewcommand{\arraystretch}{0.97}
\begin{tabular}{@{}lp{2.4in}p{0.72in}p{0.6in}p{1.3in}@{}}
\toprule
Code & Endpoint & Category & Preference & Reported unit \\
\midrule
E01 & Kinetic solubility (ASAP) & Absorption & Maximize & $\mu$M \\
E02 & MDR1-MDCKII permeability & Absorption & Maximize & $10^{-6}$ cm/s \\
E03 & Kinetic solubility (ExpansionRx) & Absorption & Maximize & $\mu$M \\
E04 & Caco-2 Papp, A to B & Absorption & Maximize & $10^{-6}$ cm/s \\
E05 & Caco-2 efflux ratio & Absorption & Minimize & Ratio \\
E06 & LogD (ASAP) & Distribution & Target 2 & Dimensionless \\
E07 & LogD (ExpansionRx) & Distribution & Target 2 & Dimensionless \\
E08 & Mouse plasma unbound fraction & Distribution & Maximize & \% unbound \\
E09 & Mouse brain unbound fraction & Distribution & Maximize & \% unbound \\
E10 & Mouse gastrocnemius muscle unbound fraction & Distribution & Maximize & \% unbound \\
E11 & CYP3A4 inhibition, pIC$_{50}$ & Metabolism & Minimize & Dimensionless \\
E12 & CYP3A4 substrate remaining & Metabolism & Maximize & \% remaining \\
E13 & CYP2J2 substrate remaining & Metabolism & Maximize & \% remaining \\
E14 & PXR induction, pEC$_{50}$ & Metabolism & Minimize & Dimensionless \\
E15 & PXR induction, Emax & Metabolism & Minimize & $\log_2$ fold change \\
E16 & PXR Emax relative to positive control & Metabolism & Minimize & Relative control response \\
E17 & PXR induction, single concentration & Metabolism & Minimize & $\log_2$ fold change \\
\midrule
E18 & Human liver microsomal clearance (ASAP) & Clearance & Minimize & $\mu$L/min/mg \\
E19 & Mouse liver microsomal clearance (ASAP) & Clearance & Minimize & $\mu$L/min/mg \\
E20 & Human microsomal CL$_{\mathrm{int}}$ (ExpansionRx) & Clearance & Minimize & mL min$^{-1}$ kg$^{-1}$ \\
E21 & Mouse microsomal CL$_{\mathrm{int}}$ (ExpansionRx) & Clearance & Minimize & mL min$^{-1}$ kg$^{-1}$ \\
E22 & Rat microsomal CL$_{\mathrm{int}}$ (ExpansionRx) & Clearance & Minimize & mL min$^{-1}$ kg$^{-1}$ \\
E23 & NR-AR & Toxicity & Inactive & Binary (0/1) \\
E24 & NR-AR-LBD & Toxicity & Inactive & Binary (0/1) \\
E25 & NR-AhR & Toxicity & Inactive & Binary (0/1) \\
E26 & NR-aromatase & Toxicity & Inactive & Binary (0/1) \\
E27 & NR-ER & Toxicity & Inactive & Binary (0/1) \\
E28 & NR-ER-LBD & Toxicity & Inactive & Binary (0/1) \\
E29 & NR-PPAR-$\gamma$ & Toxicity & Inactive & Binary (0/1) \\
E30 & SR-ARE & Toxicity & Inactive & Binary (0/1) \\
E31 & SR-ATAD5 & Toxicity & Inactive & Binary (0/1) \\
E32 & SR-HSE & Toxicity & Inactive & Binary (0/1) \\
E33 & SR-MMP & Toxicity & Inactive & Binary (0/1) \\
E34 & SR-p53 & Toxicity & Inactive & Binary (0/1) \\
\bottomrule
\end{tabular}
\caption{ADMET endpoints, acquisition preferences, and units. LogD targets 2; tissue-binding endpoints maximize percent unbound.}
\label{tab:admet_endpoint_lookup}
\end{table}

\Cref{tab:admet_endpoint_lookup} specifies acquisition directions and units. The mouse plasma, brain and gastrocnemius muscle binding objectives maximize percent unbound; their desirability depends on the intended drug profile.

\subsection{Endpoint Transfer and Split Variability}
\label{app:admet_endpoint_transfer}

Leave-one-endpoint-out selection chooses the configuration with highest mean relative utility on source endpoints. Utilities are normalized over all 77 configurations, with catalog-order ties. The all-endpoint comparison combines continuous P95 and binary inactive fractions after normalization. Within-family comparisons restrict configurations to that surrogate family and sources to endpoints of the held-out endpoint's type.

For the 22 continuous endpoints, mean rank correlation across split seeds is 0.854 [0.805,0.891] for batch mean and 0.654 [0.592,0.711] for P95; the paired difference is $-0.200$ [$-0.248,-0.155$]. We average the three seed-pair correlations within each endpoint and use 10,000 endpoint-bootstrap resamples for descriptive 95\% intervals. At batch size 20, linearly interpolated P95 weights the second-largest observation by 0.95 and the largest by 0.05.

\section{Configuration Recommendation: Protocol and Additional Results}

\subsection{Recommendation Protocol}\label{app:llm_protocol}

We compare direct prediction and ICL for Qwen3.5-9B~\citep{qwen2026qwen35}, Qwen3.8-27B~\citep{qwen2026qwen38}, Qwen3.6-35B-A3B~\citep{qwen2026qwen36}, and Gemma-4-26B-A4B-it~\citep{google2026gemma4}. Evaluation covers 100 BindingDB targets, 34 ADMET endpoints and 100 joint pools under each of three preferences.

Both conditions receive the target description, configuration catalog and evaluation objective; ICL adds descriptions and top-15 rankings from the nearest $K=20$ source datasets. Each model returns five ordered configurations. The task-specific LLM-generated prompt and retrieval rule are fixed before evaluation. Decoding is greedy, with thinking disabled and at most 384 new tokens. Qwen3.5-9B and Qwen3.8-27B use NF4 quantization; the other two models use bfloat16. Supplementary Reproduction contains prompts, checkpoint identifiers and inference settings.

\subsection{Held-Out Datasets and Evaluation Utilities}

Each collection has a fixed ten-fold partition constructed without outcomes. Every method uses the same held-out fold and the other nine folds as sources: 90 source datasets for BindingDB/joint and 30--31 for ADMET. All preferences of a joint pool stay in the same fold, with source rankings matched to the target preference. Transfer is evaluated within each collection.

Recommendation utilities follow \Cref{tab:evaluation_map}: BindingDB uses final-batch mean binding; joint tasks use batch P95 of weighted standardized binding/ADMET; ADMET uses oriented P95 for 22 continuous endpoints and inactive fraction for 12 binary endpoints. Joint and ADMET scores average five batches, and all utilities average three split seeds. The recommendation catalogs contain 57 BindingDB and 77 ADMET/joint configurations.

\subsection{Descriptions and Retrieval}
\label{app:similarity}

Descriptions give the dataset identifier, objective and direction, candidate and unique-structure counts, protein length, structural diversity, initialization, batch size and cycle count. Molecular descriptors comprise means of molecular weight, logarithmic partition coefficient (LogP), topological polar surface area (TPSA), hydrogen-bond donors and acceptors. They also specify the scoring statistic and cycles, and distinguish joint evaluation weights from acquisition coefficients. Target property values, rankings and normalization constants are excluded.

Molecular similarity is the cosine similarity between mean Morgan fingerprints of unique canonical structures. Protein similarity is the Jaccard similarity between sets of sequence 3-mers. A third term is the square root of the product of the candidate-pool-size ratio and sequence-length ratio, where each ratio is the smaller value divided by the larger. When both sequences are available, the three terms receive equal weights. Otherwise, molecular fingerprint similarity and the candidate-pool-size ratio each receive weight one half.

ICL takes the nearest $K$ sources. Random retrieval samples without replacement using three source-selection seeds and presents each subset in similarity order. The all-source condition uses the same fixed order for both retrieval methods.

The main comparison contains 3,472 predictions, including 12 invalid Qwen3.5-9B ICL responses. Including source-count ablations gives 24,032 unique predictions and 105 invalid responses, all retained in scoring.

\subsection{Scoring, Baselines, and Uncertainty}

Let $C=|\mathcal C|$ and $r(c)=1+\#\{c':M_{d,c'}>M_{d,c}\}$, so utility ties receive their best shared rank. Hit@$k$ ($k=1,3,5$) checks overlap between the first $k$ recommendations and the top $k$ reference entries, whose utility ties are broken by catalog order. Mean rank averages $r(\hat c_1)$; BestRk@5 averages $\min_{j\leq5}r(\hat c_j)$. First-choice regret uses the first entry in the regret equation above. A valid response has five distinct admissible identifiers. Invalid responses receive zero Hit@$k$, rank $C$ for both rank metrics and regret one.

Random reports exact expectations for uniform sampling of five distinct configurations. Its Hit@1 probability is $1/C$, and its Hit@$k$ probability is $1-\binom{C-k}{k}/\binom{C}{k}$ for $k=3,5$; expected ranks account for utility ties. Global Best chooses the five configurations with lowest mean rank across all eligible training-fold sources. Frequency Voting orders configurations by the number of source datasets on which they rank first, counting exact ties, then by mean rank. Most-Similar Source transfers the nearest source's reference ranking. These methods break remaining ties by catalog order.

K20 rank aggregation averages the same 20 source top-15 lists supplied to ICL: listed configurations receive ranks 1--15 and absent configurations rank 16. It returns the five lowest averages, with catalog-order ties. Every baseline excludes the complete held-out fold.

Predictions are averaged over repetitions within each held-out dataset and then equally across datasets. Confidence intervals use 10{,}000 held-out-fold block bootstrap resamples, weighting sampled folds by their held-out dataset counts and retaining repetitions within dataset. Paired method differences use the same resampled fold indices. These descriptive intervals summarize variation under the saved cross-validation partition.

\subsection{Learned Selectors and Shortlist Diagnostics}
\label{sec:learned_protocol}

Ridge selectors use the same folds and numeric descriptions as the recommendation study: molecular descriptor means, candidate and unique-structure counts, protein length and structural diversity. Counts and protein length use $\log(1+x)$; missing protein length is set to zero with a missingness indicator. Feature standardization and constant-feature removal use only the source datasets for each fit.

All-source ridge predicts complete source ranks scaled to $[0,1]$. K20 ridge uses the same 20 top-15 lists as ICL, scaling listed positions 1--15 and censored rank 16 to $[0,1]$. Both fit one response per configuration with fixed ridge penalty 10 and an unpenalized intercept. The five lowest predicted scores form the recommendation, with catalog-order ties.

Write first-choice regret as $R_1(d)$ and best-list regret as $R_5(d)=\min_{j\leq5}\mathrm{Regret}(d,\hat c_{d,j})$. The ordering gap $G(d)=R_1(d)-R_5(d)$ measures the loss from choosing the first entry instead of the best entry according to target outcomes. Invalid lists have $R_1=R_5=1$ and $G=0$.

\subsection{Recommendation Results}

\label{app:llm_results}

\Cref{tab:llm_full_results,tab:llm_joint_results,tab:llm_joint_safety} report all six recommendation metrics for the five settings. Bold marks tied best point estimates at displayed precision within each reference, direct-prediction or ICL panel.

\begin{table}[t]
\centering\scriptsize
\setlength{\tabcolsep}{1.5pt}
\renewcommand{\arraystretch}{0.9}
\begin{tabular}{@{}l@{\hspace{6pt}}rrrrrr@{\hspace{6pt}}rrrrrr@{}}
\toprule
Method & \multicolumn{6}{c}{\textbf{BindingDB}} & \multicolumn{6}{c}{\textbf{ADMET}} \\
\cmidrule(lr){2-7}\cmidrule(lr){8-13}
 & Hit@1 & Hit@3 & Hit@5 & Rank & BestRk@5 & Regret & Hit@1 & Hit@3 & Hit@5 & Rank & BestRk@5 & Regret \\
\midrule
\multicolumn{13}{l}{\textit{A. Reference methods}} \\
Random & 1.8 & 15.2 & 37.9 & 28.49 & 9.44 & 0.581 & 1.3 & 11.4 & 29.2 & 38.40 & 12.35 & 0.281 \\
Global Best & 1.0 & 44.0 & 74.0 & 15.37 & 9.38 & 0.291 & 14.7 & 29.4 & 41.2 & 16.47 & 11.15 & 0.097 \\
Frequency Voting & 5.0 & \textbf{62.0} & \textbf{85.0} & 19.69 & 4.98 & 0.377 & 14.7 & 38.2 & 61.8 & 16.47 & 5.74 & 0.097 \\
Most-Similar Source & \textbf{11.0} & 51.0 & 79.0 & 15.63 & \textbf{4.38} & 0.293 & \textbf{17.6} & 29.4 & 47.1 & 21.12 & 8.00 & 0.139 \\
K20 rank aggregation & 8.0 & 50.0 & 79.0 & \textbf{13.79} & 6.87 & \textbf{0.271} & \textbf{17.6} & \textbf{50.0} & \textbf{73.5} & \textbf{13.71} & \textbf{4.91} & \textbf{0.092} \\
\addlinespace[1.5pt]
\multicolumn{13}{l}{\textit{B. Direct prediction}} \\
Qwen3.5-9B & 5.0 & 26.0 & 58.0 & 18.43 & \textbf{6.31} & 0.396 & 0.0 & 2.9 & 17.6 & 50.47 & 28.79 & 0.361 \\
Qwen3.8-27B & 0.0 & 0.0 & 3.0 & 37.72 & 24.15 & 0.720 & 0.0 & \textbf{23.5} & 26.5 & 29.88 & \textbf{12.59} & 0.207 \\
Qwen3.6-35B-A3B & \textbf{10.0} & \textbf{42.0} & \textbf{75.0} & 14.11 & 7.73 & 0.276 & \textbf{8.8} & 20.6 & \textbf{47.1} & 30.12 & 15.79 & 0.200 \\
Gemma-4-26B-A4B-it & 5.0 & 29.0 & 69.0 & \textbf{13.93} & 9.26 & \textbf{0.270} & 0.0 & 11.8 & 26.5 & \textbf{28.18} & 16.24 & \textbf{0.170} \\
\addlinespace[1.5pt]
\multicolumn{13}{l}{\textit{C. In-context learning ($K=20$ similar sources)}} \\
Qwen3.5-9B & 9.0 & 43.0 & 70.0 & 17.67 & 5.58 & 0.340 & 20.6 & 47.1 & 61.8 & 17.76 & 8.71 & 0.132 \\
Qwen3.8-27B & 13.0 & 55.0 & 84.0 & 15.11 & 4.31 & 0.280 & \textbf{29.4} & \textbf{58.8} & \textbf{67.6} & \textbf{14.56} & \textbf{5.68} & 0.105 \\
Qwen3.6-35B-A3B & 10.0 & 53.0 & 83.0 & 15.60 & \textbf{3.68} & 0.291 & 23.5 & 47.1 & 61.8 & \textbf{14.56} & 6.50 & \textbf{0.094} \\
Gemma-4-26B-A4B-it & \textbf{14.0} & \textbf{61.0} & \textbf{85.0} & \textbf{13.45} & 4.43 & \textbf{0.260} & 20.6 & 47.1 & 55.9 & 17.35 & 7.26 & 0.119 \\
\bottomrule
\end{tabular}
\caption{Configuration recommendations on BindingDB and ADMET, with all six recommendation metrics. Hit@$k$ is a percentage, and lower rank and regret are better. Panels compare reference methods, direct prediction, and ICL with 20 similarity-selected sources, and bold marks tied best values at the displayed precision within each panel.}
\label{tab:llm_full_results}\label{tab:llm_admet_single}
\end{table}

\begin{table}[t]
\centering\scriptsize
\setlength{\tabcolsep}{1.5pt}
\renewcommand{\arraystretch}{0.9}
\begin{tabular}{@{}l@{\hspace{6pt}}rrrrrr@{\hspace{6pt}}rrrrrr@{}}
\toprule
Method & \multicolumn{6}{c}{\textbf{Joint balanced}} & \multicolumn{6}{c}{\textbf{Joint binding-heavy}} \\
\cmidrule(lr){2-7}\cmidrule(lr){8-13}
 & Hit@1 & Hit@3 & Hit@5 & Rank & BestRk@5 & Regret & Hit@1 & Hit@3 & Hit@5 & Rank & BestRk@5 & Regret \\
\midrule
\multicolumn{13}{l}{\textit{A. Reference methods}} \\
Random & 1.3 & 11.4 & 29.2 & 38.77 & 12.76 & 0.348 & 1.3 & 11.4 & 29.2 & 38.77 & 12.79 & 0.351 \\
Global Best & \textbf{6.0} & 34.0 & 53.0 & \textbf{13.46} & 6.94 & \textbf{0.082} & 5.0 & 40.0 & 62.0 & \textbf{14.44} & 6.09 & \textbf{0.092} \\
Frequency Voting & \textbf{6.0} & 33.0 & \textbf{63.0} & 22.25 & \textbf{5.23} & 0.138 & 4.0 & 38.0 & \textbf{72.0} & 20.46 & 5.11 & 0.126 \\
Most-Similar Source & 2.0 & 24.0 & 45.0 & 19.34 & 9.12 & 0.115 & \textbf{7.0} & 31.0 & 53.0 & 22.19 & 8.96 & 0.138 \\
K20 rank aggregation & 5.0 & \textbf{38.0} & 62.0 & 14.94 & 6.04 & 0.090 & 5.0 & \textbf{48.0} & \textbf{72.0} & 15.12 & \textbf{4.62} & 0.096 \\
\addlinespace[1.5pt]
\multicolumn{13}{l}{\textit{B. Direct prediction}} \\
Qwen3.5-9B & 2.0 & 6.0 & 8.0 & 51.30 & 42.30 & 0.519 & 3.0 & 9.0 & 16.0 & 41.30 & 35.48 & 0.377 \\
Qwen3.8-27B & 0.0 & 8.0 & 33.0 & 34.03 & 11.33 & 0.238 & 0.0 & 17.0 & 31.0 & 33.32 & 11.11 & 0.236 \\
Qwen3.6-35B-A3B & 2.0 & 23.0 & 36.0 & 24.32 & 11.30 & 0.174 & 2.0 & 12.0 & 27.0 & 40.25 & 16.98 & 0.339 \\
Gemma-4-26B-A4B-it & \textbf{6.0} & \textbf{31.0} & \textbf{56.0} & \textbf{15.81} & \textbf{7.91} & \textbf{0.093} & \textbf{5.0} & \textbf{37.0} & \textbf{57.0} & \textbf{16.68} & \textbf{7.03} & \textbf{0.104} \\
\addlinespace[1.5pt]
\multicolumn{13}{l}{\textit{C. In-context learning ($K=20$ similar sources)}} \\
Qwen3.5-9B & \textbf{6.0} & 24.0 & 50.0 & 21.36 & 9.72 & 0.148 & 5.0 & 30.0 & 53.0 & 19.64 & 8.84 & 0.130 \\
Qwen3.8-27B & 5.0 & 29.0 & 63.0 & 18.54 & 6.01 & 0.109 & 7.0 & \textbf{38.0} & 56.0 & 17.55 & 7.47 & 0.108 \\
Qwen3.6-35B-A3B & 3.0 & 30.0 & 59.0 & \textbf{18.49} & 6.15 & \textbf{0.101} & \textbf{8.0} & 35.0 & \textbf{64.0} & 18.57 & \textbf{6.64} & 0.105 \\
Gemma-4-26B-A4B-it & 5.0 & \textbf{32.0} & \textbf{65.0} & 18.70 & \textbf{5.71} & 0.104 & 6.0 & 36.0 & 62.0 & \textbf{16.11} & 6.68 & \textbf{0.099} \\
\bottomrule
\end{tabular}
\caption{Configuration recommendations on the joint binding--ADMET pools under the balanced and binding-heavy objective preferences, with all six recommendation metrics. Hit@$k$ is a percentage, and lower rank and regret are better. Panels compare reference methods, direct prediction, and ICL with 20 similarity-selected sources, and bold marks tied best values at the displayed precision within each panel.}
\label{tab:llm_joint_results}\label{tab:llm_joint_balanced}\label{tab:llm_joint_binding}
\end{table}

\begin{table}[t]
\centering\scriptsize
\setlength{\tabcolsep}{1.5pt}
\renewcommand{\arraystretch}{0.9}
\begin{tabular}{@{}l@{\hspace{6pt}}rrrrrr@{\hspace{6pt}}rrrrrr@{}}
\toprule
Method & \multicolumn{6}{c}{\textbf{Joint ADMET-heavy}} \\
\cmidrule(lr){2-7}
 & Hit@1 & Hit@3 & Hit@5 & Rank & BestRk@5 & Regret \\
\midrule
\multicolumn{7}{l}{\textit{A. Reference methods}} \\
Random & 1.3 & 11.4 & 29.2 & 38.77 & 12.70 & 0.320 \\
Global Best & \textbf{15.0} & 48.0 & \textbf{82.0} & 11.21 & 3.74 & 0.055 \\
Frequency Voting & \textbf{15.0} & \textbf{54.0} & 76.0 & 11.21 & 3.92 & 0.055 \\
Most-Similar Source & 3.0 & 24.0 & 59.0 & 16.98 & 6.01 & 0.084 \\
K20 rank aggregation & 10.0 & 51.0 & 79.0 & \textbf{9.84} & \textbf{3.72} & \textbf{0.050} \\
\addlinespace[1.5pt]
\multicolumn{7}{l}{\textit{B. Direct prediction}} \\
Qwen3.5-9B & 0.0 & 1.0 & 2.0 & 64.68 & 53.59 & 0.681 \\
Qwen3.8-27B & 0.0 & 2.0 & 36.0 & 29.70 & 10.46 & 0.177 \\
Qwen3.6-35B-A3B & 1.0 & \textbf{24.0} & \textbf{46.0} & 23.94 & \textbf{10.17} & 0.147 \\
Gemma-4-26B-A4B-it & \textbf{3.0} & 21.0 & 34.0 & \textbf{17.09} & 11.46 & \textbf{0.084} \\
\addlinespace[1.5pt]
\multicolumn{7}{l}{\textit{C. In-context learning ($K=20$ similar sources)}} \\
Qwen3.5-9B & 4.0 & 32.0 & 59.0 & 20.16 & 11.39 & 0.155 \\
Qwen3.8-27B & \textbf{8.0} & \textbf{45.0} & \textbf{75.0} & \textbf{12.06} & 4.11 & \textbf{0.061} \\
Qwen3.6-35B-A3B & 7.0 & 44.0 & 73.0 & 12.79 & 4.09 & \textbf{0.061} \\
Gemma-4-26B-A4B-it & 7.0 & 41.0 & 74.0 & 12.93 & \textbf{3.88} & 0.064 \\
\bottomrule
\end{tabular}
\caption{Configuration recommendations on the joint binding--ADMET pools under the ADMET-heavy objective preference, with all six recommendation metrics. Hit@$k$ is a percentage, and lower rank and regret are better. Panels compare reference methods, direct prediction, and ICL with 20 similarity-selected sources, and bold marks tied best values at the displayed precision within each panel.}
\label{tab:llm_joint_safety}
\end{table}

Supplementary Source Data contains the 40 paired estimates underlying \Cref{fig:llm_usefulness}(a,b). Some descriptive 95\% intervals include zero: Gemma-4-26B-A4B-it's binding-heavy Hit@5 gain is 5 percentage points [$-7,17$], and all four ADMET first-choice-regret differences between ICL and K20 rank aggregation include zero.

\subsection{Sensitivity to Exact Ties and Shortlist Scoring}
\label{app:llm_sensitivity}

Tie-aware Hit@$k$ includes all configurations tied at the reference cutoff. We rescore the original direct and $K=20$ ICL lists with this rule and with Regret@5.

Including cutoff ties leaves Hit@5 unchanged, and ICL reduces Regret@5 in all 20 model-setting comparisons (\Cref{tab:llm_sensitivity,tab:llm_sensitivity_joint,tab:llm_sensitivity_safety}). Tie-aware Hit@3 improves in 19 comparisons; Gemma-4-26B-A4B-it on binding-heavy joint selection changes from 37\% to 36\%.

Grouping configurations with exactly equal aggregate utilities on every dataset yields 53 BindingDB, 69 ADMET and 68 joint groups per preference. Valid ICL lists cover fewer than five groups in 24/136 ADMET, 9/398 balanced-joint, 3/399 binding-heavy and 26/392 ADMET-heavy cases, versus 0/399 BindingDB cases. Every valid direct-prediction list covers five groups. These are post hoc groups of recorded outcomes; primary scores retain the original identifiers. Supplementary Source Data gives the per-model counts and group assignments.

\begin{table}[t]
\centering\scriptsize
\setlength{\tabcolsep}{2.4pt}
\begin{tabular}{@{}l@{\hspace{9pt}}rrrrrr@{\hspace{9pt}}rrrrrr@{}}
\toprule
 & \multicolumn{6}{c}{\textbf{BindingDB}} & \multicolumn{6}{c}{\textbf{ADMET}} \\
\cmidrule(lr){2-7}\cmidrule(lr){8-13}
 & \multicolumn{2}{c}{\shortstack{Tie-aware\\Hit@3}} & \multicolumn{2}{c}{\shortstack{Tie-aware\\Hit@5}} & \multicolumn{2}{c}{Regret@5} & \multicolumn{2}{c}{\shortstack{Tie-aware\\Hit@3}} & \multicolumn{2}{c}{\shortstack{Tie-aware\\Hit@5}} & \multicolumn{2}{c}{Regret@5} \\
\cmidrule(lr){2-3}\cmidrule(lr){4-5}\cmidrule(lr){6-7}\cmidrule(lr){8-9}\cmidrule(lr){10-11}\cmidrule(lr){12-13}
Model & Direct & ICL & Direct & ICL & Direct & ICL & Direct & ICL & Direct & ICL & Direct & ICL \\
\midrule
Q9 & 26.0 & 44.0 & 58.0 & 70.0 & 0.163 & 0.140 & 2.9 & 47.1 & 17.6 & 61.8 & 0.190 & 0.062 \\
Q27 & 0.0 & 55.0 & 3.0 & 84.0 & 0.593 & 0.097 & 26.5 & 58.8 & 26.5 & 67.6 & 0.090 & 0.035 \\
Q35 & 42.0 & 53.0 & 75.0 & 83.0 & 0.167 & 0.095 & 20.6 & 47.1 & 47.1 & 61.8 & 0.100 & 0.039 \\
G4 & 29.0 & 61.0 & 69.0 & 85.0 & 0.192 & 0.091 & 11.8 & 47.1 & 26.5 & 55.9 & 0.114 & 0.047 \\
\bottomrule
\end{tabular}
\caption{Sensitivity of saved recommendations to exact utility ties and to scoring the best of five entries (Regret@5). Hit rates are percentages, lower Regret@5 is better, and ICL uses 20 similarity-selected sources. Q9, Q27, Q35 and G4 denote the four language models of \Cref{sec:llm_study}.}
\label{tab:llm_sensitivity}
\end{table}

\begin{table}[t]
\centering\scriptsize
\setlength{\tabcolsep}{2.4pt}
\begin{tabular}{@{}l@{\hspace{9pt}}rrrrrr@{\hspace{9pt}}rrrrrr@{}}
\toprule
 & \multicolumn{6}{c}{\textbf{Joint balanced}} & \multicolumn{6}{c}{\textbf{Joint binding-heavy}} \\
\cmidrule(lr){2-7}\cmidrule(lr){8-13}
 & \multicolumn{2}{c}{\shortstack{Tie-aware\\Hit@3}} & \multicolumn{2}{c}{\shortstack{Tie-aware\\Hit@5}} & \multicolumn{2}{c}{Regret@5} & \multicolumn{2}{c}{\shortstack{Tie-aware\\Hit@3}} & \multicolumn{2}{c}{\shortstack{Tie-aware\\Hit@5}} & \multicolumn{2}{c}{Regret@5} \\
\cmidrule(lr){2-3}\cmidrule(lr){4-5}\cmidrule(lr){6-7}\cmidrule(lr){8-9}\cmidrule(lr){10-11}\cmidrule(lr){12-13}
Model & Direct & ICL & Direct & ICL & Direct & ICL & Direct & ICL & Direct & ICL & Direct & ICL \\
\midrule
Q9 & 6.0 & 24.0 & 8.0 & 50.0 & 0.373 & 0.087 & 9.0 & 30.0 & 16.0 & 53.0 & 0.296 & 0.066 \\
Q27 & 8.0 & 29.0 & 33.0 & 63.0 & 0.067 & 0.042 & 17.0 & 38.0 & 31.0 & 56.0 & 0.072 & 0.048 \\
Q35 & 23.0 & 30.0 & 36.0 & 59.0 & 0.077 & 0.043 & 12.0 & 35.0 & 27.0 & 64.0 & 0.123 & 0.044 \\
G4 & 31.0 & 32.0 & 56.0 & 65.0 & 0.053 & 0.040 & 37.0 & 36.0 & 57.0 & 62.0 & 0.054 & 0.041 \\
\bottomrule
\end{tabular}
\caption{Sensitivity of saved recommendations to exact utility ties and to scoring the best of five entries (Regret@5). Hit rates are percentages, lower Regret@5 is better, and ICL uses 20 similarity-selected sources. Q9, Q27, Q35 and G4 denote the four language models of \Cref{sec:llm_study}.}
\label{tab:llm_sensitivity_joint}
\end{table}

\begin{table}[t]
\centering\scriptsize
\setlength{\tabcolsep}{2.4pt}
\begin{tabular}{@{}l@{\hspace{9pt}}rrrrrr@{\hspace{9pt}}rrrrrr@{}}
\toprule
 & \multicolumn{6}{c}{\textbf{Joint ADMET-heavy}} \\
\cmidrule(lr){2-7}
 & \multicolumn{2}{c}{\shortstack{Tie-aware\\Hit@3}} & \multicolumn{2}{c}{\shortstack{Tie-aware\\Hit@5}} & \multicolumn{2}{c}{Regret@5} \\
\cmidrule(lr){2-3}\cmidrule(lr){4-5}\cmidrule(lr){6-7}
Model & Direct & ICL & Direct & ICL & Direct & ICL \\
\midrule
Q9 & 1.0 & 32.0 & 2.0 & 59.0 & 0.450 & 0.110 \\
Q27 & 2.0 & 45.0 & 36.0 & 75.0 & 0.050 & 0.026 \\
Q35 & 24.0 & 44.0 & 46.0 & 73.0 & 0.061 & 0.024 \\
G4 & 21.0 & 41.0 & 34.0 & 74.0 & 0.055 & 0.023 \\
\bottomrule
\end{tabular}
\caption{Sensitivity of saved recommendations to exact utility ties and to scoring the best of five entries (Regret@5). Hit rates are percentages, lower Regret@5 is better, and ICL uses 20 similarity-selected sources. Q9, Q27, Q35 and G4 denote the four language models of \Cref{sec:llm_study}.}
\label{tab:llm_sensitivity_safety}
\end{table}

\subsection{Source-Count and Retrieval Ablation}

\Cref{tab:icl_ablation,tab:icl_ablation_admet,tab:icl_ablation_joint_balanced,tab:icl_ablation_joint_binding,tab:icl_ablation_joint_safety} vary source count and retrieval for Qwen3.5-9B and Qwen3.6-35B-A3B, using the same folds and utilities as the main comparison. The $K=20$ similarity rows reuse the main predictions. All-source uses 90 BindingDB/joint pools or 30--31 ADMET endpoints. Bold marks tied best values within each setting, model and retrieval group.

\begin{table}[t]
\centering\scriptsize\setlength{\tabcolsep}{1.5pt}
\begin{tabular}{@{}l@{\hspace{6pt}}rrrrrr@{\hspace{6pt}}rrrrrr@{}}
\toprule
 & \multicolumn{6}{c}{Similarity retrieval} & \multicolumn{6}{c}{Random retrieval} \\
\cmidrule(lr){2-7}\cmidrule(lr){8-13}
$K$ & Hit@1 & Hit@3 & Hit@5 & Rank & BestRk@5 & Regret & Hit@1 & Hit@3 & Hit@5 & Rank & BestRk@5 & Regret \\
\midrule
\multicolumn{13}{l}{\textbf{BindingDB}} \\
\multicolumn{13}{l}{\textit{Qwen3.6-35B-A3B}} \\
1 & 9.0 & 47.0 & 78.0 & 16.13 & 4.58 & 0.302 & 6.3 & 34.3 & 63.7 & 19.21 & 9.26 & 0.395 \\
3 & 10.0 & 52.0 & 79.0 & 17.45 & 4.76 & 0.323 & 7.7 & 38.7 & 72.7 & 17.42 & 6.85 & 0.349 \\
5 & 10.0 & 47.0 & 79.0 & 16.32 & 4.33 & 0.309 & 7.7 & 43.3 & 73.7 & 17.45 & 6.17 & 0.342 \\
10 & 9.0 & 50.0 & 82.0 & 17.07 & 3.87 & 0.312 & 7.3 & 43.3 & 76.7 & 16.05 & 5.70 & 0.319 \\
20 & 10.0 & \textbf{53.0} & \textbf{83.0} & 15.60 & \textbf{3.68} & 0.291 & 8.3 & 48.7 & 76.3 & 15.72 & 5.88 & 0.312 \\
50 & \textbf{13.0} & 52.0 & 82.0 & \textbf{15.55} & 4.30 & \textbf{0.284} & \textbf{11.0} & \textbf{50.3} & \textbf{79.3} & \textbf{14.67} & \textbf{5.08} & \textbf{0.288} \\
All & 8.0 & 49.0 & 76.0 & 16.45 & 5.64 & 0.316 & 8.0 & 49.0 & 76.0 & 16.45 & 5.64 & 0.316 \\
\multicolumn{13}{l}{\textit{Qwen3.5-9B}} \\
1 & \textbf{10.0} & \textbf{50.0} & \textbf{79.0} & \textbf{15.77} & 4.43 & \textbf{0.297} & 6.7 & \textbf{38.7} & \textbf{65.7} & 19.92 & 8.60 & 0.384 \\
3 & \textbf{10.0} & 48.0 & 73.0 & 17.78 & 4.25 & 0.344 & 9.3 & 36.7 & 65.0 & 18.13 & 7.07 & 0.354 \\
5 & 9.0 & 44.0 & 75.0 & 17.22 & \textbf{4.16} & 0.339 & \textbf{11.0} & 36.0 & 64.7 & 17.74 & 6.87 & \textbf{0.349} \\
10 & 7.0 & 47.0 & 73.0 & 16.60 & 4.70 & 0.328 & 9.3 & 33.0 & 63.3 & 17.79 & 6.51 & 0.359 \\
20 & 9.0 & 43.0 & 70.0 & 17.67 & 5.58 & 0.340 & 7.3 & 33.3 & 64.7 & \textbf{17.47} & \textbf{6.39} & 0.355 \\
50 & \textbf{10.0} & 41.0 & 65.0 & 16.41 & 6.13 & 0.330 & 7.0 & 34.0 & 62.3 & 18.10 & 7.72 & 0.352 \\
All & 6.0 & 30.0 & 59.0 & 18.52 & 6.91 & 0.374 & 6.0 & 30.0 & 59.0 & 18.52 & 6.91 & 0.374 \\
\midrule
\bottomrule
\end{tabular}
\caption{Source-count and retrieval ablation for each setting and the two ablation models. Random retrieval averages three source-selection seeds, and the all-source condition ($K$=All) reuses one prediction for both retrieval modes. Hit@$k$ is a percentage; lower rank and regret are better.}
\label{tab:icl_ablation}
\end{table}

\begin{table}[t]
\centering\scriptsize\setlength{\tabcolsep}{1.5pt}
\begin{tabular}{@{}l@{\hspace{6pt}}rrrrrr@{\hspace{6pt}}rrrrrr@{}}
\toprule
 & \multicolumn{6}{c}{Similarity retrieval} & \multicolumn{6}{c}{Random retrieval} \\
\cmidrule(lr){2-7}\cmidrule(lr){8-13}
$K$ & Hit@1 & Hit@3 & Hit@5 & Rank & BestRk@5 & Regret & Hit@1 & Hit@3 & Hit@5 & Rank & BestRk@5 & Regret \\
\midrule
\multicolumn{13}{l}{\textbf{ADMET}} \\
\multicolumn{13}{l}{\textit{Qwen3.6-35B-A3B}} \\
1 & 17.6 & 32.4 & 50.0 & 21.18 & 7.71 & 0.139 & 6.9 & 24.5 & 38.2 & 30.50 & 12.71 & 0.217 \\
3 & \textbf{23.5} & 35.3 & 58.8 & 16.03 & 7.94 & 0.109 & 8.8 & 25.5 & 45.1 & 30.07 & 12.00 & 0.207 \\
5 & \textbf{23.5} & 41.2 & \textbf{64.7} & 15.26 & 6.50 & 0.094 & 10.8 & 29.4 & 50.0 & 25.61 & 9.10 & 0.179 \\
10 & \textbf{23.5} & \textbf{50.0} & 61.8 & \textbf{13.15} & \textbf{6.29} & \textbf{0.088} & 10.8 & 36.3 & 52.9 & 22.69 & 8.04 & 0.153 \\
20 & \textbf{23.5} & 47.1 & 61.8 & 14.56 & 6.50 & 0.094 & 11.8 & \textbf{42.2} & \textbf{58.8} & 20.84 & 7.91 & 0.143 \\
All & 17.6 & 38.2 & 47.1 & 17.41 & 7.88 & 0.116 & \textbf{17.6} & 38.2 & 47.1 & \textbf{17.41} & \textbf{7.88} & \textbf{0.116} \\
\multicolumn{13}{l}{\textit{Qwen3.5-9B}} \\
1 & 17.6 & 32.4 & 47.1 & 21.76 & 8.71 & 0.157 & 3.9 & 21.6 & 41.2 & 35.69 & 17.05 & 0.268 \\
3 & 17.6 & 35.3 & 55.9 & 21.85 & 10.09 & 0.138 & 12.7 & 32.4 & 48.0 & 26.21 & 11.94 & 0.190 \\
5 & 23.5 & 41.2 & 55.9 & 18.35 & \textbf{8.00} & 0.113 & 12.7 & 34.3 & \textbf{55.9} & 23.00 & 10.96 & 0.175 \\
10 & \textbf{26.5} & \textbf{47.1} & 58.8 & 17.18 & 8.56 & 0.122 & 13.7 & 34.3 & 52.0 & 26.57 & 12.79 & 0.207 \\
20 & 20.6 & \textbf{47.1} & \textbf{61.8} & 17.76 & 8.71 & 0.132 & 18.6 & 36.3 & 52.9 & 20.20 & 9.29 & 0.149 \\
All & 20.6 & 38.2 & 55.9 & \textbf{15.26} & \textbf{8.00} & \textbf{0.097} & \textbf{20.6} & \textbf{38.2} & \textbf{55.9} & \textbf{15.26} & \textbf{8.00} & \textbf{0.097} \\
\midrule
\bottomrule
\end{tabular}
\caption{Source-count and retrieval ablation for each setting and the two ablation models. Random retrieval averages three source-selection seeds, and the all-source condition ($K$=All) reuses one prediction for both retrieval modes. Hit@$k$ is a percentage; lower rank and regret are better.}
\label{tab:icl_ablation_admet}
\end{table}

\begin{table}[t]
\centering\scriptsize\setlength{\tabcolsep}{1.5pt}
\begin{tabular}{@{}l@{\hspace{6pt}}rrrrrr@{\hspace{6pt}}rrrrrr@{}}
\toprule
 & \multicolumn{6}{c}{Similarity retrieval} & \multicolumn{6}{c}{Random retrieval} \\
\cmidrule(lr){2-7}\cmidrule(lr){8-13}
$K$ & Hit@1 & Hit@3 & Hit@5 & Rank & BestRk@5 & Regret & Hit@1 & Hit@3 & Hit@5 & Rank & BestRk@5 & Regret \\
\midrule
\multicolumn{13}{l}{\textbf{Joint balanced}} \\
\multicolumn{13}{l}{\textit{Qwen3.6-35B-A3B}} \\
1 & 2.0 & 25.0 & 44.0 & 19.34 & 8.86 & 0.115 & 3.0 & 23.3 & 45.7 & 22.84 & 9.59 & 0.155 \\
3 & 3.0 & 27.0 & 56.0 & 18.83 & 7.33 & 0.110 & 2.7 & 23.7 & 50.7 & 19.51 & 8.40 & 0.118 \\
5 & 3.0 & 29.0 & \textbf{66.0} & 19.40 & 6.52 & 0.108 & 2.0 & \textbf{27.0} & 54.7 & 19.38 & 7.53 & 0.119 \\
10 & \textbf{4.0} & 29.0 & 64.0 & 19.21 & \textbf{5.87} & 0.105 & 2.7 & 26.3 & \textbf{55.0} & 18.27 & 7.14 & 0.109 \\
20 & 3.0 & 30.0 & 59.0 & 18.49 & 6.15 & 0.101 & 3.7 & 26.3 & 54.3 & 19.42 & 7.04 & 0.115 \\
50 & 1.0 & \textbf{32.0} & 60.0 & \textbf{18.17} & 5.90 & \textbf{0.099} & 3.3 & \textbf{27.0} & 53.7 & \textbf{17.54} & \textbf{6.49} & \textbf{0.104} \\
All & \textbf{4.0} & 22.0 & 54.0 & 19.07 & 7.02 & 0.109 & \textbf{4.0} & 22.0 & 54.0 & 19.07 & 7.02 & 0.109 \\
\multicolumn{13}{l}{\textit{Qwen3.5-9B}} \\
1 & 2.0 & 19.0 & 40.0 & 25.71 & 14.49 & 0.190 & 2.3 & 19.0 & 38.7 & 25.58 & 14.19 & 0.190 \\
3 & 2.0 & 23.0 & 49.0 & 24.10 & 11.56 & 0.172 & 1.0 & 22.3 & 44.3 & 23.50 & 12.49 & 0.174 \\
5 & 4.0 & 25.0 & 47.0 & 22.85 & 9.55 & 0.149 & 3.0 & 23.7 & 47.7 & 21.68 & 10.07 & 0.150 \\
10 & \textbf{6.0} & \textbf{26.0} & 49.0 & 22.47 & 9.43 & 0.156 & 2.3 & \textbf{25.3} & \textbf{52.7} & 22.05 & 8.52 & 0.141 \\
20 & \textbf{6.0} & 24.0 & 50.0 & 21.36 & 9.72 & 0.148 & 3.7 & 24.7 & 51.0 & 20.57 & 8.17 & 0.133 \\
50 & 4.0 & 21.0 & 46.0 & \textbf{20.89} & 8.84 & \textbf{0.129} & 1.7 & 23.3 & 52.3 & \textbf{19.35} & 7.78 & \textbf{0.123} \\
All & 5.0 & 21.0 & \textbf{52.0} & 25.34 & \textbf{7.73} & 0.158 & \textbf{5.0} & 21.0 & 52.0 & 25.34 & \textbf{7.73} & 0.158 \\
\midrule
\bottomrule
\end{tabular}
\caption{Source-count and retrieval ablation for each setting and the two ablation models. Random retrieval averages three source-selection seeds, and the all-source condition ($K$=All) reuses one prediction for both retrieval modes. Hit@$k$ is a percentage; lower rank and regret are better.}
\label{tab:icl_ablation_joint_balanced}
\end{table}

\begin{table}[t]
\centering\scriptsize\setlength{\tabcolsep}{1.5pt}
\begin{tabular}{@{}l@{\hspace{6pt}}rrrrrr@{\hspace{6pt}}rrrrrr@{}}
\toprule
 & \multicolumn{6}{c}{Similarity retrieval} & \multicolumn{6}{c}{Random retrieval} \\
\cmidrule(lr){2-7}\cmidrule(lr){8-13}
$K$ & Hit@1 & Hit@3 & Hit@5 & Rank & BestRk@5 & Regret & Hit@1 & Hit@3 & Hit@5 & Rank & BestRk@5 & Regret \\
\midrule
\multicolumn{13}{l}{\textbf{Joint binding-heavy}} \\
\multicolumn{13}{l}{\textit{Qwen3.6-35B-A3B}} \\
1 & 7.0 & 31.0 & 58.0 & 22.19 & 8.37 & 0.138 & 2.3 & 22.0 & 48.3 & 24.22 & 9.95 & 0.161 \\
3 & 5.0 & 31.0 & 60.0 & 21.01 & 7.65 & 0.128 & 3.7 & 25.3 & 50.0 & 20.75 & 8.79 & 0.132 \\
5 & 6.0 & 31.0 & \textbf{64.0} & 20.76 & 7.29 & 0.121 & 3.0 & 28.7 & 54.3 & 20.38 & 8.45 & 0.133 \\
10 & \textbf{9.0} & \textbf{35.0} & 62.0 & 19.65 & 7.00 & 0.119 & \textbf{5.0} & 31.0 & 57.3 & 19.82 & 7.85 & 0.127 \\
20 & 8.0 & \textbf{35.0} & \textbf{64.0} & 18.57 & \textbf{6.64} & \textbf{0.105} & 4.3 & \textbf{33.0} & 59.3 & 19.37 & 7.63 & 0.119 \\
50 & 8.0 & 34.0 & 62.0 & \textbf{17.71} & 7.21 & 0.107 & 3.7 & 31.3 & \textbf{63.0} & \textbf{18.16} & \textbf{6.60} & \textbf{0.109} \\
All & 2.0 & 25.0 & 38.0 & 30.61 & 12.57 & 0.254 & 2.0 & 25.0 & 38.0 & 30.61 & 12.57 & 0.254 \\
\multicolumn{13}{l}{\textit{Qwen3.5-9B}} \\
1 & 5.0 & 25.0 & 43.0 & 28.38 & 14.66 & 0.211 & 4.3 & 20.3 & 38.0 & 28.10 & 16.21 & 0.224 \\
3 & 6.0 & 25.0 & 47.0 & 24.79 & 12.12 & 0.171 & 3.3 & 24.7 & 47.7 & 24.22 & 11.45 & 0.179 \\
5 & 6.0 & 23.0 & 44.0 & 24.71 & 12.19 & 0.183 & 4.0 & 29.0 & 48.7 & 22.51 & 10.31 & 0.160 \\
10 & \textbf{7.0} & 29.0 & \textbf{53.0} & 21.28 & 8.91 & 0.148 & \textbf{5.7} & 29.7 & \textbf{51.7} & 20.64 & 9.11 & 0.137 \\
20 & 5.0 & 30.0 & \textbf{53.0} & 19.64 & 8.84 & 0.130 & 5.3 & 29.3 & 51.3 & 20.32 & 9.32 & 0.136 \\
50 & \textbf{7.0} & \textbf{32.0} & 50.0 & 19.66 & 9.59 & 0.127 & 4.0 & 29.0 & 49.3 & 20.91 & 9.63 & 0.142 \\
All & 5.0 & 31.0 & 45.0 & \textbf{18.08} & \textbf{8.24} & \textbf{0.109} & 5.0 & \textbf{31.0} & 45.0 & \textbf{18.08} & \textbf{8.24} & \textbf{0.109} \\
\midrule
\bottomrule
\end{tabular}
\caption{Source-count and retrieval ablation for each setting and the two ablation models. Random retrieval averages three source-selection seeds, and the all-source condition ($K$=All) reuses one prediction for both retrieval modes. Hit@$k$ is a percentage; lower rank and regret are better.}
\label{tab:icl_ablation_joint_binding}
\end{table}

\begin{table}[t]
\centering\scriptsize\setlength{\tabcolsep}{1.5pt}
\begin{tabular}{@{}l@{\hspace{6pt}}rrrrrr@{\hspace{6pt}}rrrrrr@{}}
\toprule
 & \multicolumn{6}{c}{Similarity retrieval} & \multicolumn{6}{c}{Random retrieval} \\
\cmidrule(lr){2-7}\cmidrule(lr){8-13}
$K$ & Hit@1 & Hit@3 & Hit@5 & Rank & BestRk@5 & Regret & Hit@1 & Hit@3 & Hit@5 & Rank & BestRk@5 & Regret \\
\midrule
\multicolumn{13}{l}{\textbf{Joint ADMET-heavy}} \\
\multicolumn{13}{l}{\textit{Qwen3.6-35B-A3B}} \\
1 & 3.0 & 24.0 & 59.0 & 16.98 & 6.01 & 0.084 & 3.7 & 27.3 & 56.0 & 18.73 & 7.05 & 0.107 \\
3 & 4.0 & 35.0 & 69.0 & 15.52 & 4.62 & 0.077 & 3.7 & 35.0 & 65.3 & 16.10 & 5.92 & 0.083 \\
5 & 5.0 & 36.0 & 66.0 & 14.41 & 4.84 & 0.068 & 5.7 & 37.0 & 64.0 & 15.86 & 5.79 & 0.082 \\
10 & 3.0 & 34.0 & 69.0 & 13.89 & 4.69 & 0.066 & \textbf{11.0} & 46.7 & \textbf{70.3} & 13.16 & \textbf{4.58} & 0.064 \\
20 & 7.0 & 44.0 & \textbf{73.0} & 12.79 & \textbf{4.09} & 0.061 & 10.0 & \textbf{49.7} & 68.7 & 12.46 & 4.59 & 0.060 \\
50 & 5.0 & \textbf{48.0} & 71.0 & 13.38 & 4.52 & 0.067 & 7.7 & 44.3 & 66.3 & \textbf{12.12} & 5.12 & 0.060 \\
All & \textbf{11.0} & 46.0 & 61.0 & \textbf{12.50} & 5.92 & \textbf{0.058} & \textbf{11.0} & 46.0 & 61.0 & 12.50 & 5.92 & \textbf{0.058} \\
\multicolumn{13}{l}{\textit{Qwen3.5-9B}} \\
1 & 2.0 & 21.0 & 51.0 & 23.35 & 12.02 & 0.162 & 4.3 & 23.0 & 46.0 & 25.52 & 14.72 & 0.184 \\
3 & 5.0 & 26.0 & 57.0 & 17.65 & 6.78 & 0.090 & 4.7 & 26.7 & 51.3 & 21.81 & 10.00 & 0.142 \\
5 & 4.0 & 28.0 & 57.0 & 17.28 & \textbf{6.11} & 0.095 & 5.3 & 29.7 & \textbf{58.3} & 19.52 & 8.53 & 0.122 \\
10 & 4.0 & \textbf{33.0} & 53.0 & 16.95 & 8.32 & 0.110 & 6.7 & \textbf{31.0} & 55.0 & 18.89 & 9.60 & 0.121 \\
20 & 4.0 & 32.0 & 59.0 & 20.16 & 11.39 & 0.155 & 4.3 & 27.3 & 51.3 & 19.92 & 12.18 & 0.147 \\
50 & 6.0 & 32.0 & \textbf{60.0} & 17.17 & 9.98 & 0.126 & 6.0 & 28.3 & 51.3 & 20.04 & 11.94 & 0.151 \\
All & \textbf{9.0} & 30.0 & 52.0 & \textbf{14.31} & 7.83 & \textbf{0.066} & \textbf{9.0} & 30.0 & 52.0 & \textbf{14.31} & \textbf{7.83} & \textbf{0.066} \\
\bottomrule
\end{tabular}
\caption{Source-count and retrieval ablation for each setting and the two ablation models. Random retrieval averages three source-selection seeds, and the all-source condition ($K$=All) reuses one prediction for both retrieval modes. Hit@$k$ is a percentage; lower rank and regret are better.}
\label{tab:icl_ablation_joint_safety}
\end{table}

\section{Additional Configuration-Selection Analyses}
\label{app:jmlr_diagnostics}

\Cref{tab:learned_paired} compares the ridge selectors with K20 rank aggregation; negative differences favor ridge. \Cref{tab:ordering_decomposition} separates first-choice regret into best-list regret and the ordering gap, averaging the four models within target before averaging targets. All analyses use the scoring and fold-bootstrap conventions above.

\begin{table}[t]
\centering\scriptsize\setlength{\tabcolsep}{1.8pt}
\begin{tabular}{@{}ll@{\hspace{8pt}}rrr@{\hspace{8pt}}rrr@{}}
\toprule
 & & \multicolumn{3}{c}{$\Delta R_1$ [95\% interval]} & \multicolumn{3}{c}{$\Delta R_5$ [95\% interval]} \\
\cmidrule(lr){3-5}\cmidrule(lr){6-8}
\midrule
BindingDB & all-source ranks & $-0.0205$ & $[-0.0582,$ & $0.0183]$ & $-0.0346$ & $[-0.0625,$ & $-0.0060]$ \\
BindingDB & K20 top-15 & $-0.0334$ & $[-0.0632,$ & $-0.0030]$ & $-0.0521$ & $[-0.0819,$ & $-0.0227]$ \\
\midrule
ADMET & all-source ranks & $0.0009$ & $[-0.0137,$ & $0.0147]$ & $0.0154$ & $[0.0074,$ & $0.0243]$ \\
ADMET & K20 top-15 & $-0.0080$ & $[-0.0277,$ & $0.0083]$ & $0.0037$ & $[-0.0056,$ & $0.0119]$ \\
\midrule
Joint balanced & all-source ranks & $-0.0014$ & $[-0.0173,$ & $0.0147]$ & $0.0018$ & $[-0.0061,$ & $0.0090]$ \\
Joint balanced & K20 top-15 & $0.0124$ & $[-0.0053,$ & $0.0298]$ & $-0.0023$ & $[-0.0093,$ & $0.0043]$ \\
\midrule
Joint binding-heavy & all-source ranks & $-0.0040$ & $[-0.0168,$ & $0.0117]$ & $0.0132$ & $[0.0028,$ & $0.0240]$ \\
Joint binding-heavy & K20 top-15 & $-0.0062$ & $[-0.0180,$ & $0.0046]$ & $0.0083$ & $[0.0008,$ & $0.0175]$ \\
\midrule
Joint ADMET-heavy & all-source ranks & $0.0067$ & $[0.0011,$ & $0.0123]$ & $-0.0009$ & $[-0.0036,$ & $0.0019]$ \\
Joint ADMET-heavy & K20 top-15 & $0.0032$ & $[-0.0015,$ & $0.0085]$ & $-0.0024$ & $[-0.0086,$ & $0.0042]$ \\
\bottomrule
\end{tabular}
\caption{Learned selector minus K20 rank aggregation in first-choice regret $\Delta R_1$ and Regret@5 $\Delta R_5$, where negative values favor ridge. Intervals are descriptive 95\% paired held-out-fold block bootstrap intervals.}
\label{tab:learned_paired}
\end{table}

\begin{table}[t]
\centering\scriptsize\setlength{\tabcolsep}{3pt}
\begin{tabular}{@{}lrrrlrrl@{}}
\toprule
 & & \multicolumn{3}{c}{Direct} & \multicolumn{3}{c}{ICL} \\
\cmidrule(lr){3-5}\cmidrule(lr){6-8}
Setting & $n$ & $R_1$ & $R_5$ & $G$ [95\% interval] & $R_1$ & $R_5$ & $G$ [95\% interval] \\
\midrule
BindingDB & 100 & 0.416 & 0.279 & 0.137 [0.126, 0.151] & 0.293 & 0.106 & 0.187 [0.160, 0.215] \\
ADMET & 34 & 0.234 & 0.124 & 0.111 [0.099, 0.121] & 0.113 & 0.046 & 0.067 [0.042, 0.091] \\
Joint, balanced & 100 & 0.256 & 0.142 & 0.113 [0.106, 0.123] & 0.116 & 0.053 & 0.063 [0.053, 0.074] \\
Joint, binding-heavy & 100 & 0.264 & 0.136 & 0.128 [0.116, 0.140] & 0.111 & 0.050 & 0.061 [0.051, 0.071] \\
Joint, ADMET-heavy & 100 & 0.272 & 0.154 & 0.118 [0.108, 0.129] & 0.085 & 0.046 & 0.040 [0.035, 0.045] \\
\bottomrule
\end{tabular}
\caption{First-choice regret $R_1$, Regret@5 $R_5$ and ordering gap $G=R_1-R_5$ for direct prediction and ICL. Values are means across targets; brackets, where shown, are descriptive 95\% held-out-fold intervals.}
\label{tab:ordering_decomposition}
\end{table}

ICL has lower mean Regret@5 than Global Best in 14 of 20 model-setting comparisons and K20 rank aggregation in eight; its first-choice regret is lower in three and one, respectively. Grouping identical recorded utilities leaves regret unchanged and preserves ICL's better best-list rank than direct prediction in all 20 comparisons. Gains vary non-monotonically with source similarity: BindingDB first-choice-regret reductions are 0.114, 0.094 and 0.164 across low, middle and high nearest-source-similarity groups. ADMET reductions average 0.143 on binary and 0.111 on continuous endpoints. Supplementary Source Data provides subgroup definitions, counts and paired comparisons.

\subsection{Fixed-List Ordering Controls}
\label{app:ordering_controls}
\label{sec:ordering_controls}
The controls in \Cref{tab:same_list_rerank} keep each five-entry list fixed. Uniform ordering has expected first-choice regret equal to the mean regret of its entries. Source reranking sorts entries by their summed positions in the same 20 source top-15 lists used by ICL, assigning absent entries rank 16 and breaking ties by catalog order. Target-fold outcomes are excluded; invalid lists retain regret one.

\begin{table}[t]
\centering\scriptsize\setlength{\tabcolsep}{1.8pt}
\begin{tabular}{@{}l@{\hspace{8pt}}rrrrrrr@{\hspace{8pt}}rrrrrrr@{}}
\toprule
 & \multicolumn{7}{c}{\textbf{BindingDB}} & \multicolumn{7}{c}{\textbf{ADMET}} \\
\cmidrule(lr){2-8}\cmidrule(lr){9-15}
Model & Orig. & Unif. & Source & Best5 & \multicolumn{3}{c}{$\Delta$ [95\% interval]} & Orig. & Unif. & Source & Best5 & \multicolumn{3}{c}{$\Delta$ [95\% interval]} \\
\midrule
Q9 & 0.340 & 0.353 & 0.318 & 0.140 & $-0.022$ & $[-0.052,$ & $0.009]$ & 0.132 & 0.186 & 0.114 & 0.062 & $-0.018$ & $[-0.058,$ & $0.011]$ \\
Q27 & 0.280 & 0.273 & 0.249 & 0.097 & $-0.031$ & $[-0.080,$ & $0.015]$ & 0.105 & 0.154 & 0.101 & 0.035 & $-0.004$ & $[-0.032,$ & $0.021]$ \\
Q35 & 0.291 & 0.276 & 0.270 & 0.095 & $-0.022$ & $[-0.078,$ & $0.031]$ & 0.094 & 0.149 & 0.088 & 0.039 & $-0.006$ & $[-0.031,$ & $0.017]$ \\
G4 & 0.260 & 0.283 & 0.263 & 0.091 & $0.003$ & $[-0.038,$ & $0.047]$ & 0.119 & 0.158 & 0.092 & 0.047 & $-0.027$ & $[-0.063,$ & $0.007]$ \\
\midrule
 & \multicolumn{7}{c}{\textbf{Joint balanced}} & \multicolumn{7}{c}{\textbf{Joint binding-heavy}} \\
\cmidrule(lr){2-8}\cmidrule(lr){9-15}
Model & Orig. & Unif. & Source & Best5 & \multicolumn{3}{c}{$\Delta$ [95\% interval]} & Orig. & Unif. & Source & Best5 & \multicolumn{3}{c}{$\Delta$ [95\% interval]} \\
\midrule
Q9 & 0.148 & 0.184 & 0.139 & 0.087 & $-0.008$ & $[-0.019,$ & $0.004]$ & 0.130 & 0.167 & 0.124 & 0.066 & $-0.006$ & $[-0.027,$ & $0.014]$ \\
Q27 & 0.109 & 0.103 & 0.094 & 0.042 & $-0.015$ & $[-0.027,$ & $-0.003]$ & 0.108 & 0.114 & 0.095 & 0.048 & $-0.012$ & $[-0.025,$ & $-0.002]$ \\
Q35 & 0.101 & 0.100 & 0.094 & 0.043 & $-0.007$ & $[-0.019,$ & $0.003]$ & 0.105 & 0.117 & 0.091 & 0.044 & $-0.014$ & $[-0.022,$ & $-0.007]$ \\
G4 & 0.104 & 0.099 & 0.095 & 0.040 & $-0.009$ & $[-0.024,$ & $0.007]$ & 0.099 & 0.107 & 0.096 & 0.041 & $-0.004$ & $[-0.016,$ & $0.010]$ \\
\midrule
 & \multicolumn{7}{c}{\textbf{Joint ADMET-heavy}} \\
\cmidrule(lr){2-8}
Model & Orig. & Unif. & Source & Best5 & \multicolumn{3}{c}{$\Delta$ [95\% interval]} \\
\midrule
Q9 & 0.155 & 0.188 & 0.141 & 0.110 & $-0.014$ & $[-0.039,$ & $0.003]$ \\
Q27 & 0.061 & 0.067 & 0.054 & 0.026 & $-0.008$ & $[-0.014,$ & $-0.002]$ \\
Q35 & 0.061 & 0.067 & 0.055 & 0.024 & $-0.006$ & $[-0.013,$ & $0.000]$ \\
G4 & 0.064 & 0.068 & 0.052 & 0.023 & $-0.012$ & $[-0.020,$ & $-0.004]$ \\
\bottomrule
\end{tabular}
\caption{Same-list ordering controls for ICL. Intervals are descriptive 95\% paired held-out-fold block bootstrap intervals. Q9, Q27, Q35 and G4 denote the four language models of \Cref{sec:llm_study}.}
\label{tab:same_list_rerank}
\end{table}

Original ICL order has lower mean first-choice regret than uniform ordering in 15 of 20 comparisons. Uniform ordering has lower means in the other five, whose paired intervals include zero. Five source-reranking differences have descriptive 95\% intervals wholly below zero. Supplementary Source Data also includes direct-prediction and uniform-order paired comparisons.

\subsection{Selection across Campaign Splits}
\label{app:seed_stability}
For each target, two split seeds select the highest-utility configuration from the catalog or an existing shortlist; the third evaluates it. We rotate the held-out seed and average within target before resampling folds. Regret uses each held-out split's catalog range, whereas primary recommendation scores use three-seed-averaged utility. Joint diagnostics retain the 77-configuration catalog.

\Cref{tab:seed_oracle_stability} compares the two-seed catalog winner and, averaging the four ICL models within target, the original first entry, held-out-best list entry and two-seed-selected list entry. The two-seed gain is the original first-entry regret minus the two-seed-selected regret. \Cref{tab:seed_selector_comparison} evaluates source-based and learned selectors' original predictions under the same held-out-seed scoring.

Near-optimal sets have normalized regret at most $\epsilon\in\{0.01,0.05,0.10\}$ under each split's own catalog range. \Cref{tab:seed_near_set_sensitivity} reports Jaccard overlap, the fraction of two-seed near-optimal configurations that remain near-optimal (Transfer), and the fraction of rotations in which the two-seed catalog winner is near-optimal (Near-hit).

\begin{table}[t]
\centering\scriptsize\setlength{\tabcolsep}{3pt}
\begin{tabular}{@{}lrrrrrr@{}}
\toprule
Setting & \shortstack{Two-seed\\winner} & \shortstack{Near-set\\Jaccard} & First & \shortstack{Hindsight\\five} & \shortstack{Two-seed\\five} & \shortstack{Two-seed\\gain} \\
\midrule
BindingDB & 0.236 & 0.152 & 0.334 & 0.127 & 0.265 & 0.070 \\
ADMET & 0.145 & 0.187 & 0.149 & 0.071 & 0.131 & 0.018 \\
Joint balanced & 0.104 & 0.191 & 0.145 & 0.065 & 0.121 & 0.024 \\
Joint binding-heavy & 0.102 & 0.192 & 0.140 & 0.063 & 0.118 & 0.021 \\
Joint ADMET-heavy & 0.089 & 0.260 & 0.114 & 0.061 & 0.101 & 0.013 \\
\bottomrule
\end{tabular}
\caption{Campaign-seed stability of empirical choices. Values are means across targets; brackets, where shown, are descriptive 95\% held-out-fold intervals.}
\label{tab:seed_oracle_stability}
\end{table}

\begin{table}[t]
\centering\scriptsize\setlength{\tabcolsep}{3pt}
\begin{tabular}{@{}lrrrrrr@{}}
\toprule
Setting & \shortstack{Global\\Best} & \shortstack{K20\\aggregate} & \shortstack{Ridge\\all} & \shortstack{Ridge\\K20} & Direct & ICL \\
\midrule
BindingDB & 0.330 & 0.313 & 0.296 & 0.287 & 0.440 & 0.334 \\
ADMET & 0.137 & 0.130 & 0.131 & 0.123 & 0.259 & 0.149 \\
Joint balanced & 0.115 & 0.121 & 0.121 & 0.133 & 0.270 & 0.145 \\
Joint binding-heavy & 0.125 & 0.127 & 0.124 & 0.123 & 0.280 & 0.140 \\
Joint ADMET-heavy & 0.086 & 0.080 & 0.087 & 0.083 & 0.282 & 0.114 \\
\bottomrule
\end{tabular}
\caption{First-choice regret under campaign-seed scoring. Values are means across targets; brackets, where shown, are descriptive 95\% held-out-fold intervals.}
\label{tab:seed_selector_comparison}
\end{table}

\begin{table}[t]
\centering\scriptsize\setlength{\tabcolsep}{3pt}
\begin{tabular}{@{}lrrrrrrrrr@{}}
\toprule
 & \multicolumn{3}{c}{$\epsilon=0.01$} & \multicolumn{3}{c}{$\epsilon=0.05$} & \multicolumn{3}{c}{$\epsilon=0.10$} \\
\cmidrule(lr){2-4}\cmidrule(lr){5-7}\cmidrule(lr){8-10}
Setting & Jaccard & Transfer & Near-hit & Jaccard & Transfer & Near-hit & Jaccard & Transfer & Near-hit \\
\midrule
BindingDB & 0.154 & 0.168 & 0.163 & 0.152 & 0.219 & 0.233 & 0.214 & 0.315 & 0.370 \\
ADMET & 0.098 & 0.135 & 0.157 & 0.187 & 0.297 & 0.333 & 0.314 & 0.453 & 0.461 \\
Joint balanced & 0.097 & 0.139 & 0.147 & 0.191 & 0.302 & 0.350 & 0.383 & 0.511 & 0.593 \\
Joint binding-heavy & 0.075 & 0.100 & 0.107 & 0.192 & 0.303 & 0.327 & 0.370 & 0.518 & 0.580 \\
Joint ADMET-heavy & 0.076 & 0.107 & 0.100 & 0.260 & 0.360 & 0.377 & 0.508 & 0.613 & 0.697 \\
\bottomrule
\end{tabular}
\caption{Near-optimal-set stability at three regret thresholds $\epsilon$. Values are means across targets; brackets, where shown, are descriptive 95\% held-out-fold intervals.}
\label{tab:seed_near_set_sensitivity}
\end{table}

The two-seed catalog winner has mean held-out regret 0.236 on BindingDB versus 0.330 for Global Best. Global Best has slightly lower means on ADMET and ADMET-heavy joint selection. Near-optimal-set Jaccard overlap at $\epsilon=0.05$ ranges from 0.152 to 0.260 across settings.

\subsection{Catalog Sensitivity}
\label{app:catalog_sensitivity}

\begin{table}[t]
\centering\scriptsize
\setlength{\tabcolsep}{3pt}
\begin{tabular}{@{}l@{\hspace{10pt}}r@{\hspace{10pt}}rrr@{\hspace{10pt}}rrr@{}}
\toprule
Model & \shortstack{First excluded\\(Direct/ICL)} & \multicolumn{3}{c}{$\Delta$ first retained regret} & \multicolumn{3}{c}{$\Delta$ best retained regret} \\
\cmidrule(lr){3-5}\cmidrule(lr){6-8}
\midrule
\multicolumn{8}{l}{\textit{BindingDB}} \\
G4 & 0/52 & $0.046$ & $[0.013,$ & $0.076]$ & $0.070$ & $[0.036,$ & $0.099]$ \\
Q9 & 100/80 & $-0.390$ & $[-0.473,$ & $-0.306]$ & $-0.419$ & $[-0.509,$ & $-0.327]$ \\
Q35 & 0/52 & $-0.011$ & $[-0.054,$ & $0.040]$ & $0.016$ & $[-0.029,$ & $0.061]$ \\
Q27 & 0/49 & $-0.508$ & $[-0.565,$ & $-0.453]$ & $-0.434$ & $[-0.499,$ & $-0.368]$ \\
\addlinespace[2pt]
\multicolumn{8}{l}{\textit{ADMET}} \\
G4 & 0/1 & $-0.065$ & $[-0.105,$ & $-0.019]$ & $-0.077$ & $[-0.108,$ & $-0.043]$ \\
Q9 & 34/3 & $-0.816$ & $[-0.892,$ & $-0.743]$ & $-0.887$ & $[-0.963,$ & $-0.808]$ \\
Q35 & 0/2 & $-0.139$ & $[-0.176,$ & $-0.097]$ & $-0.081$ & $[-0.108,$ & $-0.056]$ \\
Q27 & 0/2 & $-0.137$ & $[-0.171,$ & $-0.097]$ & $-0.054$ & $[-0.081,$ & $-0.029]$ \\
\addlinespace[2pt]
\multicolumn{8}{l}{\textit{Joint balanced}} \\
G4 & 0/2 & $0.013$ & $[-0.004,$ & $0.035]$ & $-0.017$ & $[-0.030,$ & $-0.008]$ \\
Q9 & 100/9 & $-0.771$ & $[-0.833,$ & $-0.712]$ & $-0.845$ & $[-0.906,$ & $-0.781]$ \\
Q35 & 0/1 & $-0.086$ & $[-0.126,$ & $-0.047]$ & $-0.054$ & $[-0.070,$ & $-0.038]$ \\
Q27 & 0/3 & $-0.153$ & $[-0.178,$ & $-0.130]$ & $-0.031$ & $[-0.045,$ & $-0.016]$ \\
\addlinespace[2pt]
\multicolumn{8}{l}{\textit{Joint binding-heavy}} \\
G4 & 0/2 & $-0.007$ & $[-0.042,$ & $0.026]$ & $-0.015$ & $[-0.030,$ & $0.000]$ \\
Q9 & 100/7 & $-0.822$ & $[-0.857,$ & $-0.788]$ & $-0.889$ & $[-0.923,$ & $-0.852]$ \\
Q35 & 3/3 & $-0.248$ & $[-0.297,$ & $-0.195]$ & $-0.094$ & $[-0.113,$ & $-0.076]$ \\
Q27 & 0/5 & $-0.141$ & $[-0.163,$ & $-0.118]$ & $-0.024$ & $[-0.040,$ & $-0.010]$ \\
\addlinespace[2pt]
\multicolumn{8}{l}{\textit{Joint ADMET-heavy}} \\
G4 & 0/1 & $-0.023$ & $[-0.045,$ & $-0.003]$ & $-0.041$ & $[-0.061,$ & $-0.025]$ \\
Q9 & 100/2 & $-0.803$ & $[-0.854,$ & $-0.746]$ & $-0.860$ & $[-0.911,$ & $-0.804]$ \\
Q35 & 5/0 & $-0.096$ & $[-0.130,$ & $-0.065]$ & $-0.052$ & $[-0.074,$ & $-0.030]$ \\
Q27 & 0/2 & $-0.169$ & $[-0.197,$ & $-0.145]$ & $-0.034$ & $[-0.047,$ & $-0.024]$ \\
\bottomrule
\end{tabular}
\caption{Sensitivity of the original recommendation lists after excluding the twelve XGB leaf-index hybrids from the catalog. Differences are ICL minus Direct, so negative values favor ICL, and brackets are paired descriptive 95\% fold-block intervals. Q9, Q27, Q35 and G4 denote the four language models of \Cref{sec:llm_study}.}
\label{tab:catalog_sensitivity}
\end{table}

\Cref{tab:catalog_sensitivity} removes the twelve XGB leaf-statistic configurations, leaving catalogs of 45 BindingDB or 65 ADMET/joint entries. Original lists are filtered without replacement, retaining zero to five entries; regret uses the reduced catalog's utility range. Invalid or empty lists receive regret one. This evaluates filtered predictions, not recommendations generated from a reduced catalog.

The excluded configurations account for every original utility maximum on 44/100 BindingDB, 2/34 ADMET, 3/100 balanced, 6/100 binding-heavy and 1/100 ADMET-heavy targets. ICL has lower best-retained-entry regret than direct prediction in 18/20 comparisons; the exceptions are Gemma-4-26B-A4B-it and Qwen3.6-35B-A3B on BindingDB. First-excluded counts record removal of the original first choice. 

\section{Limitations and Responsible Use}
\label{app:limitations}

Configuration rankings depend on the dataset collection, representations, budgets and source-target split. Predicted XTT scores concern one cell-viability assay; therapeutic safety requires separate assessment. Prospective campaigns also involve assay noise, synthesis constraints and project-specific costs, while measured ADME desirability depends on the endpoint's screening objective.

\section{Data and Code Availability}
\label{app:availability}
The five supplementary ZIP archives are available for review at \supplementlink{}. We will replace this review link with a permanent repository link upon acceptance.

Supplementary Inputs contains the candidate tables. Supplementary Raw Results contains campaign archives, recommendation records and associated metadata. Supplementary Source Data supplies figure/table inputs and generation routines. Supplementary Reproduction provides study-specific simulation, recommendation and analysis workflows; Supplementary Software provides the reusable \sciselect{} library, documentation, examples and tests.

To regenerate figures and tables or recompute the supplied numerical analyses, start with Supplementary Software, Supplementary Reproduction and Supplementary Source Data. Supplementary Inputs supports campaign reconstruction, and Supplementary Raw Results supports reconstruction of campaign metrics and rescoring of saved recommendations. The raw-results archive contains separately compressed campaign parts that can be read without extracting all simulation records. Package documentation specifies dependencies and the distinction between analysis of saved results and fresh simulation runs.

Project-owned software is distributed under the MIT licence; author-generated campaign records, utilities, recommendation records and metadata use CC BY 4.0. Third-party datasets and models retain their original terms.

\clearpage
\bibliography{gym}
\end{document}